\documentclass[12pt]{article}

\newcommand{\blambda}{\boldsymbol{\lambda}}
\newcommand{\balpha}{\boldsymbol{\alpha}}
\newcommand{\bLambda}{\boldsymbol{\Lambda}}
\newcommand{\boeta}{\boldsymbol{\eta}}
\newcommand{\bPhi}{\boldsymbol{\Phi}}
\newcommand{\bmu}{\boldsymbol{\mu}}
\newcommand{\bSigma}{\boldsymbol{\Sigma}}

\newcommand{\btheta}{\boldsymbol{\theta}}
\newcommand{\bxi}{\boldsymbol{\xi}}

\newcommand{\Cov}{\mr{Cov}}
\newcommand{\bL}{\mathbf{L}}
\newcommand{\bbR}{\mathbb{R}}
\newcommand{\what}[1]{\widehat{#1}}
\newcommand{\new}[1]{#1^{new}}
\newcommand{\cur}[1]{#1^{cur}}
\newcommand{\prop}[1]{#1^{prop}}
\newcommand{\bA}{\mathbf{A}}

\newcommand{\bF}{\mathbf{F}}
\newcommand{\bM}{\mathbf{M}}
\newcommand{\bom}{\mathbf{m}}

\newcommand{\bW}{\mathbf{W}}
\newcommand{\bN}{\mathbf{N}}

\newcommand{\bNtil}{\widetilde{\bN}}

\newcommand{\R}{\mathtt{R}}
\newcommand{\bn}{\mathbf{n}}
\newcommand{\bntil}{\widetilde{\bn}}

\newcommand{\bX}{\mathbf{X}}

\newcommand{\bx}{\mathbf{x}}
\newcommand{\bY}{\mathbf{Y}}
\newcommand{\by}{\mathbf{Y}}
\newcommand{\bZ}{\mathbf{Z}}
\newcommand{\bU}{\mathbf{U}}

\newcommand{\bD}{\mathbf{D}}
\newcommand{\bp}{\mathbf{p}}

\newcommand{\bP}{\mathbf{P}}
\newcommand{\bz}{\mathbf{z}}
\newcommand{\e}{\mathrm{e}}
\newcommand{\E}{\mathrm{E}}

\newcommand{\bs}[1]{\boldsymbol{#1}}

\newcommand{\mcI}{\mathcal{I}}
\newcommand{\mcS}{\mathcal{S}}

\newcommand{\mcV}{\mathcal{V}}

\newcommand{\bV}{\mathbf{V}}
\newcommand{\mr}[1]{\mathrm{#1}}
\newcommand{\mb}[1]{\mathbf{#1}}
\newcommand{\diag}{\mr{diag}}
\newcommand{\rmd}{\mr{d}}
\newcommand{\dt}{\rmd t}
\newcommand{\logit}{\mr{logit}}

\newcommand{\pdiv}[2]{\frac{\partial#1}{\partial#2}}
\newcommand{\doLNA}{\texttt{doLNA}}

\newcommand{\doElliptSS}{\texttt{doElliptSS}}
\newcommand{\ind}[1]{\mathds{1}_{\left \lbrace#1\right \rbrace}}

\usepackage{fancyhdr}
\usepackage[noend]{algpseudocode}
\usepackage{algorithm}
\usepackage{amsmath}       
\usepackage{amsfonts}      
\usepackage{authblk}
\usepackage{booktabs}
\usepackage[margin=1in]{geometry}
\usepackage{dsfont}        
\usepackage{nicefrac}
\usepackage[font=small,margin=0.5in,labelfont=bf,tableposition=top]{caption}
\DeclareCaptionLabelFormat{andtable}{#1~#2  \&  \tablename~\thetable}
\usepackage{kbordermatrix}
\usepackage{float}
\usepackage{multirow}
\usepackage{hyperref}       
\usepackage[comma,authoryear]{natbib}
\usepackage{tikz}          
\usepackage{tikzscale}
\usetikzlibrary{arrows,shapes,positioning,external}
\usepackage{url}

\usepackage{xcolor}
\usepackage{ulem}
\newcommand\redsout{\bgroup\markoverwith{\textcolor{red}{\rule[0.5ex]{2pt}{1.4pt}}}\ULon}

\newcommand\bluesout{\bgroup\markoverwith{\textcolor{blue}{\rule[0.5ex]{2pt}{1.4pt}}}\ULon}

\newcommand\greensout{\bgroup\markoverwith{\textcolor{green}{\rule[0.5ex]{2pt}{1.4pt}}}\ULon}

\newcommand{\cmmnt}[1]{\ignorespaces}

\usepackage{etoolbox}

\makeatletter

\@ifundefined{protected@iwrite}{%
	\let\protected@iwrite\protected@write
	\patchcmd{\protected@iwrite}{\write}{\immediate\write}{}{}%
	\def\efloat@iwrite#1{\expandafter\protected@iwrite\csname efloat@post#1\endcsname{}}%
}{}

\appto{\appendix}{%
	\setcounter{postfigure}{0}%
	\efloat@iwrite{fff}{%
		\unexpanded{\unexpanded{%
				\renewcommand{\thefigure}{S\arabic{figure}}^^J%
				\setcounter{figure}{0}^^J%
		}}%
	}%
	\setcounter{posttable}{0}%
	\efloat@iwrite{ttt}{%
		\unexpanded{\unexpanded{%
				\renewcommand{\thetable}{S\arabic{table}}^^J%
				\setcounter{table}{0}^^J%
		}}%
	}%
}

\usepackage[figuresright]{rotating}

\begin{document}
\def\spacingset#1{\renewcommand{\baselinestretch}%
{#1}\small\normalsize} \spacingset{1}


\begin{center}
{\LARGE A Linear Noise Approximation for Stochastic Epidemic Models Fit to Partially Observed Incidence Counts}\\   
\ \\
Jonathan Fintzi\textsuperscript{1},
Jon Wakefield\textsuperscript{2,*},
Vladimir N. Minin\textsuperscript{3,*}\\
{\footnotesize
\textsuperscript{1}Biostatistics Research Branch, National Institute of Allergy and Infectious Diseases, Rockville, Maryland, U.S.A.\\
\textsuperscript{2}Departments of Biostatistics and Statistics, University of Washington, Seattle, Washington, U.S.A.\\
\textsuperscript{3}Department of Statistics, University of California, Irvine, California, U.S.A.\\
\textsuperscript{*}corresponding authors: \url{jonno@uw.edu} and \url{vminin@uci.edu}}
\end{center}
%
%

\renewcommand{\abstractname}{Abstract}
\begin{abstract}
	Stochastic epidemic models (SEMs) fit to incidence data are critical to elucidating outbreak dynamics, shaping response strategies, and preparing for future epidemics. SEMs typically represent counts of individuals in discrete infection states using Markov jump processes (MJPs), but are computationally challenging as imperfect surveillance, lack of subject--level information, and temporal coarseness of the data obscure the true epidemic. Analytic integration over the latent epidemic process is impossible, and integration via Markov chain Monte Carlo (MCMC) is cumbersome due to the dimensionality and discreteness of the latent state space. Simulation--based computational approaches can address the intractability of the MJP likelihood, but are numerically fragile and prohibitively expensive for complex models. A linear noise approximation (LNA) that approximates the MJP transition density with a Gaussian density has been explored for analyzing prevalence data in large--population settings, but requires modification for analyzing incidence counts without assuming that the data are normally distributed. We demonstrate how to reparameterize SEMs to appropriately analyze incidence data, and fold the LNA into a data augmentation MCMC framework that outperforms deterministic methods, statistically, and simulation-based methods, computationally. Our framework is computationally robust  when the model dynamics are complex and applies to a broad class of SEMs.  We evaluate our method in simulations that reflect Ebola, influenza, and SARS-CoV-2 dynamics, and apply our method to national surveillance counts from the 2013--2015 West Africa Ebola outbreak.
\end{abstract}

\noindent%

\spacingset{1.0}

\section{Introduction}
\label{sec:intro}

Incidence count data reported by public health surveillance systems help us to understand outbreak transmission dynamics, quantify uncertainty about the trajectory of an outbreak, and design interventions to interrupt transmission. Incidence counts reflect the number of new cases accumulated in each inter-observation interval, and are distinct from prevalence counts, which record the number of infected individuals at each observation time. Imperfect surveillance and asymptomatic cases result in systematic under-reporting and make it difficult to disentangle whether the data arose from a severe outbreak, observed with low fidelity, or a mild outbreak where most cases were detected. The absence of subject-level data and the temporal coarseness of observations further reduce the amount of available information. 

Outbreak data are commonly modeled using mechanistic compartmental models in which individuals transition between discrete infection states \citep{allen2017primer}. The parameters that govern the epidemic dynamics are of interest because they are informative about the mechanistic aspects of disease transmission. For example, the basic and effective reproduction numbers, $R_0$ and $R_{eff}$, are measures of the expected number of secondary cases per index case. These quantities, respectively, inform us about the likelihood that an outbreak will take off and persist, and provide insights into how interventions might short circuit transmission since they depend on population susceptibility, infectiousness, and contact rates.

Stochastic epidemic models (SEMs) commonly  represent the epidemic as a density-dependent Markov jump process (MJP) that evolves on a discrete state space of compartment counts with times between elementary transition events, such as infections and recoveries, taken to be exponentially distributed \citep{allen2017primer}. The challenge in fitting a SEM to partially observed incidence is that we must sum over all epidemic paths from which the data could have arisen. This is difficult since the set of possible paths is enormous, even in small populations. Hence, the observed data likelihood of a SEM is usually intractable.

Computation and inference for SEMs in the presence of under-reporting typically relies on either simulation-based methods or data augmentation (DA), often in combination with an approximation of the target MJP \citep{oneill2010}. Simulation-based methods generate realizations of the epidemic process that form the basis for inference. This class of methods has been referred as ``plug-and-play" since inference requires only the ability to simulate epidemic paths \citep{breto2009time}. The particle marginal Metropolis-Hastings (PMMH) algorithm of \citep{andrieu2010particle} stands out as a state-of-the-art method for Bayesian inference with a fast and robust implementation for fitting epidemic models as part of the \texttt{pomp} \texttt{R} package \citep{pomp}. Despite their flexibility, simulation-based methods can be prohibitively expensive for models with complex dynamics and may fail entirely in the absence of an adequate model from which to simulate epidemic paths \citep{dukic2012}. 

Data augmentation facilitates Bayesian inference by targeting the joint posterior distribution of the latent epidemic process and SEM parameters. Modern DA algorithms can be more computationally robust than simulation-based methods, especially in the absence of subject-level data or when the SEM dynamics are complex \citep{pooley2015,fintzi2017efficient,nguyen2020stochastic}. However, repeatedly evaluating the MJP likelihood, which is a product of exponential waiting time densities, at each iteration of a Markov chain Monte Carlo (MCMC) algorithm is prohibitively expensive in large populations. Hence, it is advantageous to approximate the MJP with a process whose likelihood is more tractable. 


Markov jump processes are commonly approximated with ordinary and stochastic differential equations (ODEs, SDEs) (\citet{allen2017primer}; \hyperref[sec:weba]{Web Appendix A}). Deterministic ODE models, which are derived as the infinite-population functional limits of density dependent MJPs, are attractive for their computational tractability. These models also lend themselves to analytic characterizations of the outbreak, e.g., we can relate the final outbreak size to the basic reproductive number \mbox{\citep{miller2012note}}. However, deterministic models are known to underestimate uncertainty about the epidemic process \citep{king2015avoidable} and cannot address questions that are inherently stochastic, e.g., distributional questions about the size or duration of an outbreak. Although SDEs reasonably approximate the stochastic aspects of the MJP in large population settings, SDE transition densities are typically unavailable in closed form. Therefore, computation for SDE models relies on simulation based methods or simplification of the SEM.

We develop a computational framework for fitting SEMs to partially observed incidence based on a linear noise approximation (LNA) of MJP transition densities. The LNA approximates the MJP transition density with a Gaussian density whose moments are obtained by numerically solving systems of ODEs. Reviews of the LNA and other MJP approximations can be found in \citet{wilkinson2011stochastic} and \citet{schnoerr2017approximation}. The LNA has been applied in the analysis of gene regulatory networks \citep{komorowski2009, finkenstadt2013quantifying} and in outbreak modeling \citep{ross2009parameter, ross2012parameter, fearnhead2014}. The LNA has not been used to analyze incidence data. Rather, its application to outbreak modeling has been limited to prevalence or cumulative incidence data under an assumption of Gaussian disease counts \mbox{\citep{zimmer2017likelihood}}, or has relied on simulation-based methods to allow for non-Gaussian surveillance models \citep{golightly2015delayed}. 

Our contributions in this work are threefold. First, we demonstrate how SEMs should be parameterized, in general, to appropriately analyze incidence data. In doing so, we aim to correct an unfortunately common error in transmission modeling where incidence counts are wrongly conflated with prevalence data. Second, we demonstrate how the LNA can be folded into a computationally robust Bayesian data augmentation framework that leverages cutting edge MCMC algorithms and can be used to fit a broad class of SEMs. Critically, these tools absolve us of the \textit{de facto} requirement that the data follow a Gaussian distribution for the sake of computational efficiency, the alternative being computationally intensive particle filter methods. 
Finally, we provide guidance on optimal parameterizations that massively improve MCMC sampling efficiency for LNA paths and model parameters. 
In particular, we introduce a non-centered parameterization (NCP) for LNA transition densities that enables us to sample LNA paths via elliptical slice sampling \mbox{\citep{murray2010}}, which is a simple, computationally robust, and efficient MCMC algorithm free of tuning parameters. Our framework enables us to fit models with complex dynamics that would otherwise be impossible to fit without compromising the model, even with cutting edge computational tools such as particle MCMC.

\section{Stochastic Epidemic Models for Incidence Data}
\label{sec:model}

For clarity, we present the LNA framework in the context of fitting a susceptible-infected-recovered (SIR) model to negative binomial distributed incidence counts. The SIR model describes the transmission dynamics of an outbreak in a closed, homogeneously mixing population of $ P $ exchangeable individuals who are either susceptible $ (S) $, infected/infectious, $ (I) $, or recovered $ (R) $. The infection states relate to transmission, so, e.g., individuals transition from $ I $ to $ R $ when they no longer have infectious contact with others, not when they cease to experience symptoms of illness. Our framework can easily be generalized beyond the SIR model and used to fit more complex SEMs. In simulations, we will explore models with susceptible-exposed-infected-recovered (SEIR) dynamics, which have recently found widespread use in transmission models for SARS-CoV-2. These models add a compartment for latent infection where people are exposed but not yet infectious. In our application, we will fit a more complex multi-country model for the spread of Ebola in West Africa with country-specific SEIR dynamics and explicit cross-border transmission.

\subsection{Surveillance Model and Data}
\label{subsec:lna_measproc}

Incidence data, $ \bY = \lbrace Y_1,\dots,Y_L\rbrace $, are reported counts  of new infections in time intervals, $ \mcI = \lbrace\mcI_1,\dots,\mcI_L:\ \mcI_\ell = (t_{\ell-1},t_\ell]\rbrace $. The observed incidence may reflect a fraction of the true incidence due to imperfect surveillance or asymptomatic cases. Or, cases may be over-reported if non-specific symptoms lead to over-diagnosis. Let $ \bN^c = (N^c_{SI}, N^c_{IR}) $ denote the counting process for the cumulative numbers of infections ($ S\rightarrow I $ transitions) and recoveries ($ I\rightarrow R $ transitions), and let $ \Delta \bN^c(t_\ell) = \bN^c(t_\ell) - \bN^c(t_{\ell-1})$ denote the change in cumulative transitions over $ \mcI_\ell $; so, $ \Delta N^c_{SI}(t_\ell)$ is true unobserved incidence over $ (t_{\ell-1},t_\ell] $. Heterogeneities in case detection rates could lead to over-dispersion in the reporting distribution. Hence, we model the number of observed cases as a negative binomial sample of the true incidence with mean case detection rate $ \rho $ and over-dispersion parameter $ \phi $:
\begin{align}
	\label{eqn:incidence_emitprob}
	Y_\ell|\Delta N^c_{SI}(t_\ell),\rho,\phi &\sim \mr{Neg.Binom.}(\mu_\ell = \rho\Delta N^c_{SI}(t_\ell),\ \sigma^2_\ell = \mu_\ell + \mu_\ell^2/\phi).
\end{align}

\subsection{Latent Epidemic Process}
\label{subsec:lna_epid_proc}

The SIR model describes the evolution of compartment counts, $ \bX^c = \lbrace S^c,I^c,R^c\rbrace $, in continuous time on the state space $$ \mcS_X^c = \left \lbrace (l,m,n):l,m,n\in\lbrace0,\dots,P\rbrace,\ l+m+n=P\right \rbrace, $$ where $ P $ is the population size. We take the waiting times between state transitions to be exponentially distributed, which implies that $ \bX^c $ evolves according a MJP. Now, if our data had consisted of prevalence counts, we could approximate the MJP transition densities of $ \bX^c $ as in \citep{ross2009parameter, fearnhead2014}. However, incidence data reflect the new infections in each inter-observation interval as the surveillance model (\ref{eqn:incidence_emitprob}) depends on the change in $ N_{SI}^c $, not $ I $, over $ (t_{\ell-1},t_\ell] $. Hence, it is here that we first diverge from the approaches previously taken in the LNA literature. It would be incorrect to treat incidence as though it were a change in prevalence. For instance, there could be a positive number of infections but prevalence might not change due to an equal number of recoveries over the inter-observation interval. To correctly specify the likelihood for incidence data, we must construct the LNA that approximates transition densities of $ \bN^c $.

\subsection{Cumulative Incidence of Infections and Recoveries}
\label{subsec:lna_motivation}

The cumulative incidence of infections and recoveries, $ \bN^c $, is a MJP with state space $$ \mcS_N^c = \left \lbrace (j, k):j,k\in\lbrace0,\dots,P\rbrace,\bX(\bN^c(j,k)) \in \mcS_X\right \rbrace, $$ which is the set of cumulative infections and recoveries that do not lead to invalid prevalence paths (e.g., if there more recoveries than infections). Let $ \beta $ denote the per-contact infection rate, and $ \mu $ the recovery rate. The infinitesimal transition probability from state $ \bn $ to $ \bn^{\prime}$ is
\begin{equation}
	\label{eqn:lna_sir_rates}
	\mathrm{Pr}_{\bn,\bn^\prime}(\Delta t;\bX^c) = \left \lbrace \begin{array}{ll}
			\beta S I \Delta t + o(\Delta t), & \bn = (n_{SI},n_{IR}),\ 
			(n_{SI}+1,n_{IR}),\\
			\mu I\Delta t + o(\Delta t), &  \bn = (n_{SI},n_{IR}),\  \bn^\prime = (n_{SI},n_{IR}+1), \\
			1 - (\beta S I + \mu I)\Delta t + o(\Delta t), & \bn = \bn^\prime,\\
			o(\Delta t), & \text{otherwise},
		\end{array} \right.
\end{equation}
and we define the transition intensity from $ \bn $ to $ \bn^\prime $ as $ \blambda_{\bn,\bn^\prime}(\bX^c) = \lim_{\Delta t \rightarrow0}\mathrm{Pr}_{\bn,\bn^\prime}(\Delta t;\bX^c) $.

We seek to infer the posterior distribution of $ \bN^c $ and $ \btheta = \lbrace R_0, 1/\mu,\rho,\phi,\bX_0\rbrace$, where $ R_0 = \beta P /\mu $ is the basic reproduction number and $ 1/\mu $ is mean infectious period duration. The basic reproduction number takes the same form for the SEIR model, which also adds a mean latent period with duration $ 1/\gamma $, where $ \gamma $ is the rate at which an individual transitions from $ E $ to $ I $. By the Markov property and standard hidden Markov model conditional independence assumptions, the complete data likelihood is a product of surveillance densities and transition densities, 
\begin{align}
	\label{eqn:complete_data_likelihood}
	\pi(\btheta,\bN^c \mid \bY)\ &\propto\   L(\bY \mid \bN^c,\btheta)\pi(\bN^c \mid \btheta)\pi(\btheta)\nonumber\\
	&= \prod_{\ell=1}^L \left [ \Pr\left (Y_\ell \mid \Delta\bN_{SI}^c(t_\ell),\btheta\right ) \times \pi\left (\bN^c(t_\ell) \mid \bN^c(t_{\ell-1}),\btheta\right )\right ]\pi(\btheta),
\end{align}
where $ \pi(\btheta) $ is the prior distribution of $ \btheta $ and $ \bN(t_0) = \bs{0} $.
The challenge in sampling from this posterior is that the transition densities of $ \bN^c $ are intractable due to the dimensionality of its state space. Moreover, we cannot analytically integrate over the latent epidemic process, except in trivial cases that are of little practical interest. In the following subsections, we use the LNA to approximate transition densities of $ \bN^c $, turning (\ref{eqn:complete_data_likelihood}) into a more tractable product of Gaussian transition densities and arbitrary surveillance densities. This will facilitate the use of efficient algorithms for sampling from the approximate posterior. 

\subsection{Diffusion Approximation}
\label{subsec:diff_approx}

We approximate the integer-valued MJPs, $ \bX^c $ and $ \bN^c $, with the real-valued diffusion processes, $ \bX $ and $ \bN $, that satisfy an SDE, referred to as the chemical Langevin equation (CLE), whose drift and diffusion terms are chosen to match the approximate moments of MJP path increments in infinitesimal time intervals \citep{wilkinson2011stochastic,golightly2013simulation}. Additional details regarding the diffusion approximation are given in \hyperref[sec:weba]{Web Appendix A}, and we refer to \citet{gillespie2000chemical} and \citet{fuchs2013inference} for comprehensive discussions.

The state spaces of $ \bX $  and $ \bN $ for the SIR model respectively are
\begin{align*}
	\mcS_X^R &= \lbrace (l,m,n):l,m,n \in [0,P],\ l+m+n=P\rbrace,  \text{ and } \\
	\mcS_N^R &= \lbrace (j,k): j,k \in [0,P],\ \bX(\bN(j,k))\in\mcS_X^R \rbrace.
\end{align*}
In words, $ \mcS_X^R $ is the real-valued set of non-negative compartment volumes that sum to the population size, and $ \mcS_N^R $ is the real-valued set of non-decreasing and non-negative incidence paths, constrained to yield valid prevalence paths. Let $ \blambda(\bX(t)) $ be the vector of transition rates at time $ t $, e.g., $ \blambda(\bX(t)) =\left (\beta S(t) I(t), \mu I(t)\right ) $, and $ \bLambda(\bX(t)) = \diag(\blambda(t)) $. 
For now, we ignore the constraints on $ \mcS_N^R $ and $ \mcS_X^R $, and approximate changes in cumulative incidence of infections and recoveries in an infinitesimal time step with the CLE,
\begin{equation}
	\label{eqn:sir_cle_X}
	\rmd \bN(t) = \blambda(\bX(t))\dt + \bLambda(\bX(t))^{1/2}\rmd\bW_t, 
\end{equation}
where the vector $ \bW_t $ is distributed as a bivariate Brownian motion with independent components and  $ \bLambda(\bX(t))^{1/2} $ is any matrix square root of the diffusion matrix. 

Following \citet{breto2011compound} and \citet{ho2018direct}, we reparameterize $ \bX (t)$ in terms of $ \bN(t) $, with initial conditions $ \bX(t_0) = \bx_0 $ and $ \bN(t_0) = \bs{0} $. Let $ \bA $ denote the stoichiometric matrix whose rows specify changes in the compartment counts per elementary transition event, e.g., per infection or recovery:
\begin{equation}
	\label{eqn:sir_stoich}
	\bA = \kbordermatrix{& S & I &  R\\
		S\rightarrow I& -1& 1 & 0\\
		I \rightarrow R & 0& -1 & 1
	}.
\end{equation}
Now, $ \bX $ is coupled to $ \bN $ via 
\begin{equation}
	\label{eqn:incid2prev}
	\bX(t) = \bx_0 + \bA^T\bN(t).
\end{equation}
For the SIR model, 
\begin{align}
	\left (\begin{array}{c}
		S(t) \\
		I(t) \\
		R(t)
	\end{array}\right ) &= \left (\begin{array}{c}
		S_0 - N_{SI}(t) \\
		I_0 + N_{SI}(t) - N_{IR}(t) \\
		R_0 + N_{IR}(t)
	\end{array}\right ),
\end{align}
so we rewrite (\ref{eqn:sir_cle_X}) as
\begin{align}
	\label{eqn:sir_cle_N}
	\rmd\bN(t)&= \blambda(\bx_0 + \bA^T\bN(t))\dt + \bLambda(\bx_0 + \bA^T\bN(t))^{1/2}\rmd\bW_t.
\end{align}

Changes in compartment volumes affect transition rates, and hence increments in incident events, multiplicatively. Therefore, perturbations about the drift in (\ref{eqn:sir_cle_N}) should be symmetric on a multiplicative, not an additive scale. This leads us to log transform of (\ref{eqn:sir_cle_N}). Let $ \bNtil = \log(\bN + \bs{1})$, so $\bN = \exp(\bNtil) - \bs{1}$. By It\^{o}'s lemma \citep{oksendal2003stochastic}, the SDE for $ \bNtil $ is 
\begin{align}
	\rmd\bNtil(t) &= \underbrace{\diag\left (\exp(-\bNtil(t)) - 0.5\exp(-2\bNtil(t))\right )
		\hspace{0.0in}\blambda\left (\bx_0 + \bA^T(\exp(\bNtil(t))-\bs{1} )\right )}_{\boeta(t)}\dt\ + \nonumber\\
	&\hspace{0.5in}\left.\underbrace{\diag\left (\exp(-\bNtil(t))\right )\bLambda\left (\bx_0 + \bA^T(\exp(\bNtil(t))-\bs{1})\right )^{1/2}}_{\bPhi(t)^{1/2}}\right.\rmd\bW_t,
	\label{eqn:sir_log_cle_gen}
\end{align}
where $ \boeta(t) $ is the drift vector and $ \bPhi(t) $ is the diffusion matrix of the SDE.

\subsection{Linear Noise Approximation}
\label{subsec:sir_lna}

The LNA is a multivariate normal (MVN) approximation for the transition density of $ \bNtil $ over the interval $ (t_{\ell-1},t_\ell] $ and is given by
\begin{align}
	\label{eqn:lna_transition_density}
	\bNtil(t_\ell) \mid \bntil(t_{\ell-1}), \bx(t_{\ell-1}),\btheta &\sim MVN\left (\bmu(t_\ell) + \bom(\bntil(t_{\ell-1}) - \bmu(t_{\ell-1})), \bSigma(t_\ell)\right ),
\end{align}
where $ \bmu(\cdot) $, $ \bom(\cdot) $, and $ \bSigma(\cdot) $ are solutions to the coupled, non-autonomous system of ODEs,
\begin{align}
	\label{eqn:lna_ode_drift}
	\frac{\rmd\bmu(t)}{\dt} &= \boeta(\bmu(t)),\\
	\label{eqn:lna_ode_resid}
	\frac{\rmd\bom(t)}{\dt} &= \bF(t)\bom(t),\\
	\label{eqn:lna_ode_diffusion}
	\frac{\rmd\bSigma(t)}{\dt} &= \bF(t)\bSigma(t) + \bSigma(t)\bF(t)^T + \bPhi(t)^{1/2}\left (\bPhi(t)^{1/2}\right )^T,
\end{align}
subject to initial conditions $ \bN(t_{\ell-1}) = \bs{0} $,\ $ \bX(t_{\ell-1}) = \bx(t_{\ell-1}),\ \bom(t_{\ell-1}) = \bs{0}$, and $ \bSigma(t_{\ell-1}) = \bs{0} $, and where $ \boeta(t) $ and $ \bPhi(t) $ are given in (\ref{eqn:sir_log_cle_gen}) and $ \bF(t) = \left (\pdiv{\boeta_i(\bmu(t))}{\bmu_j(t)}\right )_{i,j\in{1,\dots,|\bNtil|}} $ is the Jacobian of $ \boeta(t) $. 
The solutions to \ref{eqn:lna_ode_drift})--(\ref{eqn:lna_ode_diffusion}) can be viewed as a mean vector that solves a nonlinear system of ODEs, an autoregressive term, and the covariance for a Gaussian approximation to the transition density of $ \bNtil $. In practice, we need not solve (\ref{eqn:lna_ode_resid}) since $ \bom(t_{\ell-1}) = \bs{0} $ implies the right hand side of (\ref{eqn:lna_ode_resid}) always equals zero.

The LNA is derived by writing $ \bNtil $ as the sum of its deterministic ODE limit, which is the solution to (\ref{eqn:lna_ode_drift}), and a stochastic residual. These terms are substituted into (\ref{eqn:sir_log_cle_gen}), which is then Taylor expanded around its deterministic limit, with higher order terms discarded, to obtain a linear SDE for the stochastic residual term. The solution of this linear SDE is a Gaussian random variable with transition density (\ref{eqn:lna_transition_density}). We refer readers to \mbox{\citet{golightly2013simulation}} and \mbox{\citet{wallace2010simplified}} for derivations of the LNA. The LNA reasonably approximates the stochastic aspects of a density dependent MJP over time intervals where the CLE approximation, (\ref{eqn:sir_log_cle_gen}), is valid (see \hyperref[sec:weba]{Web Appendix A} and \mbox{\citet{wallace2012linear}}). Over long time periods, the LNA may diverge from the MJP as stochastic perturbations to the system accumulate. The approximation can be improved by phase-correcting the LNA via a time-transformation \mbox{\citep{minas2017long}}, or by restarting the LNA at the start of each inter-observation interval and approximating the one-step conditional MJP transition densities \mbox{\citep{fearnhead2014}}. We have adopted the latter strategy for its simplicity. Under the restarting formulation of LNA, the initial conditions in (\ref{eqn:lna_ode_drift})--(\ref{eqn:lna_ode_diffusion}) are reset at the start of each inter-observation interval.

The LNA posterior factorizes as a Gaussian state space model,
\begin{align}
	\label{eqn:lna_approximate_posterior} 
	&\pi(\bNtil,\btheta \mid \bY) \propto L(\bY \mid \bNtil,\btheta)\pi(\bNtil \mid \btheta)\ind{\bN\in\mcS_N^R}\ind{\bX\in\mcS_X^R}\pi(\btheta) \nonumber\\
	&= \prod_{\ell=1}^{L}\Pr(Y_\ell \mid \Delta\bN(t_\ell),\btheta) \pi(\bNtil(t_\ell) \mid \bntil(t_{\ell-1}),\bx(t_{\ell-1}),\btheta) \times \ind{\bN(t_\ell)\in\mcS_N^R} \ind{\bX(t_\ell)\in\mcS_X^R} \pi(\btheta).
\end{align}
We exponentiate LNA sample paths when computing the surveillance densities in (\ref{eqn:lna_approximate_posterior}) since these densities depend on incidence, not log-incidence. We also explicitly include indicators for whether an LNA path respects the positivity and monotonicity constraints of the original MJP. We do this for two reasons: first, to more faithfully approximate the MJP, and second, to avoid numerical instabilities that arise when $ \bN $ or $ \bX $ are negative. 

\subsection{Inference via the Linear Noise Approximation}
\label{subsec:lna_inference}

We sample LNA paths using the elliptical slice sampling (ElliptSS) algorithm of \citet{murray2010}, which is an efficient and computationally robust MCMC algorithm free of tuning parameters. ElliptSS can be used to sample a latent variable of interest, $ \bZ $, when the posterior decomposes into a Gaussian prior for $ \bZ $ and an arbitrary likelihood, $ L(\bY|\bZ,\btheta) $, i.e.,
\begin{equation}
	\label{eqn:eliptss_posterior_decomp}
	\pi(\btheta,\bZ|\bY)\propto L(\bY|\bZ,\btheta)MVN(\bZ;\mu_\bZ,\bSigma_\bZ).
\end{equation}

ElliptSS cannot na\"{i}vely be used to sample LNA paths if we restart the LNA ODEs at the start of each inter-observation interval since the mean of $ \bN(t_\ell) $ depends non-linearly on the value of $ \bN(t_{\ell-1}) $. Therefore, paths of $ \bNtil $ are not jointly Gaussian. To facilitate the use of ElliptSS, we introduce a non-centered parameterization (NCP) that maps standard normal random variables onto LNA paths (see Algorithm \ref{alg:doLNA} for pseudo-code). Let $ \bZ(t_\ell)\sim MVN(\bs{0},\mb{I}) $, and $ \bNtil(t_\ell)\sim MVN\left (\bmu(t_\ell),\bSigma(t_\ell)\right ) $, where $ \bmu(t_\ell) $ and $ \bSigma(t_\ell) $ solve the LNA ODEs over $ (t_{\ell-1},t_\ell] $. The NCP is $ (\btheta,\bZ) $, and maps $ \bZ $ to $\bNtil(t_\ell)$ via $\bNtil(t_\ell) = \bmu(t_\ell) + \bSigma(t_\ell)^{1/2}\bZ(t_\ell)$.

Our MCMC targets the joint posterior of the parameters and non-centered LNA draws,
\begin{align}
	\label{eqn:lna_noncentered_posterior}
	\pi(\btheta,\bZ \mid \bY) &\propto L(\bY \mid \bN(\bZ,\btheta,\mcI)) \pi(\bZ)\ind{\bN(\bZ,\btheta,\mcI)\in\mcS_N^R}\ind{\bX(\bZ,\btheta,\mcI)\in\mcS_X^R}\pi(\btheta),
\end{align}
where $ \bN(\bZ,\btheta,\mcI) $ and $ \bX(\bZ,\btheta,\mcI) $ denote sample paths obtained by centering the LNA draws. 

The NCP also helps to alleviate issues of poor MCMC mixing that arise when fitting hierarchical latent variable models using a data augmentation MCMC framework (\hyperref[subsec:cp_vs_ncp]{Web Appendix B} and Figure \ref{fig:lna_combined_traces}). In weak data settings, such as ours, MCMC samples can become severely autocorrelated when alternately sampling latent variables and model parameters \citep{bernardo2003non,papaspiliopoulos2007general}. This can be traced to the centered parameterization (CP) for the LNA in (\ref{eqn:lna_approximate_posterior}).  Under the CP, updates to $ \btheta \mid \bNtil,\bY $ are made conditionally on a \textit{fixed} LNA path and are accepted if they are concordant with the data \textit{and} the current path. This dynamic limits the magnitude of perturbations that can be made to the model parameters at each MCMC iteration and results in severe autocorrelation. In contrast, the NCP locates a sample LNA path within the transition densities induced by a set of proposed model parameters, thus allowing us to make more meaningful perturbations to model parameters at each MCMC iteration.

\subsection{Parameter updates}
\label{subsec:lna_param_updates}
In each MCMC iteration, we alternate between sampling $ \bZ \mid \btheta,\bY $ and  $ \btheta \mid \bZ,\bY $. We sample $ \btheta \mid \bZ,\bY$ using either a global adaptive random walk Metropolis (GA-RWM) sampler when fitting simple SEMs, (Algorithm 4 in \citet{andrieu2008tutorial}) or an adaptive multivariate normal slice sampler (MVNSS) when the dynamics are more complex (\hyperref[subsec:mvnss]{Web Appendix B}). The GA-RWM algorithm is faster per-iteration, though we have found the MVNSS algorithm to be more robust in complex settings. When $ \bX_0 $, is estimated, we assign an informative prior $ \bX_0 \sim TMVN_{\mcS_X^R}\left (P\mb{p}, P(\diag(\mb{p}) - \mb{p}\mb{p}^T)\right ) $, which is a truncated multivariate normal (TMVN) approximation to a multinomial distribution on the state space of compartment volumes with initial state probabilities, $ \mb{p} $. We update $ \bX_0 $ via ElliptSS (\hyperref[subsec:lna_init_volumes]{Web Appendix B}).

\section{Simulation Studies}
\label{sec:lna_simulations}

\subsection{Comparison with Common MJP Approximations in Typical Settings}
\label{subsec:3dis_sim}

The fidelity of the LNA as a MJP approximation has been well established in more general settings \mbox{\citep{grima2012study, wallace2012linear}}. Here, we sought to evaluate the inferential performance of the LNA in the specific use-case that is the focus of this work: transmission models fit to incidence count data in medium- to large-population settings. We explored three scenarios where the outbreak dynamics were set to resemble pathogens of particular interest: influenza, Ebola, and SARS-CoV-2. The three simulation scenarios varied in their dynamics and detection rates (Table \ref{tab:threedis_setup}). The population sizes ranged from $ P=2,000 $, below which subject-level data requiring different statistical methods might reasonably be available, to $P=50,000$. As the population size increases, the outbreak dynamics tend to become increasingly deterministic \mbox{\citep{wallace2012linear}}. For each scenario, we fit SEIR models to weekly incidence data from 1,000 outbreaks simulated from a MJP via Gillespie's direct algorithm \mbox{\citep{gillespie1976general}}. Each dataset consisted of at least twelve weeks worth of data, and we truncated each time series after four consecutive weeks of fewer than five cases. Our priors were agnostic about surveillance model parameters and weakly informative about the transmission dynamics, though somewhat more diffuse than might be used in practice (Table \ref{tab:3dis_priors}). Additional details are provided in \hyperref[sec:3dis_supp]{Web Appendix C}.

We compared the LNA with the deterministic ODE representation of the latent epidemic process and with a discrete-time approximation where epidemic paths were simulated within a particle marginal Metropolis-Hastings (PMMH) framework \citep{andrieu2010particle} using a multinomial modification of the $ \tau $-leaping algorithm (MMTL) \citep{breto2011compound} and $ \tau $-leap intervals of either one hour or one day. The MMTL/PMMH approximation was chosen because of its implementation in the popular \texttt{pomp}  \texttt{R} package \citep{pompjss}. Given a sufficiently fine  $ \tau $-leap interval, the MMTL approximation is, arguably, a more faithful approximation to the target MJP vis-a-vis the LNA since it preserves the discreteness of the latent state space. The ODE approximation, which does not allow for stochasticity in the latent process, solves (\ref{eqn:lna_ode_drift}) with $ \bSigma(t) = 0 $ in (\ref{eqn:lna_ode_diffusion}), and has been ubiquitous in models for SARS-CoV-2 as it is easily implemented and not computationally burdensome.

\begin{table}[htbp]
	\caption{Setup for simulating outbreaks with dynamics similar to Ebola, influenza, and SARS-CoV-2.}
	\label{tab:threedis_setup}
	\footnotesize
	\centering
	\begin{tabular}{cccc}	
		& \textbf{Ebola}&\textbf{Influenza} &\textbf{SARS-CoV-2}\\
		\hline	
		Population size ($ P $) & 2,000 & 50,000 & 10,000 \\
		Basic reproduction number ($ R_0 $) & 1.8 & 1.3 & 2.5 \\
		Latent period ($ 1/\omega $) & 10 days & 2 days & 4 days \\
		Infectious period  ($ 1/\mu $) & 7 days & 2.5 days & 10 days \\
		Detection rate $ (\rho) $ & 0.5 & 0.02 & 0.1 \\
		Neg. bin. overdispersion $ (\phi) $ & 36 & 36 & 36\\
		\hline
	\end{tabular} 
\end{table}

\begin{figure}[htbp]
	\centering
	\includegraphics[width=\linewidth]{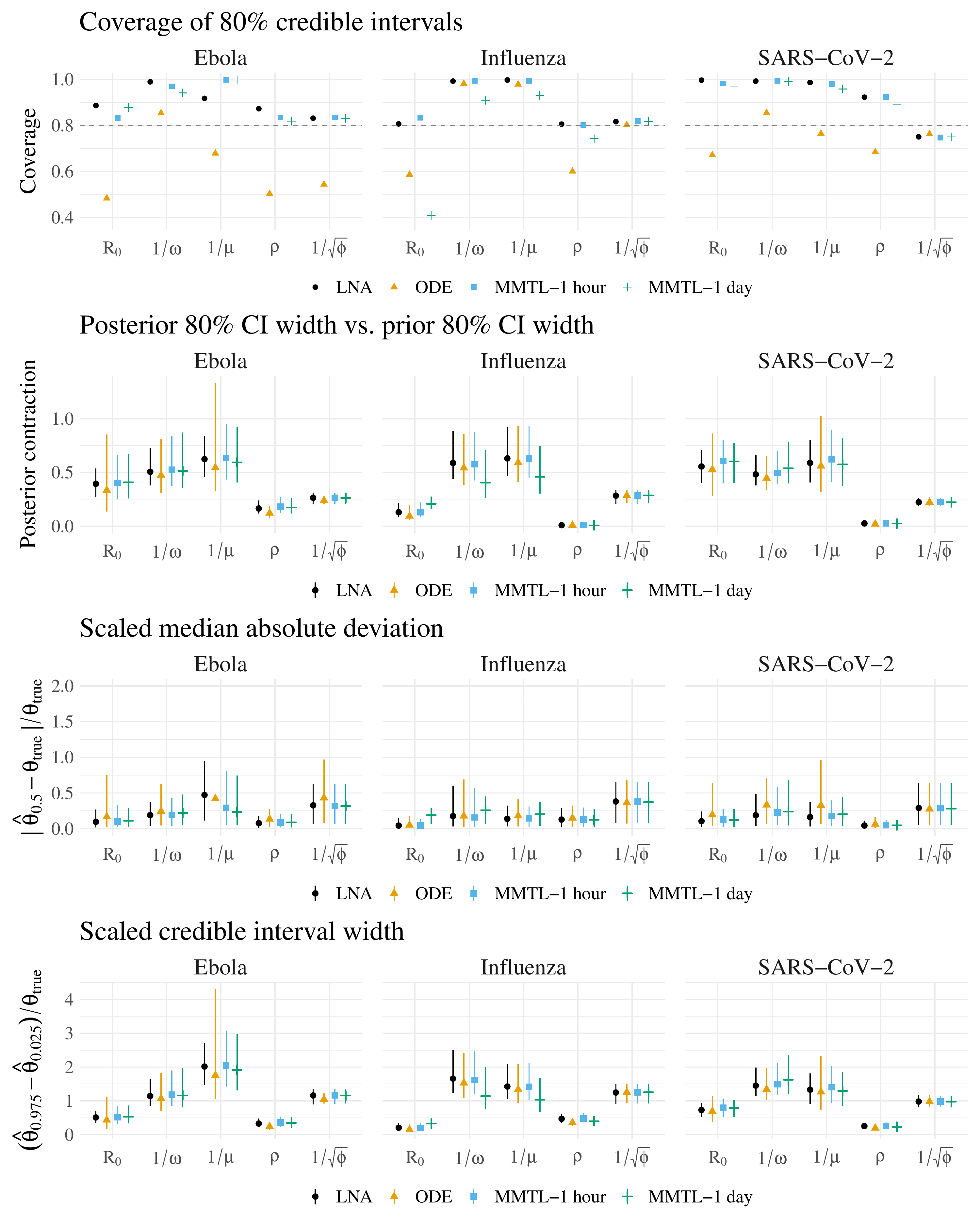}
	\caption{Simulation results for outbreaks simulated under Ebola-, influenza-, and SARS-CoV-2 dynamics in populations of size 2000, 100000, and 10000, respectively. $ R_0 $ is the basic reproduction number, $ 1/\omega $ in the mean latent period, $ 1/\mu $ is the mean infectious period, $ \rho $ is the mean case detection rate, and $ 1/\sqrt{\phi} $ is (loosely) the excess dispersion of the negative binomial surveillance model on the standard deviation scale.}
	\label{fig:threedis_res}
\end{figure}

Models fit via the LNA recovered the model parameters in all three scenarios (Figure \mbox{\ref{fig:threedis_res}}). Coverage of Bayesian credible intervals (BCIs) was good, if somewhat conservative due to the diffuse priors and the limited extent of the data. We see decent posterior contraction, along with scaled posterior median absolute deviations and scaled BCIs that together indicate that the model parameters are reliably recovered. The MMTL approximation was sensitive to the choice of $ \tau $-leap interval. Models fit using an interval of one hour yielded comparable inferences to the LNA, albeit at much higher computational cost (Table \mbox{\ref{tab:threedis_compres}}). MMTL models fit using a one day $ \tau $-leap interval were comparable to the LNA in computational performance, but exhibited poor coverage for the $ R_0 $ in the influenza scenario, likely because the discretization interval was long relative to the generation interval of the outbreak. The ODE models struggled to reliably recover model parameters in all three scenarios, and exhibited low coverage for $ R_0 $ and $ \rho $ despite the lack of structural model misspecification. ODE posteriors had higher median absolute errors and narrower BCIs compared with LNA and MMTL posteriors. This is in agreement with results of \mbox{\citet{king2015avoidable}}, who found that ODE models underestimated uncertainty about epidemic dynamics. The ODE was faster than the LNA or MMTL/PMMH, suggesting the ODE may be useful as an exploratory tool, but not as a basis for inference. 


\begin{sidewaystable}[htbp]
	\caption{Computational statistics for SEIR models fit to data simulated under Ebola, Influenza, and SARS-CoV-2 dynamics. We report 50\% (10\%, 90\% quantiles) of total CPU time for five MCMC chains and total log-posterior effective sample size (LP-ESS). The MMTL models were fit using a leap interval of either one hour or one day. Additional details are provided in \hyperref[subsec:three_dis_supplement]{Web Appendix D}.}
	\label{tab:threedis_compres}
	\centering\small
	\begin{tabular}{lcccc}
		\toprule
		\addlinespace[0.3em]
		& \textbf{LNA} & \textbf{ODE} & \textbf{MMTL (}\textbf{1 hour)} & \textbf{MMTL (}\textbf{1 day)}\\
		\cmidrule{2-5}\textbf{Ebola} &&&& \\
		\hspace{1em}\textit{CPU time (hours)} & 3.8 (3.4, 4.3) & 0.072 (0.066, 0.081) & 210 (180, 240) & 12 (11, 14)\\
		\hspace{1em}\textit{Total  LP-ESS / total CPU time} & 160 (88, 270) & 380000 (240000, 570000) & 0.095 (0.065, 0.14) & 31 (22, 45)\\
		\addlinespace[0.3em]
		\textbf{Influenza}& &&&\\
		\hspace{1em}\textit{Total CPU time (hours)} & 3.4 (3, 3.7) & 0.059 (0.054, 0.063) & 170 (150, 190) & 9.3 (8.5, 10)\\
		\hspace{1em}\textit{Total LP-ESS / total CPU time}  & 390 (300, 500) & 390000 (260000, 570000) & 0.15 (0.1, 0.23) & 59 (42, 78)\\
		\addlinespace[0.3em]
		\textbf{SARS-CoV-2}&&&& \\
		\hspace{1em}\textit{Total CPU time (hours)} & 3.4 (3.1, 3.8) & 0.06 (0.056, 0.064) & 170 (150, 180) & 9.7 (8.1, 10)\\
		\hspace{1em}\textit{Total LP-ESS / total CPU time}  & 380 (290, 470) & 530000 (340000, 770000) & 0.15 (0.11, 0.21) & 48 (35, 72)\\
		\bottomrule
	\end{tabular}
\end{sidewaystable}

\subsection{Supporting Simulations}
\label{subsec:supp_sims}
In addition to providing further details about the simulations in the previous section, \hyperref[sec:3dis_supp]{Web Appendix C} presents results from analogous simulations in which we truncated each time series at one year or after eight consecutive weeks of zero cases. The LNA performed substantially worse with so much of the data coming from the tail of the outbreak. We do not recommend the LNA as a process approximation when interest lies in stochastic extinction or emergence of outbreaks for two reasons. First, because the leap conditions that underpin the LNA are likely to be violated in these settings (\hyperref[sec:weba]{Web Appendix A}). Moreover, the LNA approximates MJP transition densities with a unimodal Gaussian distribution. For this reason, it is incapable of capturing the bimodal dynamics and ``false starts'' that are typical in the early stages of outbreaks, and lacks a point mass at zero new infections to capture epidemic extinction. Still, it may be possible to adapt our method for use in a hybrid framework that switches between the LNA and an exact MJP, as in \mbox{\citep{rebuli2017hybrid}}.

In \hyperref[sec:est_scale_sim]{Web Appendix D}, we used the LNA, ODE, and MMTL with a one-hour $ \tau $-leap interval to fit SIR models to data from outbreaks in populations of size $ P = 2,000 $, $ P = 10,000 $, and $ P = 50,000 $. We simulated outbreaks conditional on parameters drawn from a set of prior distributions that reflected outbreak dynamics and detection rates that are typical of many contact driven outbreaks. We fit the models using the priors from which we drew parameters, ensuring that Bayesian inference would be properly calibrated up to the approximation error vis-a-vis the MJP used to simulate the data. Results obtained with the LNA and MMTL were comparable, and BCI coverage probabilities for all parameters were close to their nominal levels. In contrast, BCI coverage probabilities were too low in ODE models. 

In \hyperref[sec:est_scale_sim]{Web Appendix E}, we discuss how MCMC performance can be improved through  reparameterizations that reflect how the parameters interact to shape the outbreak dynamics. For the simpler models considered in this paper, we recommend parameterizing the estimation scale on which a Markov chain explores the posterior in terms of the (log) basic reproduction number, (log) mean sojourn period durations, and (logit) mean case detection rate. This approach also allows us to more naturally set priors about transmission and surveillance parameters. In more complex settings, we can use deterministic ODEs to speed up the task of identifying good MCMC estimation scales. In our application to modeling the spread of Ebola in Guinea, Liberia, and Sierra Leone, we allow for the effective size of the population at risk in each country to be estimated as a parameter in the model. The implications of this choice for identifiability and MCMC parameterization are explored in \hyperref[sec:effpop_identifiability]{Web Appendix F}. 

Finally, \hyperref[sec:ebola_mods]{Web Appendix G} presents a simulation confirming that our framework remains computationally robust when fitting the multi-country SEIR model that we apply to the West Africa Ebola outbreak. We successfully fit the model to simulated data and recover the true parameter values via the LNA. In contrast, the MMTL/PMMH algorithm failed to yield a convergent MCMC run when given a similar computational budget, despite the absence of any structural model misspecification. Our inability to obtain a valid posterior sample in the ``easy" setting where we knew the true data generating mechanism, even with great expenditure of time and computational resources, led us to abandon PMMH as a computational strategy for analyzing real-world data from the West Africa outbreak.

\section{Modeling the Spread of Ebola in West Africa}
\label{sec:ebola_application}

\begin{figure}[htbp]
	\centering
	\includegraphics[width=\linewidth]{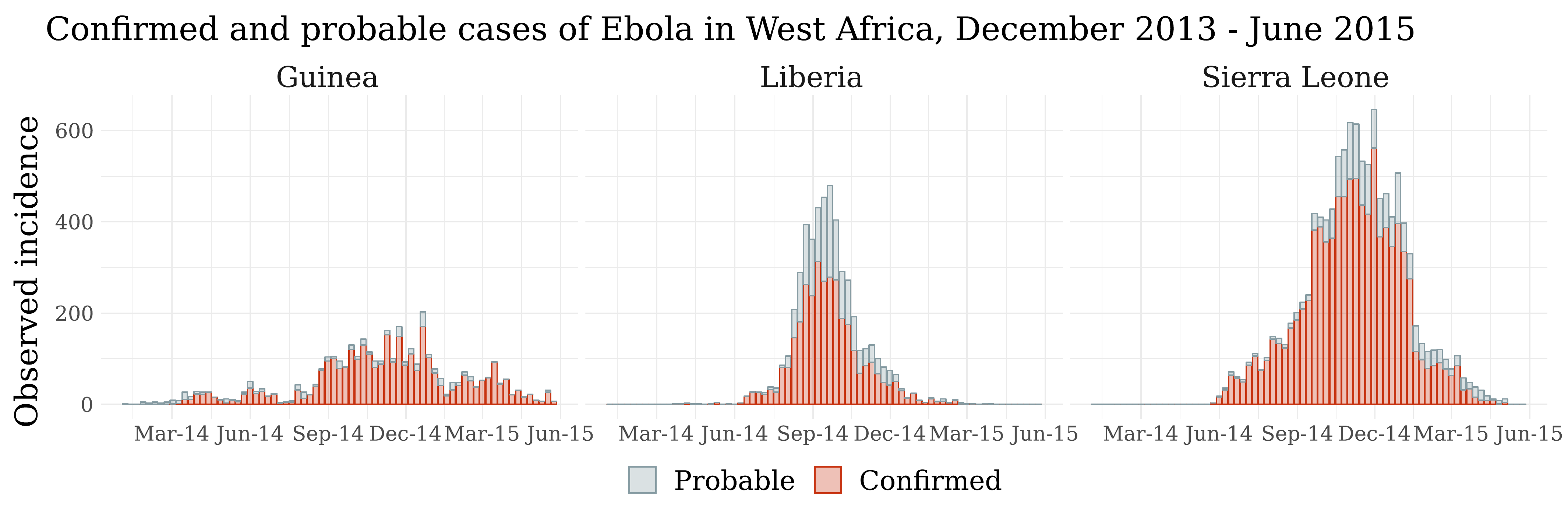}
	\caption{Weekly incidence of confirmed and probable cases of Ebola in Guinea, Liberia, and Sierra Leone. The total incidence was 3,627 in Guinea, 4,994 in Liberia, and 11,317 in Sierra Leone, and the estimated population sizes were 11.8 million, 4.4 million, and 7.1 million, respectively  (United Nations, 2017).}
	\label{fig:eboladat}
\end{figure}

We now turn our attention to modeling the 2013--2015 Ebola outbreak in Guinea, Liberia, and Sierra Leone. Our objective will be to describe the transmission dynamics of the outbreak, and in particular to estimate the basic reproductive numbers for each country. The data, shown in Figure \ref{fig:eboladat}, consist of national case counts from the World Health Organization patient database consisting of weekly confirmed and probable Ebola cases \citep{who2016eboladat}. Probable cases were defined as patients with Ebola-like symptoms who were epidemiologically linked to a suspected, probable, or confirmed case, or to a sick or dead animal, or who had died and were linked to confirmed cases \mbox{\citep{who2014casedef}}. Cases were confirmed if they tested positive for Ebola virus RNA or IgM Ebola antibodies \citep{coltart2017ebola}. For illustrative purposes, we aggregated confirmed and probable cases.

\begin{figure}[htbp]
	\centering
	\includegraphics[width=\linewidth]{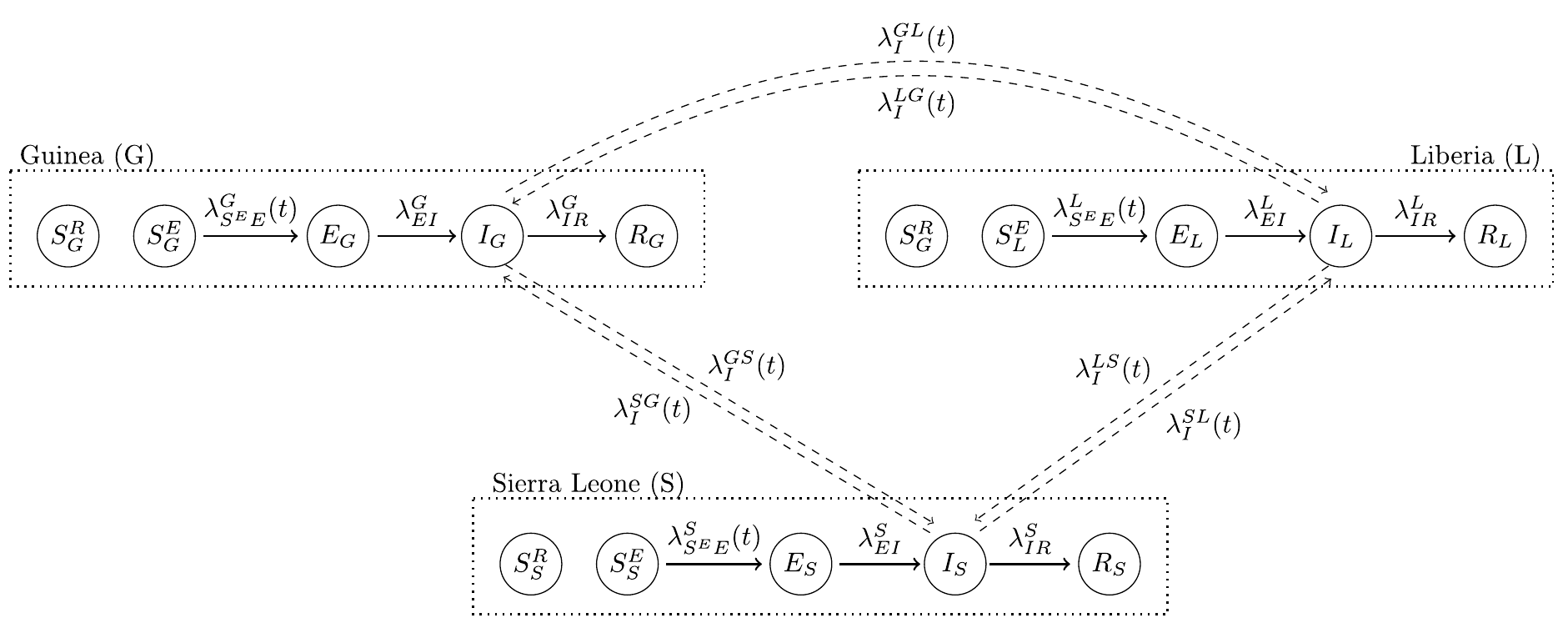}
	\caption{Diagram of state transitions for a joint model for Ebola transmission in Guinea, Liberia, and Sierra Leone. Dotted boxes denote countries, nodes in circles denote the model compartments: susceptible but removed from infectious contact $ (S^R) $, susceptible but exposed to infectious contact $ (S^E) $, exposed $ (E) $, infectious $ (I) $, recovered $ (R) $. Compartments  are subscripted with country indicators. Solid lines with arrows indicate stochastic transitions between model compartments, which occur continuously in time. Dashed lines indicate that infected individuals in one country contribute to the force of infection in another country. Rates at which individuals transition between compartments are denoted by $ \lambda $ and are subscripted by compartments and superscripted by countries, e.g., $ \lambda_{S^EE}^L $ is the rate at which susceptible individuals become exposed in Liberia. Transmission in Liberia and Sierra Leone was assumed to commence at 10 and 19 weeks, respectively. Full expressions for the rates are given in Table S12\cmmnt{\ref{tab:ebola_mod_rates}}.}
	\label{fig:ebola_mod_diag}
\end{figure}

\nocite{un2017popdat} We used the LNA and ODE approximations to fit a multi-country model with country-specific SEIR dynamics and cross-border transmission between each of Guinea, Liberia, and Sierra Leone (Figure \ref{fig:ebola_mod_diag}). Transmission was assumed to commence in Liberia on March 2, 2014, and in Sierra Leone on May 4, 2014, corresponding to three weeks prior to the first cases in those countries. The observed incidence was modeled as a negative binomial sample of the true incidence. The total incidence in each country was small relative to the population size, suggesting that only a fraction of the population was geographically or socially linked to ongoing transmission. Hence, we also estimated the effective size of the population at risk in each country. Priors for model parameters were informed by published estimates of Ebola transmission dynamics (Table \ref{tab:ebola_priors};  \hyperref[subsec:effpop_initdist_priors]{Web Appendix G}). We also fit a set of simplified models to data from each country independently to understand the benefit of explicitly modeling cross-border transmission (\hyperref[sec:ebola_mods]{Web Appendix G}). Results from the single-country models did not differ substantively from results obtained with the joint model (Table \ref{tab:ebola_lna_vs_ode_ests}). 

\begin{figure}[htbp]
	\centering
	\includegraphics[width=0.9\linewidth]{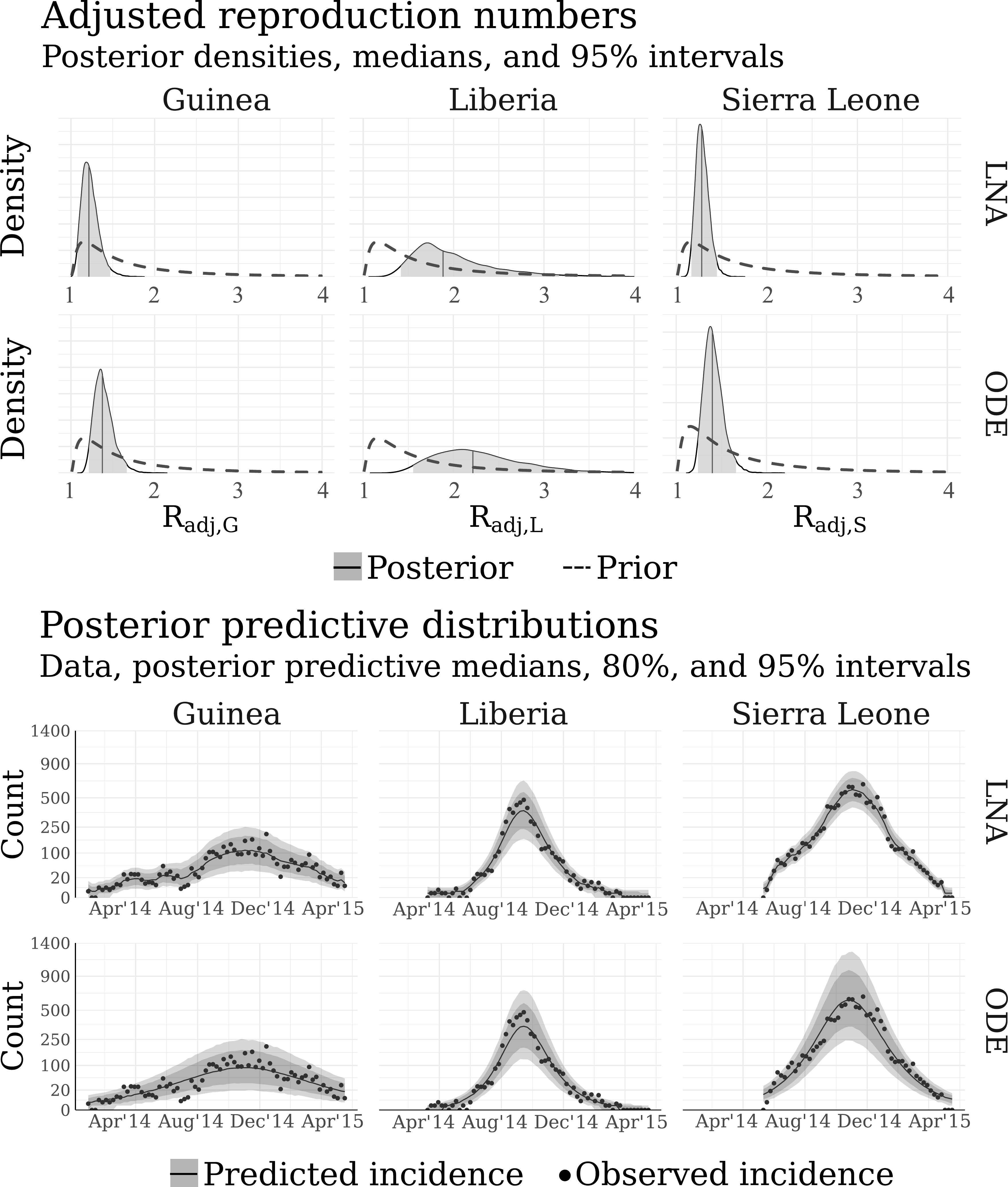}
	\caption{(Top panel) Posterior distributions (shaded densities) of effective population size adjusted basic reproduction numbers under the LNA and ODE models, with prior densities (dashed lines) plotted over the posterior ranges. Solid vertical lines are posterior medians, and shaded regions correspond to 95\% Bayesian credible intervals. The adjusted basic reproduction number is defined with respect to the effective population size. For example, $R_{adj,G} = P_{eff,G} \beta_G / \mu_G$, where $P_{eff,G}$, $\beta_G$, and $\mu_G$ are the effective population size, within-country per-contact infection rate, and recovery rate for Guinea, respectively. (Bottom panel) Posterior predicted incidence in West Africa under the LNA and ODE models, marginalizing over the joint posterior distributions of the latent incidence and model parameters. Solid lines are pointwise posterior predictive median incidence, shaded bands correspond to 80\% and 95\% pointwise posterior predictive incidence distributions, and dots are the observed incidence.}
	\label{fig:ebolaplots}
\end{figure}

The estimated transmission dynamics under the LNA model (Table \ref{tab:ebola_lna_vs_ode_ests}) were consistent with published estimates obtained with stochastic models fit to aggregate incidence data \citep{chretien2015mathematical}. The posterior median (95\% BCI) basic reproduction numbers, adjusted by the estimated effective population sizes, were 1.2 (1.1, 1.5) in Guinea, 1.9 (1.4, 3.2) in Liberia, and 1.3 (1.2, 1.4) in Sierra Leone. Adjusted basic reproduction numbers estimated under the ODE tended to be higher than estimates obtained under the LNA (top panel, Figure \ref{fig:ebolaplots}). Posterior distributions of latent and infectious period durations in Guinea and Liberia largely recovered the priors, while the estimated durations for Sierra Leone were somewhat shorter, though not unreasonably so. Estimated latent and infectious period durations under the ODE were longer than under the LNA, especially in Guinea and Liberia, where they were much longer than expected. Though we allowed for cross-border transmission, we were not able to resolve its contribution from the data and exactly recovered the priors for extrinsic basic reproduction numbers between each pair of countries. Additional results, including diagnostics, are provided in \hyperref[sec:ebola_mods]{Web Appendix H}. 

Compared with the deterministic ODE approximation, the LNA provides a better fit  and more appropriately accounts for uncertainty about the outbreak. The LNA posterior predictive distribution, which integrates over the joint posterior of the model parameters and latent epidemic process, better matches the observed incidence than does the ODE posterior predictive distribution (bottom panel, Figure \ref{fig:ebolaplots}). ODE posterior predictive intervals (PPIs) are wider than their LNA counterparts, and ODE posterior predictive p-values (PPPs) are more extreme (Figure \ref{fig:ebolappicomp}). The poor performance of the ODE is due to its being forced to account for all observations with a single deterministic path, which inevitably leaves many observations poorly explained and results in higher estimates of the negative binomial overdispersion parameter. In contrast, the stochastic nature of the LNA, and in particular, the restarting formulation explored in this work, allows the model to capture local variability in the epidemic process. 

\section{Discussion}
\label{sec:discuss}

We have presented a broadly applicable framework for fitting stochastic epidemic models to partially observed incidence counts. We demonstrated how the LNA could be reparameterized to properly analyze incidence data and leverage state-of-the-art MCMC machinery. Unlike previous applications of the LNA to epidemic count data, our framework accommodates non-Gaussian surveillance models without relying on simulation-based computational tools. We have verified that the LNA is statistically and computationally performant in simulations designed to reflect important use-cases for SEMs. We have also highlighted instances where modelers should use caution. Our framework is not panacea for weak data and cannot conjure up precise estimates for aspects of the model that are unresolvable in the data. Nor is the LNA an adequate model for stochastic emergence and extinction of outbreaks, and we have demonstrated in simulations that applying it in these cases can lead to poor performance. Using data from the 2013--2015 outbreak of Ebola in West Africa, we demonstrated that our LNA framework is capable of fitting complex, geographically stratified SEMs. 

For clarity of exposition, this work has focused on fitting simple SEMs to incidence data in closed, homogeneously mixing populations. In practice, we can improve the realism of our models by incorporating demography and allowing for heterogeneous mixing among subgroups in the population, e.g., age \mbox{\citep{li2008continuous}} or geographic strata \mbox{\citep{van2008spatial}}. The computational cost of the LNA increases as the number of model compartments increases. One interesting extension of this work would be to model transmission within each stratum as conditionally independent given correlated hyperparameters, e.g., spatially structured reproduction numbers. This would allow for LNA paths for each stratum to be updated in parallel within each MCMC iteration. 

It is frequently unreasonable to assume that outbreak dynamics are constant as the spatial and temporal extent of an outbreak increases. We are often interested in assessing the effects of interventions \mbox{\citep{anderson2020}}, seasonal factors \mbox{\citep{koepke2016predictive}}, or changes in disease surveillance \mbox{\citep{karcher2015}}, all of which require us to incorporate possibly time-varying covariates into our models. Another class of extensions involves modeling different aspects of transmission and surveillance semi-parametrically \mbox{\citep{xu2016bayesian}}.  

Mechanistic SEMs provide us with a framework for integrating multiple data streams into a coherent model. This can help to resolve multimodality in the posterior and stabilize inference for aspects of the model that are only weakly identifiable from incidence data. Incidence and mortality data have been combined in models for SARS-CoV-2 \mbox{\citep{fintzi2020using}} and influenza \mbox{\citep{shubin2014estimating,de2019analysing}}. In work that was concurrent to the methods presented in this paper, the LNA has also been used to jointly analyze incidence data and genetic data \mbox{\citep{tang2019fitting}}. Our framework can easily be extended to accommodate multiple data streams by adding surveillance model components and appropriately tracking incidence or prevalence in different compartments.



\section*{Acknowledgements}

J.F., J.W., and V.N.M. were supported by the NIH grant U54 GM111274. J.W. was supported by the NIH grant R01 AI029168. V.N.M. was supported by the NIH grant R01 AI107034. The authors thank Aaron King for his help fitting models in \texttt{pomp} and Jing Wang for her help with simulations. This work utilized the NIH HPC Biowulf cluster (http://hpc.nih.gov).
\vspace*{-8pt}


\section*{Supporting Information}

The algorithms for fitting LNA and ODE models are implemented in the \texttt{stemr} $ \R $ package, which is available along with code for reproducing the results in this manuscript at \texttt{\url{https://github.com/fintzij/stemr}}. Web Appendices A through G, referenced throughout this work, are available in the supporting material.\vspace*{-8pt}

\pagebreak
\bibliographystyle{plainnat}
\bibliography{incidence_lna}
\clearpage


\renewcommand{\thefigure}{S\arabic{figure}}%
\setcounter{figure}{0}%

\renewcommand{\thetable}{S\arabic{table}}%
\setcounter{table}{0}%

	\begin{center}
		\Large
		Supporting Information for\\
		``A Linear Noise Approximation for Stochastic Epidemic Models Fit to Partially Observed Incidence Counts by Jonathan Fintzi, Jon Wakefield, and Vladimir N. Minin.''
	\end{center}
	\section{Web Appendix A: Large Population Approximations for Markov Jump Processes}
	\label{sec:weba}
	There are a variety of ways to arrive at the diffusion approximation for a MJP \citep{fuchs2013inference}. We outline an intuitive, though somewhat informal, construction of a stochastic differential equation (SDE), referred to as the chemical Langevin equation (CLE), where the drift and diffusion terms are chosen to match the approximate moments of MJP path increments in infinitesimal time intervals \citep{wilkinson2011stochastic,golightly2013simulation}, and refer to \citet{gillespie2000chemical} and \citet{fuchs2013inference} for more detailed presentations. 
	
	Denote the compartment counts at time $ t $ by $ \bX^c(t) = \bx^c_t $. We want to approximate the numbers of infections and recoveries in a small time interval, $ (t, t+\dt] $, i.e., $ \bN^c(t+\dt) - \bN^c(t)$. Suppose that we can choose $ \dt $ so that the following two \textit{leap} conditions hold:
	\begin{enumerate}
		\item $ \dt $ is sufficiently \textit{small} that the $ \bX^c $ is essentially unchanged over $ (t,t+\dt] $, so that the rates of elementary transition events, e.g., infections and recoveries, are approximately constant: 
		\begin{equation}\label{eqn:tau_cond_1}
			\blambda(\bX^c(t^\prime)) \approx \blambda(\bx^c(t)),\ \forall t^\prime \in (t,t+\dt].
		\end{equation}
		\item $ \dt $ is sufficiently \textit{large} that we can expect many disease state transitions of each type:
		\begin{equation}\label{eqn:tau_cond_2}
			\blambda(\bx^c(t))\dt \gg \bs{1}.
		\end{equation}
	\end{enumerate}
	Condition (\ref{eqn:tau_cond_1}), holds when $ \dt $ is small, and implies that the numbers of infections and recoveries in $ (t,t+\dt] $ are essentially independent since their rates of occurrence are approximately constant within the interval \citep{gillespie2000chemical}. This condition also implies that the numbers of events are Poisson random variables with rates $ \blambda(\bx^c(t)\dt) $, i.e., $ N^c_{SI}(\dt) \sim \mr{Poisson}(\beta S(t)I(t)\dt) $ and $ N^c_{IR}(t+\dt) \sim \mr{Poisson}(\mu I(t)\dt) $ \citep{wilkinson2011stochastic}. Condition (\ref{eqn:tau_cond_2}) implies that the Poisson innovations are well--approximated by Gaussian random variables. 
	
	When (\ref{eqn:tau_cond_1}) and (\ref{eqn:tau_cond_2}) hold, we can approximate the integer--valued processes, $ \bX^c $ and $ \bN^c $, with the real--valued processes, $ \bX $ and $ \bN $. The state space of $ \bX $ for the SIR model is $$ \mcS_X^R = \lbrace (l,m,n):l,m,n \in [0,P],\ l+m+n=P\rbrace, $$ and the state space  of $ \bN $ is $$ \mcS_N^R = \lbrace (j,k): j,k \in [0,P],\ \bX(\mcV_{jk})\in\mcS_X^R \rbrace. $$ In words, the state space of $ \bX $ is the set of compartment volumes that are non--negative and that sum to the population size, while the state space of $ \bN $ is the set of non--decreasing and non--negative incidence paths, constrained so that they do not lead to invalid prevalence paths (e.g., where there are more recoveries than infections and hence negative number of infected individuals). For now, we ignore the constraints on $ \mcS_N^R $ and $ \mcS_X^R $, and approximate changes in cumulative incidence of infections and recoveries in an infinitesimal time step as 
	\begin{equation}
		\bN(t+\dt) - \bN(t) \approx \blambda(\bX(t))\dt + \bLambda(\bX(t))^{1/2}\dt^{1/2}\bZ,
	\end{equation}
	where $ \bLambda = \diag\left (\blambda(\bX) \right )$ and $ \bZ\sim MVN(\bs{0},\mb{I}) $. This implies the equivalent CLE,
	\begin{equation}
		\rmd \bN(t) = \blambda(\bX(t))\dt + \bLambda(\bX(t))^{1/2}\rmd\bW_t, 
	\end{equation}
	where the vector $ \bW_t $ is distributed as independent Brownian motion, and $ \bLambda(\bX(t))^{1/2} $ is a matrix square root of $ \bLambda(\bX(t)) $. 
	
	\subsection{Validity of the Linear Noise Approximation}
	\label{subsec:tau_conds_lna}
	
	\mbox{\citet{wallace2010simplified}} and \mbox{\citet{wallace2012linear}} argue that the LNA is perhaps more appropriately viewed as an approximation to the CLE. The LNA provides us with an analytic characterization (obtained numerically) of the stochastic differential equation (SDE) transition density as the size of a chemical reaction system is scaled back from its infinite population ordinary differential equation (ODE) limit. The $ \tau $--leap conditions under which the LNA is a reasonable MJP approximation are actually the conditions that are required for the CLE to reasonably approximate a MJP. Hence, there are two questions here: First, under what conditions does the LNA approximation to the CLE deteriorate? And second, under what conditions does the CLE approximation of the MJP break down?
	
	The first of these questions is easier to answer in some generality. Given a long enough time interval, the LNA will diverge from the CLE as perturbations to the SDE accumulate. In contrast, the the deterministic drift of the LNA is determined by the initial conditions at the start of the time interval. Solutions to this issue have been proposed in \mbox{\citep{fearnhead2014}} and \mbox{\citep{minas2017long}}. In this work, we have adopted the restarting formulation of the LNA, proposed in \mbox{\citet{fearnhead2014}}. Restarting the LNA at observation times, though informal, is in a sense optimal when viewed as a filtering procedure. The restarting algorithm helps to inoculate us against divergence of the LNA from the CLE by ensuring that we are only approximating the CLE over short time intervals. Now, it is possible that a finer observation interval, and hence more frequent restarts, would lead to less error in the approximation. However, it also possible that a finer observation interval, say, using daily incidence counts, could lead to less robust inferences. For instance, incidence data are often aggregated to weakly counts in order to eliminate day-of-week effects and to mitigate fine scale problems caused by reporting delays. 
	
	To the second question, it is impossible to say in complete generality when the CLE is a good approximation for a MJP, and often this can only be answered through simulations in the context of the target application. The advice that the CLE is reasonable in large populations is an often useful heuristic for the necessary and sufficient (and vague) condition that the system is ``close to its thermodynamic limit." Even in huge populations, the CLE is not uniformly valid since there are few transition events in the incipient and terminal stages of an outbreak. This was shown in \mbox{\citep{buckingham2018gaussian}}, who demonstrated in simulations that the KL divergence between the MJP and a variety of approximations, including the LNA and several SDEs, is greatest at the start and end of an outbreak.
	
	\subsection{Infinite Population Deterministic Limit of Markov Jump Processes}
	\label{subsec:ode_limit}
	In the infinite population limit, the trajectory of a density dependent MJP converges to a deterministic trajectory that is the solution to (\ref{eqn:lna_ode_drift}) given some initial conditions, $ \bX_0 = \bx_0 $ and $ \bN = \bs{0} $, and model parameters, $ \btheta $ \mbox{\citep{allen2017primer}}. The solution to the ODE, (\ref{eqn:lna_ode_drift}), is a deterministic mapping of the parameters and initial conditions, meaning that for fixed $ \btheta,\bX_0,\bN_0 $ there is only a single path that the latent process could take. In large population settings, it is common, though ill--advised, practice to ignore stochasticity in the latent epidemic process and assume that the time--evolution of $ \bX $ and $ \bN $ is given by their infinite population functional limits. Hence, deterministic ODE models simply replace the stochastic latent epidemic process with a deterministic curve. The other aspects of the model remain unchanged, e.g., we model the observed incidence as a negative binomial sample of the true incidence, but now the negative binomial mean is conditional on increments of the ODE path obtained by solving (\ref{eqn:lna_ode_drift}). With that said, we still have to estimate $ \btheta $, and possibly $ \bX_0 $. This is accomplished by sampling the posteriors of these random variables via MCMC as outlined in Section \ref{subsec:lna_param_updates}.

	\newpage
	\section{Web Appendix B: Algorithms and Additional MCMC Details}
	\label{sec:webb}
	
	\subsection{Algorithms for Sampling LNA Paths}
	\label{subsec:lna_algs}
	
	\begin{algorithm}[htbp]
		\caption{Mapping standard normal draws onto LNA sample paths.}
		\label{alg:doLNA}
		\begin{algorithmic}[1]
			\Procedure{doLNA}{$ \bZ,\btheta,\mcI $}
			\State \textbf{initialize: }$ \bX(t_0) \gets \bX_0,\ \bN(t_0) \gets \bs{0},\ \bNtil(t_0) \gets \bs{0},$
			\State \hspace{0.62in} $ \bmu(t_0) \gets \bs{0},\ \bSigma(t_0) \gets \bs{0} $
			\For{$ \ell = 1,\dots,L $}
			\State $ \bmu(t_\ell),\ \bSigma(t_\ell) \gets $ solutions to (\ref{eqn:lna_ode_drift}) and (\ref{eqn:lna_ode_diffusion}) over $ (t_{\ell-1}, t_\ell] $
			\State $ \bNtil(t_\ell)\gets \bmu(t_\ell) + \bSigma(t_\ell)^{1/2}\bZ(t_\ell) $ \Comment{non--centered parameterization}
			\State $ \bN(t_\ell)\gets \bN(t_{\ell-1}) + \exp(\bNtil(t_\ell)) - \bs{1} $
			\State $ \bX(t_\ell) \gets \bX(t_{\ell-1}) + \bA^T(\bN(t_\ell)-\bN(t_{\ell-1}))$
			\State $\bNtil(t_\ell) \gets \bs{0},\ \bmu(t_\ell)\gets\bs{0},\ \bSigma(t_\ell)\gets\bs{0} $ \Comment{restart initial conditions}
			\EndFor
			\State \hspace{-0.2in}\Return \Comment{incidence and/or prevalence sample paths}
			\State$\bN = \left \lbrace\bN(t_0),\bN(t_1),\dots,\bN(t_L)\right \rbrace $
			\State $\bX = \left \lbrace \bX(t_0),\ \bX(t_1),\dots,\bX(t_\ell) \right \rbrace $
			\EndProcedure
		\end{algorithmic}
	\end{algorithm}
	
	\newpage
	
	\begin{algorithm}[htbp]
		\caption{Sampling LNA draws via elliptical slice sampling.}
		\label{alg:elliptss_lna}
		\begin{algorithmic}[1]
			\Procedure{\doElliptSS}{$ \bZ_{cur},\btheta,\bY,\mcI,\omega = 2\pi $}
			\State Sample ellipse: $ \bZ_{prop} \sim N(\bs{0}, \mb{I}) $
			\State Sample threshold:\vspace{-0.05in} $$ u|\bx \sim \mr{Unif}(0, L(\bY|\doLNA(\bZ_{cur},\btheta,\mcI))) $$
			\State Position the bracket and make initial proposal: \vspace{-0.05in}
			\begin{align*}
				\psi &\sim \mr{Unif}(0,\omega)\\
				L_\psi &\leftarrow -\psi;\ R_\psi \leftarrow L_\psi + \psi\\
				\phi &\sim \mr{Unif}(L_\psi,R_\psi)
			\end{align*}
			\State Set $ \bZ' \leftarrow \bZ_{cur}\cos(\phi) + \bZ_{prop}\sin(\phi) $. 
			\If{$ L(\bY|\doLNA(\bZ',\btheta,\mcI)) > u $}{ accept $ \bZ' $}
			\State\Return{ $ \bZ' $}
			\Else
			\State Shrink bracket and try a new angle:
			\State{\textbf{If:} $ \phi < 0 $}{ \textbf{then: }$ L_\phi \leftarrow\phi $ }{ \textbf{else: }$ R_\phi \leftarrow \phi $}
			\State $ \phi \sim \mr{Unif}(L_\phi, R_\phi) $
			\State \textbf{GoTo:} 5
			\EndIf
			\EndProcedure
		\end{algorithmic}
	\end{algorithm}
	
	\newpage
	\subsection{Tuning the Initial Elliptical Slice Sampling Bracket Width}
	\label{subsec:lna_init_bracket_width}
	
	When fitting SEMs with complex dynamics, e.g., when there are many strata or when the dynamics are time varying, we will be able improve the computational efficiency of our MCMC by initialing the ElliptSS bracket width at $ \omega < 2\pi $. This is motivated by the observation that when the model dynamics are complex, the ElliptSS bracket will typically need to be shrunk many times before the sampler reaches a range of acceptable angles in the proposal. Each time we propose a new angle in the ElliptSS algorithm we must solve the LNA ODEs in order to compute the observed data likelihood. Thus, if we can reduce the number of ElliptSS steps, we will be able to shorten the run time of our MCMC.
	
	We typically set the initial bracket width to a constant times the standard deviation of the accepted angles in a tuning phase. Since we do not step out the ElliptSS bracket, the initial width should not be so small as to induce additional autocorrelation in the latent process, and should also not be so wide that the bracket is contracted needlessly. We have found a bracket width of $ \omega = 2\sqrt{2\log(10)}\sigma $, corresponding to the full width at one tenth maximum for a Gaussian with standard deviation $ \sigma $, to work well in practice. In order to facilitate tuning of the initial bracket width, the elliptical slice sampling Algorithm \ref{alg:elliptss_lna} was modified slightly from the one in \citet{murray2010} with respect to the initial angle proposal so that the distribution of angles for accepted proposals would be symmetric around zero.
	
	Figure \ref{fig:esstuning} presents histograms of the number of contractions per ElliptSS update and the accepted angles before and after contracting the initial ElliptSS bracket width for the Ebola model of Section \ref{sec:ebola_application}. In this instance, we were able to substantially reduce the number of contractions, and hence likelihood evaluations, per ElliptSS update while leaving the distribution of accepted angles essentially unchanged. We like to call this a ``free lunch".
	\setcounter{figure}{0}
	\begin{figure}[htbp]
		\centering
		\includegraphics[width=\linewidth]{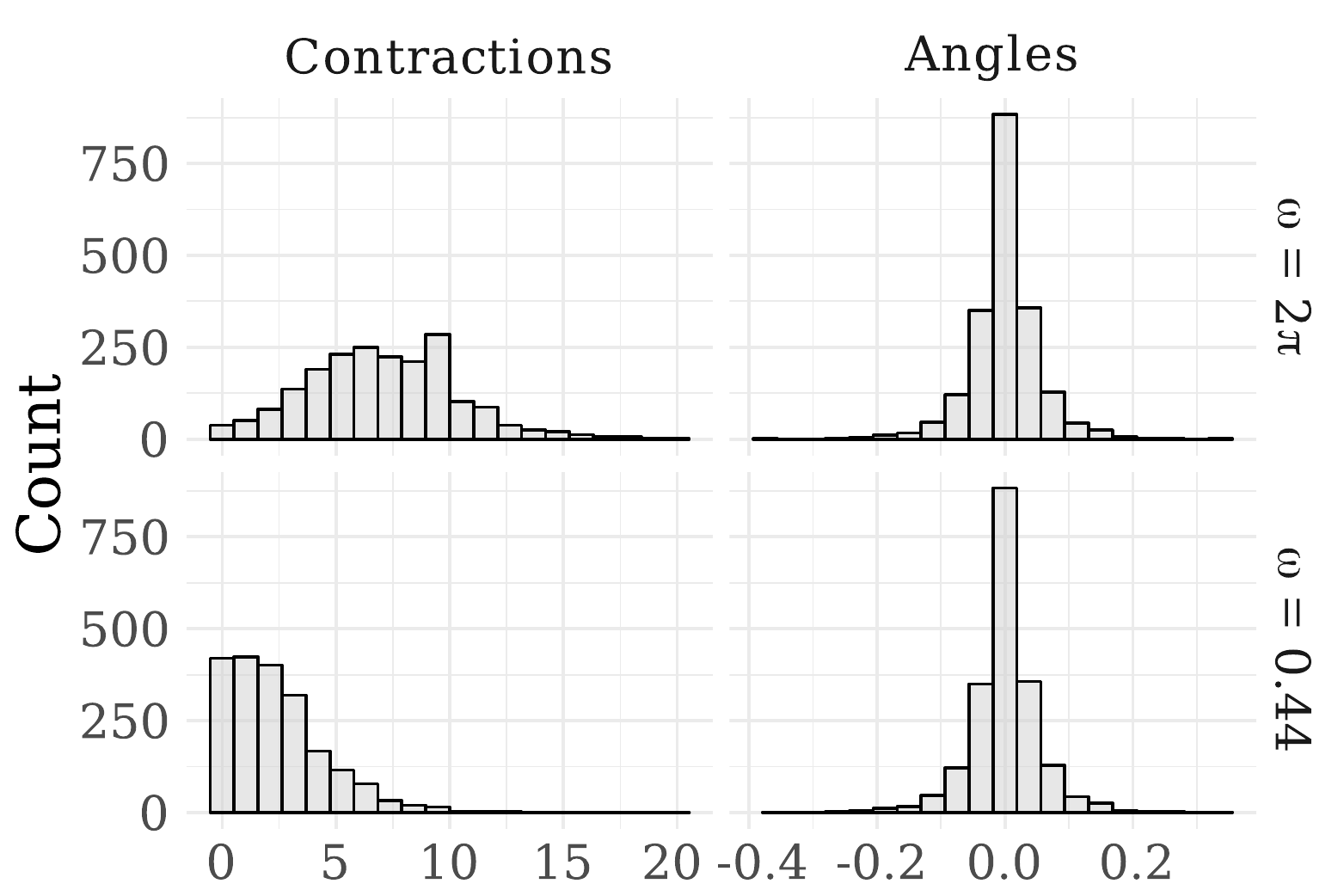}
		\caption{Distributions of the numbers of contractions per ElliptSS update and the accepted angles for an MCMC chain for the Ebola model of Section \ref{sec:ebola_application} fit to a simulated dataset. An initial bracket width of $ 2\pi $ was used for the first 5,000 iterations (top row), after which the initial bracket width was set to $ 2\sqrt{2\log(10)}\sigma_{ElliptSS} $, where $ \sigma_{ElliptSS} $ was the standard deviation of the accepted angles from the initial run (bottom row).} 
		\label{fig:esstuning}
	\end{figure}
	
	\subsection{Centered vs. Non--centered Parameterization}
	\label{subsec:cp_vs_ncp}
	
\begin{figure}[htbp]
		\centering
		\includegraphics[width=\linewidth]{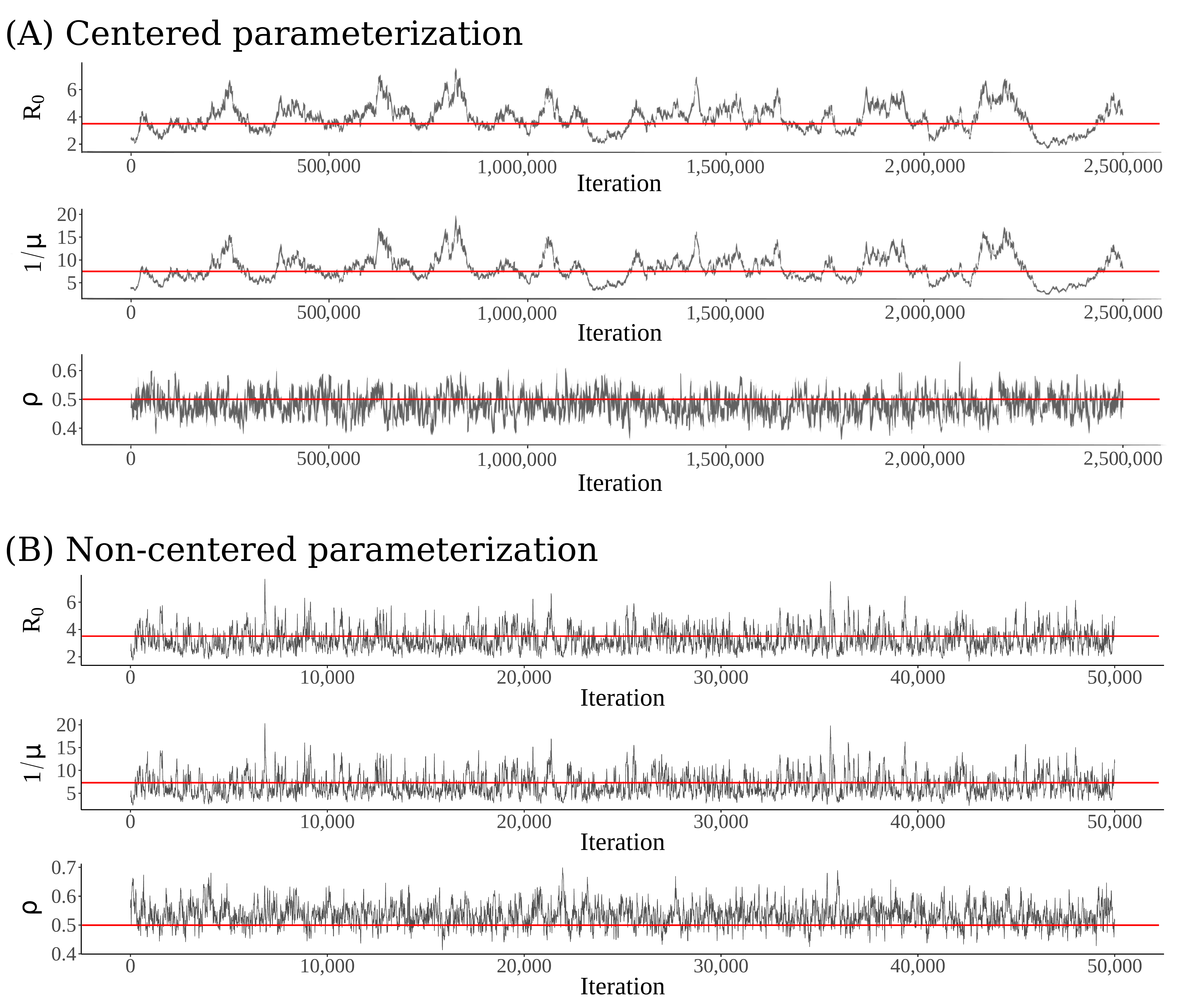}
		\caption{(A) Posterior traceplots from a single MCMC chain for an SIR model fit to Poisson distributed incidence data targeting the centered posterior (\ref{eqn:lna_approximate_posterior}). (B) Posterior traceplots from a single MCMC chain targeting the non--centered posterior (\ref{eqn:lna_noncentered_posterior}). MCMC alternated between updating the latent path via elliptical slice sampling, and updating parameters via a multivariate random walk Metropolis algorithm. $ R_0 = \beta P / \mu$ is the basic reproductive number, $ 1/\mu $ is the mean infectious period duration, and $ \rho $ is the mean case detection rate. The true values of $ R_0,\ 1/\mu,$ and $ \rho $ were 3.5, 7, and 0.5, respectively.}
		\label{fig:lna_combined_traces}
	\end{figure}

	A central computation challenge for fitting hierarchical latent variable models via DA MCMC is that samples can become autocorrelated when alternately updating latent variables and model parameters \citep{bernardo2003non,papaspiliopoulos2007general}. DA MCMC that alternately updates LNA paths and model parameters is no exception (Figure \hyperref[fig:lna_combined_traces]{2A}). 
	
	This phenomenon of poorly mixing MCMC chains can be traced to the use of a centered parameterization (CP) for the LNA in (\ref{eqn:lna_approximate_posterior}).  Under the CP, updates to $ \btheta|\bNtil,\bY $ are made conditionally on a \textit{fixed} LNA path. Therefore, parameter proposals are accepted if they are concordant with the data \textit{and} the current path. Small perturbations to model parameters can significantly shift the LNA transition densities and render the current path unlikely under the proposal. This limits the magnitude of perturbations to the model parameters at each MCMC iteration and results in severely autocorrelated posterior samples.
	
	The NCP for latent LNA paths massively improves MCMC mixing. Figure \hyperref[fig:lna_combined_traces]{2A} shows traceplots of model parameters for one of the MCMC chains for an SIR model fit to Poisson distributed incidence data using the CP. Each MCMC chain was run for 2.5 million iterations, following a tuning run of equal length, but only yielded an effective sample sizes in the low double digits for the basic reproduction number and infectious period duration. In contrast, the NCP yielded effective sample sizes per--chain of between 500--700 for each of the model parameters in only 50,000 iterations following a  tuning run of equal length. Figure \hyperref[fig:lna_combined_traces]{2B} shows traceplots for one of the NCP MCMC chains. Note that since the posterior distribution of the restarting LNA under the CP does not actually decompose into the form required to use ElliptSS, applying ElliptSS to the CP of the restarting LNA amounts to essentially further approximating the CP with Gaussian random variables. 
	
	\subsection{Multivariate Normal Slice Sampler}
	\label{subsec:mvnss}
	Univariate slice samplers can suffer from poor mixing in moderate-- to high--dimensional settings much in the same way as Gibbs samplers. One option for reducing autocorrelation is to update blocks of parameters. Methods for slice sampling in multiple dimensions are explored in \citet{neal2003slice,thompson2011slice,tibbits2014automated}. These include slice sampling in hyperrectangles, the use of adaptive Gaussian crumbs that guide slice proposals, and slice sampling along eigenvectors of the estimated posterior covariance matrix. 
	
	We present a simple method for sampling a parameter vector, $ \btheta\in\bbR^d $, where we perform univariate slice sampling updates along rays drawn from a non--isotropic angular central Gaussian distribution, which is tuned to match the covariance structure of the posterior. This helps to account for correlations among model parameters. The method, which we refer to as the multivariate normal slice sampler (MVNSS), is similar to the algorithm in \citet{ahmadian2011efficient}, although we also adapt the proposal covariance matrix using a Robbins--Monro recursion during the adaptation phase of the MCMC \citep{andrieu2008tutorial,liang2011advanced}. The computational cost of MVNSS does not increase dramatically with the dimensionality of the parameter space, though we have found that multiple MVNSS updates per MCMC iteration can, in some cases, improve performance. The algorithm is amenable to tuning of the initial bracket width as in \citet{tibbits2014automated}, which helps to reduce the number of likelihood evaluations per iteration.
	
	Suppressing the dependence on the data for notational clarity, slice sampling $ \btheta\in\bbR^d $ from its posterior $ \pi(\btheta|\by) $ is largely the same as sampling $ \btheta \sim \pi(\btheta)\propto f(\btheta)$. Let $ \bSigma = \Cov(\btheta) = \bL\bL^T $, where $ \bL $ is the lower triangular matrix of the Cholesky decomposition of $ \bL $ (any other matrix square root would do). In practice, $ \bSigma $ is approximated by $ \what{\bSigma}_n $, which is estimated over an initial MCMC run. The strategy in MVNSS is to propose $ \prop{\btheta} = \cur{\btheta} + c\bxi $, where $ \bxi = h(\bz),\ \bz\sim MVN(\mb{0},\bSigma),\ h(\bz) = \bz / ||\bz|| $, and to sample $ c $ in a univariate slice sampling update. Normalizing $ \bz $ allows us to more easily tune the initial bracket width. During an adaptation phase, we construct proposals as $ \prop{\btheta} = \cur{\btheta} + ch(w\bxi_1 + (1-w)\bxi_2) $, where $ \bxi_1 = h(\bz_1),\ \bz_1\sim MVN(\mb{0},\bSigma),\,\ \bxi_2 = h(\bz_2),\ \bz_2\sim MVN(\mb{0}, \mb{I}_d), \ h(\bz) = \bz / ||\bz|| $, and $ w\in[0,1] $. The weight given to $ \bz_2 $ is typically quite small, but helps to avoid degeneracy of the empirical covariance matrix during adaptation. We give the non--adaptive version of the algorithm below. 
	
	\begin{algorithm}[htbp]
		\caption{Multivariate normal slice sampling with stepping out.}\label{alg:mvnss}
		\begin{algorithmic}[1]
			\Procedure{MVNSS}{$ \cur{\btheta},\ \bL,\ S^\prime = (0,\omega)$}
			\State $ u \sim \mr{Unif}(0,f(\cur{\btheta})) $
			\Comment{Set threshold}
			\State $\bz\sim MVN(\mb{0},\mb{I}_d),\ \bxi\gets h(\bL\bz)$ \Comment{Propose direction}
			\State $ p\sim \mr{Unif}(0,1) $; $ L \gets - \omega p,\ U \gets L+\omega $ \Comment{Position $ S' $ around $ 0 $}
			\State $ S^\prime \gets $\textsc{StepOut}$ (u,S^\prime) $ \Comment{Step out bracket}
			\State\hspace{\algorithmicindent}\textbf{while }{$ u<f(\btheta - L\bxi) $} \textbf{do} $ L \gets L-\omega $
			\State\hspace{\algorithmicindent}\textbf{while }{$u<f(\btheta + U\bxi)$} \textbf{do} $ U \gets U+\omega $
			\State $ c \sim \mr{Unif}(L,U);\  \prop{\btheta}\gets \cur{\btheta} + c\bxi $ \Comment{Propose new value}
			\If{$ f(\prop{\btheta})>u $}{ $\new{\btheta}\gets \prop{\btheta} $}\Comment{Accept proposal}
			\State\Return{$ \new{\btheta} $}
			\Else\Comment{Shrink bracket}
			\If{$ c < 0 $}{$ \ L\gets c $}{ \textbf{else} $U\gets c $}
			\EndIf 	
			\State{\textbf{GoTo} 8} 
			\EndIf
			\EndProcedure
		\end{algorithmic}
	\end{algorithm}
	
	\subsubsection{Adapting the proposal covariance in MVNSS}
	\label{subsubsec:mvnss_adaptation}
	Adaptive MCMC algorithms aim to improve computational efficiency by using MCMC samples to learn optimal values of tuning parameters on the fly. MCMC proposal kernels can be adapted in a number of different ways, but must be adapted with care to preserve the stationarity of the target distribution. In order for an adaptive MCMC algorithm to preserve the stationary distribution, it must satisfy two conditions, vanishing adaptation, and bounded convergence \citep{andrieu2008tutorial}.
	
	The main computational tool used in the adaptive variations of the MCMC algorithms in this dissertation is the Robbins--Monro recursion, which allows us to continuously adapt the tuning parameters of an MCMC kernel. The Robbins--Monro recursion is a stochastic approximation algorithm that searches for a solution to an equation, $ f(\theta) = \alpha $, that has a unique root at $ \theta^\star $. The function $ f(\theta) $ is not directly observed. Instead, we use a noisy sequence of estimates, $ h(\what{\theta}_n) $, satisfying $ \E(h(\what{\theta}_n)) = f(\theta) $ to recursively approximate $ \theta^\star $. The recursion takes the form, $$\theta_{n+1} = \theta_n + \gamma_{n+1}(h(\what{\theta}_n) - \theta^\star).$$
	Hence, the recursion increments $ \theta $ by an amount proportional to the difference between $ h(\what{\theta}_n) $ and its target.  The gain factor sequence, $ \lbrace\gamma_{n+1}\rbrace $, is a deterministic non--increasing, positive sequence such that $$(i) \lim\limits_{n\rightarrow\infty}\gamma_n = 0,\ \hspace{0.25in}(ii)\ \sum_{n=1}^\infty \gamma_n = \infty,\hspace{0.25in} (iii)\  \sum_{n=1}^\infty\gamma_n^{1+\lambda} < \infty,\ \lambda > 0.$$
	Note that since $ \gamma_n\rightarrow0 $ and $ \E(h(\what{\theta}_n))= f(\theta) $, it follows that $ |\theta_{n+1} - \theta_n|\rightarrow0 $ as $ n\rightarrow\infty $, i.e., the recursion is constructed to satisfy diminishing adaptation. Condition $ (ii) $ ensures that the gain sequence does not decay so fast that there are values of $ \theta $ in its state space, $ \Theta $, that cannot be reached. Condition $ (iii) $ ensures bounded convergence of the sequence $ \lbrace\theta_n\rbrace $. Gain factor sequences of the form \begin{equation}
		\gamma_n = C(1+pn)^{-\alpha},\ \alpha\in(0.5,1],\ p>0
	\end{equation}  will satisfy these conditions \citep{andrieu2008tutorial,liang2011advanced}. We adapt the proposal covariance over the course of an initial MCMC tuning run, which is followed by a final run with a fixed MCMC kernel. The samples accumulated during the adaptation phase are discarded. 
	
	Let $ \mu_n $ and $ \Sigma_n $ denote the empirical mean and covariance of the posterior samples from the first $ n $ MCMC iterations, and $ \lbrace\gamma_n\rbrace $ be a sequence of gain factors. In each adaptive MCMC iteration, we sample $ \new{\btheta}|\cur{\btheta} $ via MVNSS, and update the empirical mean and covariance via the following recursions:
	\begin{align*}
		\bmu_n &= \bmu_{n-1} + \gamma_n(\new{\btheta}_n - \bmu_{n-1}), \\
		\bSigma_n &= \bSigma_{n-1} + \gamma_n\left ((\cur{\btheta}_n - \bmu_{n-1})(\new{\btheta}_n - \bmu_{n-1})^T - \bSigma_{n-1}\right ).
	\end{align*}
	
	\subsection{Initializing the LNA Draws}
	\label{subsubsec:lna_init}
	In simple models, biologically plausible parameter values will generally lead to valid LNA paths, and we can initialize the LNA draws by simply drawing $ \bZ\sim MVN(\bs{0},\mb{I}) $. However, this is not necessarily the case for complex models with many types of transition events, or when the time--series of incidence counts is long. One option is to include a resampling step after line 6 in Algorithm \ref{alg:doLNA}, in which $ \bZ(t_\ell) $ is redrawn in place until we have obtained a valid LNA path that respects the constraints on the latent state space. However, such a procedure does not sample from the correct distribution since $ \bZ $ is not distributed as a truncated multivariate Gaussian. To correct for this, we ``warm--up" the LNA path with an initial run of ElliptSS iterations in which the likelihood only consists of the indicators for whether the path is valid. Note that ElliptSS, or any other valid MCMC algorithm for updating $ \bZ|\btheta,\bY $, will never lead to an invalid LNA path being accepted if the current LNA draws and model parameters correspond to a valid path. Similarly, any valid MCMC algorithm for updating model parameters conditional on LNA draws will also preserve the validity of LNA paths.
	
	\subsection{Inference for Initial Compartment Volumes}
	\label{subsec:lna_init_volumes}
	When the initial compartment volumes are included as initial parameters in the model instead of being treated as fixed, we will model them as arising from the following truncated multivariate normal distribution: \begin{equation}
		\label{eqn:lna_initdist_prior}
		\bX_0 \sim TMVN_{\mcS_X^R}(P\bp,\alpha P(\bP - \bp\bp^T)),
	\end{equation} 
	where $ \bp $ is a vector of subject--level initial state probabilities, $ \bP = \diag(\bp) $, $ P $ is the population size, $ \alpha $ is an over--dispersion parameter, and the subscript $ \mcS_X^R $ specifies the state space of $ \bX $ (so that the compartment volumes add up to $ P $ and each compartment volume is non--negative and less than the total population size at time $ t_0 $). Thus, the initial distribution is the truncated normal approximation of either a multinomial distribution with size $ P $ and probability vector $ \bp $ if $ \alpha = 1 $, or of a dirichlet--multinomial distribution with parameters $ \balpha \implies \bp = \balpha / \balpha^T\boldsymbol{1}$, and over--dispersion $ \alpha = (P + \balpha^T\boldsymbol{1}) / (1 + \balpha^T\boldsymbol{1})$. In models with multiple strata, we will similarly model the initial compartment volumes as having independent truncated multivariate normal distributions that are each approximations of multinomial distributions over initial compartment counts within each stratum. Notation and details are completely analogous to the single stratum case, and are therefore omitted for clarity.
	
	Let $ \bM = P\bp$, $ \bV = \alpha P(\bP - \bp\bp^T) $, and $ \bV^{1/2} $ be the matrix square root of $ \bV $, which we will compute using the singular value decomposition $ \bV = \bU\bD\bU^T \implies \bV^{1/2} = \bU\bD^{1/2}$. Let $ \bZ^X $ denote the LNA draws as before, and let $ \bZ^{X_0}\sim\mr{MVN}(\bs{0},\mb{I}) $ denote the vector of draws that will be mapped to $ \bX_0 $. We will update the initial compartment volumes jointly with the LNA draws using elliptical slice sampling.
	
	\newpage 
	
	\begin{algorithm}[htbp]
		\caption{Sampling LNA draws and initial volumes via elliptical slice sampling.}
		\label{alg:elliptss_lna_initvols}
		\begin{algorithmic}[1]
			\Procedure{\doElliptSS2}{$ \bZ^X_{cur}, \bZ^{X_0}_{cur},\btheta,\bY,\mcI,\omega = 2\pi $}
			\State Sample ellipse: $ \bZ^X_{prop} \sim N(\bs{0}, \mb{I}),\ \bZ^{X_0}_{prop} \sim N(\bs{0}, \mb{I})$
			\State Sample threshold: $ u|\bx \sim \mr{Unif}(0, L(\bY|\doLNA(\bZ_{cur},\btheta,\mcI))) $
			\State Position the bracket: \vspace{-0.1in}
			\begin{align*}
				\psi &\sim \mr{Unif}(0,\omega)\\
				L_\psi &\leftarrow -\psi;\ R_\psi \leftarrow L_\psi + \psi\\
				\phi &\sim \mr{Unif}(L_\psi,R_\psi)
			\end{align*}
			\State Make the initial proposal: \vspace{-0.1in}\begin{align*}
				\bZ^{X'} &\leftarrow \bZ^X_{cur}\cos(\phi) + \bZ^X_{prop}\sin(\phi) \\ \bZ^{X_0'} &\leftarrow \bZ^{X_0}_{cur}\cos(\phi) + \bZ^{X_0}_{prop}\sin(\phi) \implies \bX_0^\prime = \bM + \bV^{1/2}\bZ^{X_0^\prime}
			\end{align*}
			\If{$ L(\bY|\doLNA(\bZ',\btheta^\prime,\mcI)) > u $}{ accept $ \bZ^{X^\prime},\bZ^{X_0^\prime} $}
			\State\Return{ $ \bZ' $}
			\Else
			\State Shrink bracket and try a new angle:
			\State{\textbf{If:} $ \phi < 0 $}{ \textbf{then: }$ L_\phi \leftarrow\phi $ }{ \textbf{else: }$ R_\phi \leftarrow \phi $}
			\State $ \phi \sim \mr{Unif}(L_\phi, R_\phi) $
			\State \textbf{GoTo:} 5
			\EndIf
			\EndProcedure
		\end{algorithmic}
	\end{algorithm}
	
	\newpage
	
	\section{Web Appendix C: Comparison with Other Common MJP Approximations}
	\label{sec:3dis_supp}
	
	We simulated 1,000 outbreaks with SEIR dynamics from a MJP via Gillespie's direct algorithm for each of the three disease settings given in Table \ref{tab:three_dis_setup_supp}. We initialized the outbreaks with 25 exposed (E) and 5 infectious (I) individuals in the influenza setting, and 5 exposed and 5 infected individuals in the Ebola and SARS-CoV-2 settings. We fixed the initial conditions at their true values instead of estimating them as parameters in the model. The data were a negative binomial sample of the true latent incidence in each week. We truncated the incidence time series at 52 weeks or after observing four consecutive counts of five or fewer cases, whichever came first. 
	
	We fit each model using the LNA, ODE, and MMTL/PMMH approximations. Five MCMC chains per model were initialized at random values near the true parameters. For models fit via the LNA and ODE approximations, an initial adaptive MCMC run of 25,000 iterations of the GA-RWM algorithm was used to estimate the empirical covariance matrix. The empirical covariance matrix was initialized at 0.1 times an identity matrix. The gain factor sequence was $ \gamma_n = (1 + n)^{-2/3} $, and a small nugget variance of 0.00001 was added during the adaptation phase. The target acceptance rate used in the adaptation was 0.234. After the adaptive run, the empirical covariance matrix was frozen and we ran the MCMC for an additional 50,000 iterations where parameters were sampled via a GA-RWM algorithm with fixed covariance matrix.  The samples from all five chains were combined to form the final MCMC posterior sample. The total run time included both the adaptive and fixed kernel phases of the MCMC run. The models were implemented using the \texttt{stemr R} package \citep{stemr}. 
	
	Inference via the MMTL approximation within PMMH was implemented using the \texttt{pomp R} package \citep{pompjss}. We used 500 particles in the PMMH algorithm, and a $ \tau $--leap interval of either one day or one hour. MCMC was initialized in the same way as in the LNA and ODE models. The empirical covariance was adapted using two initial tuning runs of 1,000 and 10,000 iterations with a gain factor sequence of $ \gamma_n = n^\alpha $, with the cooling term, $ \alpha = 0.999 $. The initial tuning run was adopted because some Markov chains were plagued by severe particle degeneracy early in the adaptation. Hence, we found it useful to check whether the adaptation had degenerated early in the MCMC run. We restarted the first adaptive MCMC phase if the number of accepted MCMC proposals was not between 100 and 800. The empirical covariance matrix from this first adaptive phase was plugged into a second adaptive phase which was restarted until we obtained an adaptive run with between 500 and 8,000 accepted MCMC proposals. The empirical covariance matrix was then frozen and we accrued an additional 50,000 samples per chain that were combined to form the final MCMC posterior sample. The total run time included the run time from the second adaptive MCMC phase and the fixed kernel MCMC phase. 
	
	\begin{table}[htbp]
		\caption{Setup for simulating outbreaks with dynamics similar to Ebola, influenza, and SARS-CoV-2.}
		\label{tab:three_dis_setup_supp}
		\footnotesize
		\centering
		\begin{tabular}{cccc}	
			& \textbf{Ebola}&\textbf{Influenza} &\textbf{SARS-CoV-2}\\
			\hline	
			Population size ($ P $) & 2,000 & 50,000 & 10,000 \\
			Basic reproduction number ($ R_0 $) & 1.8 & 1.3 & 2.5 \\
			Latent period ($ 1/\omega $) & 10 days & 2 days & 4 days \\
			Infectious period  ($ 1/\mu $) & 7 days & 2.5 days & 10 days \\
			Detection rate $ (\rho) $ & 0.5 & 0.02 & 0.1 \\
			Neg. bin. overdispersion $ (\phi) $ & 36 & 36 & 36\\
			\hline
		\end{tabular} 
	\end{table}
	
	\begin{sidewaystable}[htbp]
		\caption{Priors and estimation scales for the simulation comparing models fit via the LNA, ODE, and MMTL/PMMH approximations for SEIR models fit to data simulated under dynamics similar to Ebola, influenza, and SARS-CoV-2. Latent and infectious periods were given in weeks.}
		\label{tab:3dis_priors}
		\footnotesize
		\centering
		\begin{tabular}{cllcc}
			\toprule
			\multicolumn{5}{c}{\textbf{Ebola}}\\
			\hline
			\textbf{Parameter} & \textbf{Interpretation} & \textbf{Est. Scale} &\textbf{Prior} & \textbf{Median (95\% Interval)} \\ \hline
			$ R_0 = \beta P / \mu $ & Basic reproduction \# & $ \log(R_0) $ & LogNormal(log(1.7), 0.5) & 1.7 (0.63, 4.53) \\ 
			$ 1/\omega $ & Mean latent period & $ \log(1/\omega) $ & LogNormal(log(9/7), 0.82) & 1.3 (0.26, 6.41) \\
			$ 1/\mu $ & Mean infectious period & $ \log(1/\mu) $ & LogNormal(log(9/7), 0.82) & 1.3 (0.26, 6.41) \\
			$ \rho $ & Case detection rate & $ \logit(\rho) $& Beta(1,1) & 0.5 (0.025, 0.975) \\
			$ 1/\sqrt{\phi} $ & Neg.Binom. over--dispersion & $ \log(1/\sqrt(\phi)) $ & Exponential(rate = 3) & 0.23 (0.008, 1.23)\\
			\hline
		\end{tabular}\vspace{0.25in}
		\begin{tabular}{cllcc}
			\multicolumn{5}{c}{\textbf{Influenza}}\\
			\hline
			\textbf{Parameter} & \textbf{Interpretation} & \textbf{Est. Scale} &\textbf{Prior} & \textbf{Median (95\% Interval)} \\ \hline
			$ R_0 = \beta P / \mu $ & Basic reproduction \# & $ \log(R_0) $ & LogNormal(log(1.5), 0.5) & 1.5 (0.56, 4.0) \\ 
			$ 1/\omega $ & Mean latent period & $ \log(1/\omega) $ & LogNormal(log(2.25/7), 0.82) & 0.32 (0.064, 1.6) \\
			$ 1/\mu $ & Mean infectious period & $ \log(1/\mu) $ & LogNormal(log(2.25/7), 0.82) & 0.32 (0.064, 1.6) \\
			$ \rho $ & Case detection rate & $ \logit(\rho) $& Beta(1,1) & 0.5 (0.025, 0.975) \\
			$ 1/\sqrt{\phi} $ & Neg.Binom. over--dispersion & $ \log(1/\sqrt(\phi)) $ & Exponential(rate = 3) & 0.23 (0.008, 1.23)\\
			\hline
		\end{tabular}\vspace{0.25in}
		\begin{tabular}{cllcc}
			\multicolumn{5}{c}{\textbf{SARS-CoV-2}}\\
			\hline
			\textbf{Parameter} & \textbf{Interpretation} & \textbf{Est. Scale} &\textbf{Prior} & \textbf{Median (95\% Interval)} \\ \hline
			$ R_0 = \beta P / \mu $ & Basic reproduction \# & $ \log(R_0) $ & LogNormal(log(2.4), 0.5) & 2.4 (0.9, 6.4) \\ 
			$ 1/\omega $ & Mean latent period & $ \log(1/\omega) $ & LogNormal(log(6/7), 0.82) & 0.86 (0.17, 4.28) \\
			$ 1/\mu $ & Mean infectious period & $ \log(1/\mu) $ & LogNormal(log(9/7), 0.82) & 1.3 (0.26, 6.4) \\
			$ \rho $ & Case detection rate & $ \logit(\rho) $& Beta(1,1) & 0.5 (0.025, 0.975) \\
			$ 1/\sqrt{\phi} $ & Neg.Binom. over--dispersion & $ \log(1/\sqrt(\phi)) $ & Exponential(rate = 3) & 0.23 (0.008, 1.23)\\
			\bottomrule
		\end{tabular}
	\end{sidewaystable}
	
	\subsection{Supplementary simulation in a stochastic extinction setting}
	
	We repeated the simulation described above, but now truncated the incidence time series at 52 weeks or after eight consecutive weeks of zero observed cases, whichever came first. As discussed in Section \ref{subsec:supp_sims}, the LNA now performs poorly with so much of the data coming from the tail of the outbreak.
	
	\begin{figure}[htbp]
		\centering
		\includegraphics[width=\linewidth]{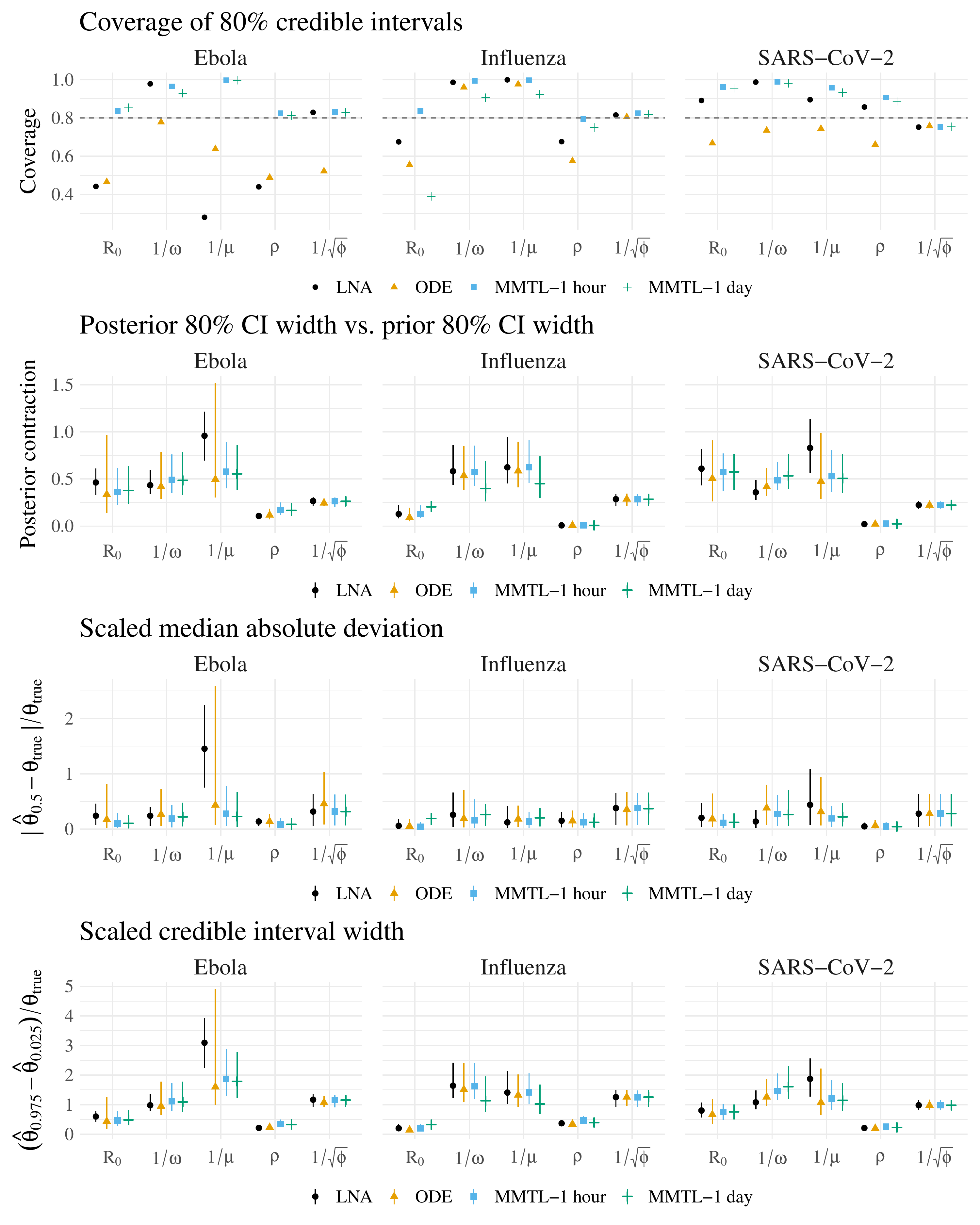}
		\caption{Supplementary simulation results for outbreaks simulated under Ebola-, influenza-, and SARS-CoV-2-like dynamics in populations of size 2000, 100000, and 10000, respectively. $ R_0 $ is the basic reproduction number, $ 1/\omega $ in the mean latent period, $ 1/\mu $ is the mean infectious period, $ \rho $ is the mean case detection rate, and $ 1/\sqrt{\phi} $ is (loosely) the excess dispersion of the negative binomial surveillance model on the standard deviation scale. In this simulation, incidence time series were truncated at either 52 weeks or after eight consecutive weeks of zero observed cases.}
		\label{fig:threedis_res_long}
	\end{figure}
	
	\newpage
	
	\section{Web Appendix D: Bayesian Coverage Simulation for SIR models}
	\label{sec:approx_comp_supp}
	
	\subsection{Simulation Setup}
	\label{subsec:lna_coverage_setup}
	
	In this simulation, repeated for each of the three different regimes of population size and initial conditions given in Table \ref{tab:lna_coverage_sim}, we simulated 1,000 datasets as follows:
	\begin{enumerate}
		\item Draw $ \log(R0 - 1),\ 1/\mu,\ \logit(\rho),\ \log(\phi) $ from the priors given in Table \ref{tab:lna_coverage_sim}.
		\item Simulate an outbreak, $ \bN|\btheta $, under SIR dynamics from the MJP via Gillespie's direct algorithm \citep{gillespie1976general}. If there were fewer than 15 cases, simulate another outbreak. 
		\item Simulate the observed incidence, $ \bY|\bN,\btheta $, as a negative binomial sample of the true incidence in each epoch, i.e., $ Y_\ell\sim\mr{Neg.Binomial(\rho(N_{SI}(t_\ell) - N_{SI}(t_{\ell-1})), \phi)} $. The incidence time series was truncated at 52 weeks or after 4 consecutive weeks of five or fewer counts.
	\end{enumerate}
	
	We fit SIR models using the LNA, ODE, and MMTL approximations. Priors for model parameters were assigned as in Table \ref{tab:lna_coverage_sim}. Five MCMC chains per model were initialized at random values near the true parameters and run for 75,000 iterations per chain. The first 25,000 iterations used an adaptive GA-RWM MCMC algorithm to estimate the empirical covariance matrix to be used in the multivariate Gaussian random walk Metropolis--Hastings proposals. The empirical covariance matrix was initialized as 0.01 times an identity matrix. After the warm--up period, the empirical covariance matrix was frozen and the final 50,000 iterations from each chain were combined to form the final MCMC sample. For models fit via the LNA and ODE approximations, the covariance matrix was adapted as in algorithm 4 of \citet{andrieu2008tutorial}. The gain factor sequence was $\gamma_n = (1 + n)^{-2/3}$, and a small nugget variance, equal to 0.00001 times an identity matrix, was added during the adaptation phase. The target acceptance rate used in the adaptation was 0.234. The models were implemented using the \text{stemr} R package \citep{stemr}.
	
	Inference via the MMTL approximation within PMMH were fit using the \texttt{pomp} R package \citep{pompjss}. We used 250 particles in the PMMH algorithm. The $ \tau $--leap interval for MMTL was set to one hour in contrast to the one week inter--observation interval. The MCMC was initialized in the same way as LNA and ODE models, but the empirical covariance matrix was adapted according to a different cooling schedule. The gain factor sequence provided by the package is of the form $ \gamma_n = n^\alpha $, where the cooling term, $ \alpha $, was set to 0.999.The initial tuning run was adopted because some Markov chains were plagued by severe particle degeneracy early in the adaptation. Hence, we found it useful to check whether the adaptation had degenerated early in the MCMC run. We restarted the first adaptive MCMC phase if the number of accepted MCMC proposals was not between 100 and 800. The empirical covariance matrix from this first adaptive phase was plugged into a second adaptive phase which was restarted until we obtained an adaptive run with between 500 and 8,000 accepted MCMC proposals. The empirical covariance matrix was then frozen and we accrued an additional 50,000 samples per chain that were combined to form the final MCMC posterior sample. The total run time included the run time from the second adaptive MCMC phase and the fixed kernel MCMC phase. 
	
	\begin{table}[htbp]
		\caption[LNA coverage simulation settings.]{Population sizes, initial conditions, and priors under which datasets were simulated for Bayesian coverage simulations with SIR models. The infectious period duration is given in weeks.}
		\label{tab:lna_coverage_sim}
		\footnotesize
		\centering
		\begin{tabular}{lccc}
			\hline
			& \textbf{Regime 1} & \textbf{Regime 2} & \textbf{Regime 3} \\\hline
			\textbf{Population size ($ N $)} & 2,000 & 10,000 & 50,000 \\ 
			\textbf{Initial infecteds ($ I_0 $)} & 1 & 5 & 25 \\
			\hline
			&&&
		\end{tabular}\vspace{0.25in} 
		
		\begin{tabular}{cllc}
			\hline
			\textbf{Parameter} & \textbf{Interpretation} & \textbf{Prior} & \textbf{Median (95\% Interval)} \\ \hline
			$ R_0-1 $ & Basic reproduction \# - 1 & LogNormal(0, 0.56) & $ \implies R_0 = $ 2.00 (1.33, 4.0) \\ 
			$ 1/\mu $ & Mean infectious period & LogNormal(0, 0.354)& 1 (0.5, 2) \\
			$ \rho $ & Case detection rate & LogitNormal(0, 1) & $ \implies \rho =$ 0.5 (0.12, 0.88) \\
			$ 1/\sqrt{\phi} $ & Neg.Binom. over--dispersion & Gamma(2, rate = 4) & 0.42, 0.06, 1.39\\
			\hline
		\end{tabular}\vspace{0.25in}
	\end{table}
	
	\begin{figure}[htbp]
		\centering
		\includegraphics[width=\linewidth]{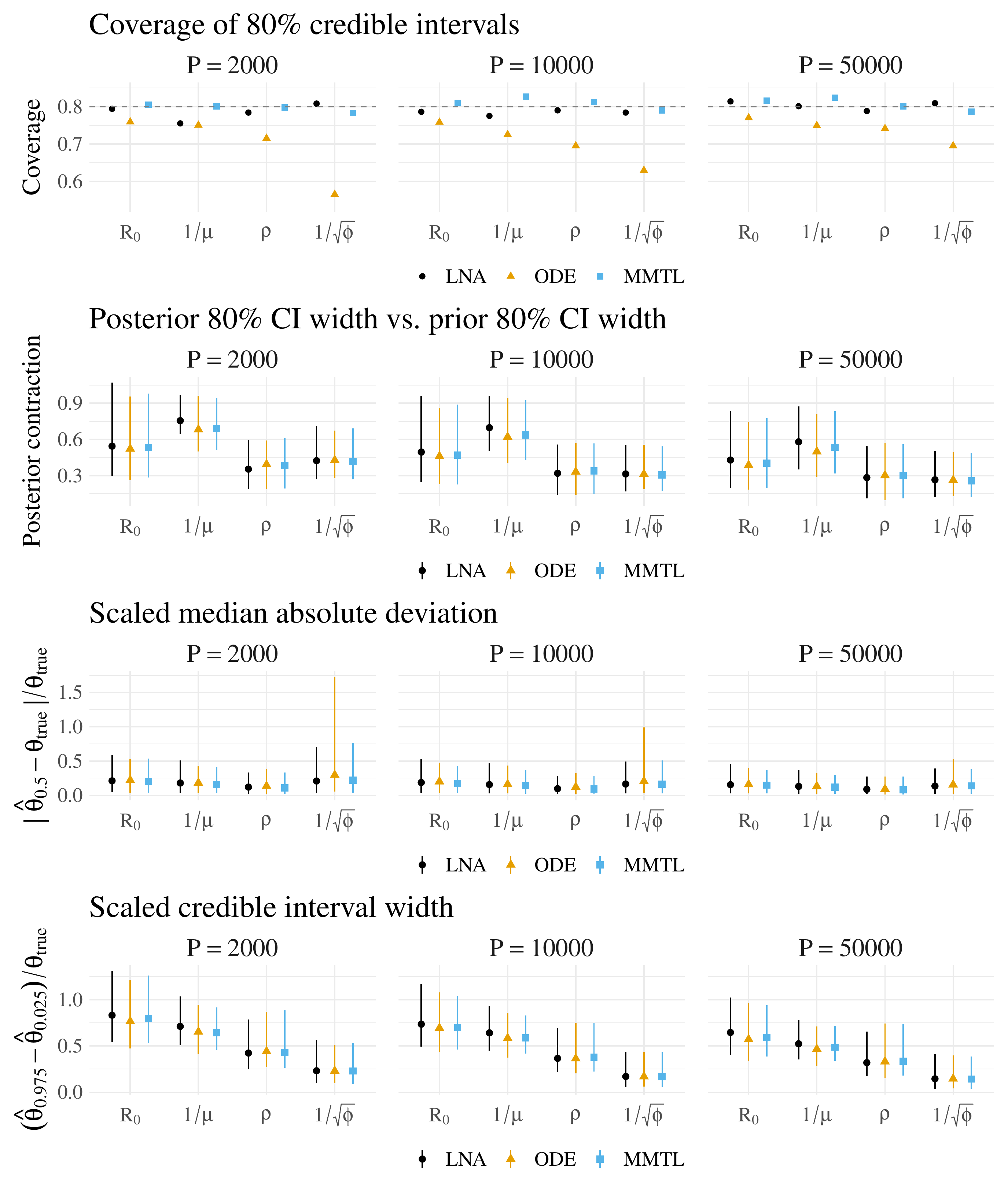}
		\caption{Supplementary simulation results for a Bayesian coverage simulation for SIR models fit via the LNA, ODE, and MMTL/PMMH. $ R_0 $ is the basic reproduction number, $ 1/mu $ is the mean infectious period, $ \rho $ is the mean case detection rate, and $ 1/\sqrt{\phi} $ is (loosely) the excess dispersion of the negative binomial surveillance model on the standard deviation scale.}
		\label{fig:lna_coverage}
	\end{figure}
	
	\newpage
	\section{Web Appendix E: Choice of Estimation Scale and Implications for MCMC}
	\label{sec:est_scale_sim}
	
	How we parameterize the estimation scale on which a Markov chain explores the posterior is critically important to its computational performance. If we can identify transformations of the model parameters that minimize strong correlations and non--linear relationships on the estimation scale, we will be able to substantially improve MCMC mixing. In our context, it will often be relatively straightforward to identify such transformations (or at least intermediate transformations that can be used in combination). As a general approach, we will try to identify transformations that reflect the ways in which model parameters jointly act on the model dynamics, and then a second set of transformations that remove any boundary conditions. 
	
\begin{figure}[htbp]
		\centering
		\includegraphics[width=\linewidth]{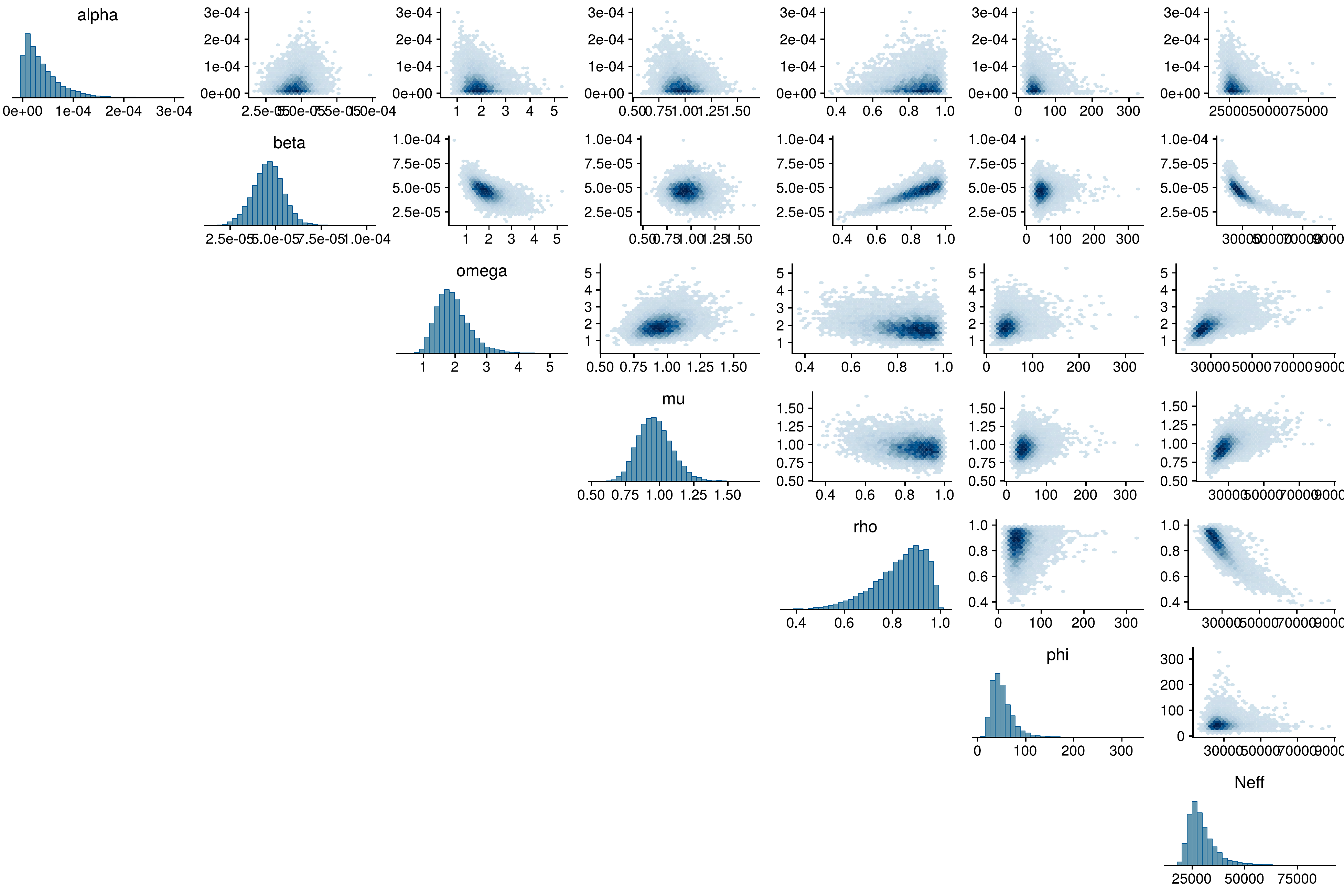}
		\caption[Posterior scatterplots for Sierra Leone SEIR model parameters on their natural scales.]{Marginal histograms and pairwise scatterplots of posterior samples for parameters for the SEIR model fit to the Sierra Leone Ebola dataset using the estimation scale in Table \ref{tab:seir_params_nat}.} 
		\label{fig:slpairs1}
	\end{figure}

	As an example, consider an SEIR model fit to incidence data from Sierra Leone. This model includes parameters for the external force of infection and the effective population size, which add complexity to the usual formulation of the SEIR dynamics as being entirely driven by endogenous contacts within a closed homogeneously mixing population. The effective population size is roughly the size of the initially susceptible population. The model parameters on their natural scales are provided in Table \ref{tab:seir_params_nat}. Each of the model parameters is obviously interpretable on its natural scale, but upon examining the pairwise scatterplots of the posterior (Figure \ref{fig:slpairs1}) it becomes obvious that the parameters interact in highly non--linear ways. We would encounter a variety of pathological computational problems if we were to naively parameterize the MCMC estimation scale without considering the ways in which the parameters interact to affect the dynamics. For example, it would be extremely difficult for any sampler that does not account for the curvature in the posterior, e.g., Hamiltonian Monte Carlo (HMC), to explore the parameter space. (An aside: we experimented with implementing the LNA in \texttt{Stan} and using HMC to sample the posterior, but repeatedly integrating the LNA ODEs along with their augmented sensitivity equations was prohibitively slow for even simple models).  
	
\begin{figure}[htbp]
		\centering
		\includegraphics[width=\linewidth]{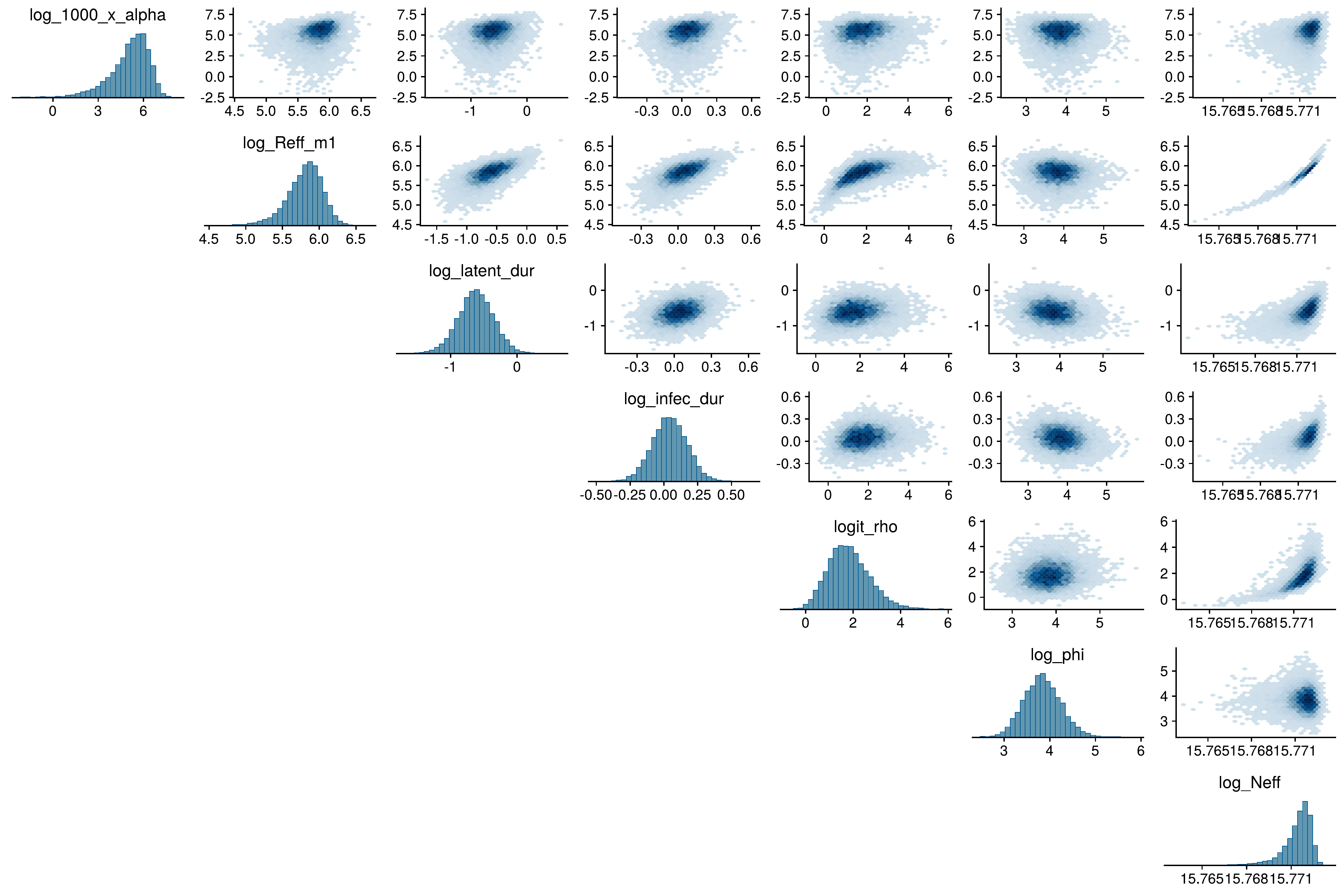}
		\caption[Posterior scatterplots for transformed Sierra Leone SEIR model parameters.]{Marginal histograms and pairwise scatterplots of posterior samples for parameters for the SEIR model fit to the Sierra Leone Ebola dataset using the estimation scale in Table \ref{tab:seir_params_est2}.} 
		\label{fig:slpairs2}
	\end{figure}

	We can mitigate the problems caused by non--linear relationships and strong correlations among parameters by parameterizing the estimation scale in terms of how the parameters jointly affect the model dynamics and then removing the boundary conditions. Table \ref{tab:seir_params_est2} provides a list of parameters on their estimation scale that are reflective of an initial first pass at how we would expect the parameters to interact. For example, the parameters governing the rates of infectious contact, $ \alpha$ and $\beta $, combine with the effective population size and the infectious period duration to produce the basic reproductive numbers with respect to initially infected individuals outside and inside the population. Figure \ref{fig:slpairs2} shows the pairwise scatterplots for posterior samples in this parameterization. We can see that there are some residual non--linear relationships between the log effective population size, the logit case detection probability, and the adjusted reproductive number.
	
	A heuristic argument for an estimation scale that further simplifies the posterior geometry proceeds by analogy with the analogous deterministic ODE model. In particular, we will consider how the functions of model parameters in Table \ref{tab:seir_params_est2} act on the model dynamics through the final size relation, and how they are informed by the data. The final size relation for the ODE model \citep{miller2012note} relates the fraction of the population that eventually becomes infected $ \pi $, with the basic reproductive number:
	
	\begin{align}
		\label{eqn:final_size_relation}
		\pi &= 1 - \e^{-R0\pi}.
	\end{align}
	
\begin{figure}[htbp]
		\centering
		\includegraphics[width=\linewidth]{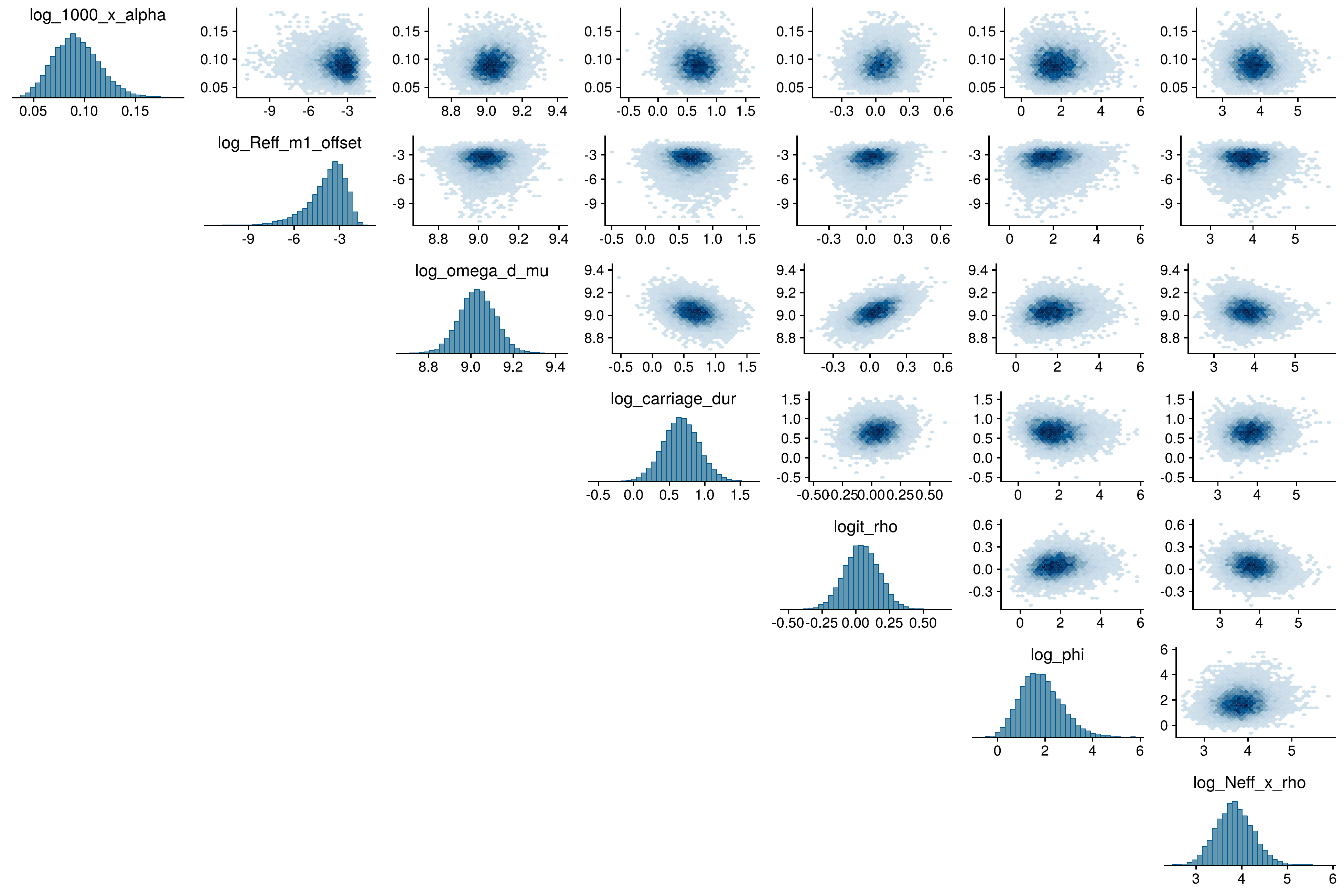}
		\caption[Posterior scatterplots for linear combinations of transformed Sierra Leone SEIR model parameters.]{Marginal histograms and pairwise scatterplots of posterior samples for parameters for the SEIR model fit to the Sierra Leone Ebola dataset using the estimation scale in Table \ref{tab:seir_params_est3}. } 
		\label{fig:slpairs3}
	\end{figure}

	As $ R0 $ increases, a larger fraction of the population becomes infected. As the effective population size, $ P_{eff}$, increases, so too does the overall scale of the outbreak, and the number of cases we would expect to detect. The expected scale of the observed outbreak is related to the quantity,  $ \pi \times \rho \times P_{eff} $, which should be concordant with the total number of observed cases. We can interpret the product, $ \rho \times P_{eff} $, as a measure of scale the observed incidence data. An implication of this is that the combination of parameters we would expect to jointly act on the model, \textit{a posteriori}, is the adjusted reproduction number offset by the case detection rate, $ R0\times\rho \times P_{eff} $, which will enter our estimation scale as $ \log(R0\times\rho\times P_{eff}) $. The other modification of the estimation scale in Table \ref{tab:seir_params_est2} replaces the log latent period with the log ratio of infectious to latent period durations. The new estimation scale is given in Table \ref{tab:seir_params_est3} and pairwise scatterplots for the posterior samples are shows in Figure \ref{fig:slpairs3}. On this estimation scale, the posterior is much better behaved, with weaker pairwise correlations and little in the way of non--linear relationships between the model parameters. 
	
	\begin{sidewaystable}[htbp]
		\caption{SEIR model parameter and their interpretation on their natural scales.}
		\label{tab:seir_params_nat}
		\footnotesize
		\centering
		\begin{tabular}{clc}
			\hline
			\textbf{Param.} & \textbf{Interpretation} & \textbf{Domain}\\
			\hline
			$\alpha$ & Rate of exogenous infection & $[ 0,\infty) $ \\
			$ \beta $ & Per--contact endogenous infection rate& $[0,\infty) $\\
			$ \omega $ & Rate of transition from $ E\rightarrow I $ & $[ 0,\infty) $\\
			$ \mu $ & Rate of transition from $ I\rightarrow R $ & $[ 0,\infty) $\\
			$ \rho $ & Mean case detection probability & $[ 0,1] $\\
			$ \phi $ & Neg. binom. overdispersion parameter & $[ 0,\infty) $ \\
			$ P_{eff} $ & Effective population size & $[0,N]$\\
			\hline
		\end{tabular}
	\end{sidewaystable}
	
	\begin{sidewaystable}[htbp]
		\caption{SEIR model parameter and their interpretation on a possible set of estimation scales.}
		\label{tab:seir_params_est2}
		\footnotesize
		\centering
		\begin{tabular}{clc}
			\hline
			\textbf{Parameter} & \textbf{Interpretation} & \textbf{Domain}\\
			\hline
			$\log(1000\alpha)$ & \multirow{1}{*}{Log effective number of additional infecteds per  1000 foreign infecteds} & $(-\infty,\infty) $ \\
			$ \log(R_{adj} - 1) = \log(\beta P_{eff}/\mu - 1) $ & \multirow{1}{*}{Log basic reproductive number for native index case and $ R_{adj} > 1 $.} & $(-\infty,\infty) $\\
			$ \log(1/\omega) $ & Log mean latent period duration & $(-\infty,\infty) $\\
			$ \log(1/\mu) $ & Log mean infectious period duration & $ (-\infty,\infty) $\\
			$ \logit(\rho) $ & Logit mean case detection probability & $(-\infty,\infty) $\\
			$ \log(\phi) $ & Log negative binomial overdispersion parameter & $ (-\infty,\infty) $ \\
			$ \log(P_{eff}) $ & Log effective population size & $(-\infty,\log(P))$\\
			\hline
		\end{tabular}
	\end{sidewaystable}
	
	\begin{sidewaystable}[htbp]
		\caption{SEIR model parameters and their interpretation on a possible set of estimation scales.}
		\label{tab:seir_params_est3}
		\footnotesize
		\centering
		\begin{tabular}{clc}
			\hline
			\textbf{Parameter} & \textbf{Interpretation} & \textbf{Support}\\
			\hline
			$\log(1000\alpha)$ & Log effective number of additional infecteds per  1000 infecteds outside the population & $(-\infty,\infty) $ \\
			$ \log(R_{adj} - 1) + \log(\rho P_{eff}) $ & Log adjusted reproductive number for native index case  with offset, and $ R_{adj} > 1 $ & $(-\infty,\infty) $\\
			$ \log(\omega/\mu) $ & Log ratio of mean latent to infectious period durations & $(-\infty,\infty) $\\
			$ \log(1/\mu) $ & Log mean infectious period duration & $ (-\infty,\infty) $\\
			$ \logit(\rho) $ & Logit mean case detection probability & $(-\infty,\infty) $\\
			$ \log(\phi) $ & Log negative binomial overdispersion parameter & $ (-\infty,\infty) $ \\
			$ \log(\rho P_{eff}) $ & Log mean case detection rate & $(-\infty,\log(P))$\\
			\hline
		\end{tabular}
	\end{sidewaystable}
	
	\newpage
	\section{Web Appendix F: Identifiability when Estimating the Effective Population Size}
	\label{sec:effpop_identifiability}
	
	When the scale of an outbreak is small relative to the population size, it may be unreasonable to assume that the entire population mixes homogeneously and participates in propagating the epidemic. An alternative is to split the population into two sub--populations, one that is effectively removed from infectious contact, and another at--risk sub--population within which infectious contacts  arise via homogeneous mixing of infectious and susceptible individuals. For example, in the case of the SIR model, we split susceptibles into two compartments, $ S^R $ and $ S^E $, where $ S^R $ is a susceptible sub--population that is effectively removed, and $ S^E $ is a susceptible population that may become exposed. In this model, the \textit{effective population size} is $ P_{eff} = S^E + I + R $, and is interpreted as the size of the population at risk of infection. The transmission model is otherwise constructed in the same way as the canonical SIR model, except with $ S^E $ replacing $ S $. 
	
	The effective population size and the mean case detection probability are weakly identifiable parameters in that we require prior information about their scales to disentangle their effects. To see why this is so, note that $ \rho $ and $ P_{eff} $ enter into the complete data likelihood, (\ref{eqn:complete_data_likelihood}), through the priors and through surveillance models that have means of the form, $ \rho\Delta N_{SI}(t_\ell) $. The incidence here should be understood as an increment in compartment concentrations scaled by the effective population size, since the LNA is a density dependent process \citep{komorowski2009,wilkinson2011stochastic,fearnhead2014}. Thus, the surveillance models have means of the form, $ \rho\left (N^\prime_{S^SI}(t_{\ell}) - N^\prime_{S^SI}(t_{\ell-1})\right )P_{eff} $, where $ \bN^\prime = \bN/P_{eff} $ is the equivalent representation of the LNA in terms of compartment concentrations. 
	
	Absent prior information, we will have difficulty identifying both the case detection probability and effective population size. Despite this, there are a few reasons that $ \rho $ and $ P_{eff} $ might be weakly identifiable, rather than completely unidentifiable. First, certain stochastic aspects of the outbreak, such as the probability of a major outbreak and persistence of transmission, depend on the population size. Furthermore, the scale of the observed incidence and the observed outbreak duration are informative about the minimal effective population size. Therefore, the fact that we observe part of the outbreak is itself informative. Finally, a rough estimate the true population size is typically available, providing an upper bound on the effective population size. By extension, the latent epidemic process is also weakly identifiable in models where the effective population size and case detection probability are estimated.
	
	In contrast, the effective detectable population at risk, $ \rho\times P_{eff} $, along with parameters governing the outbreak dynamics, are directly informed by the data and remain identifiable. It might seem paradoxical that we can infer the dynamics of an outbreak when we are unable to estimate the latent process. To understand why this is so, note that a SEM can be rewritten in terms of concentrations by dividing the compartment counts by the population size, yielding the so--called ``true mass--action model". The dynamics of this equivalent model, expressed, for example, by the basic reproductive number $ R0 $ and recovery rate for an SIR model, are known to be independent of the population size \citep{dejong1995does}. Combinations of model parameters yield latent paths, $ \bN^\prime $, expressed in terms of increments in concentrations, and are weighted in the posterior proportionally (modulo the prior) to the likelihood of scaled paths, $ \rho P_{eff}\bN^\prime $. The temporal nature of the data is important here because the curvature of scaled paths should roughly match that of the data. Thus, we are leveraging curvature in the data to make inferences, not just the pointwise surveillance probabilities. 
	
	We check that $ \rho\times P_{eff} $ and the parameters governing the outbreak dynamics are identifiable via a simple simulation. We drew 500 sets of parameters from the priors given in Table \ref{tab:effpop_coverage_settings}, and simulated an outbreak and a dataset for each set of parameters. The models were fit via the LNA under the same priors from which the parameters were drawn using the MCMC procedure used to fit models for the main coverage simulation (described in the first section of \hyperref[sec:lna_coverage_setup]{Web Appendix C}). The results are summarized in Table \ref{tab:effpop_coverage_results}. The nominal coverage rates for all model parameters is approximately correct. However, the relative widths of the posterior credible intervals for the case detection rate are substantially narrower than the relative widths of the credible intervals for $ \rho $ and $ P_{eff} $ vis--a--vis their prior intervals. This suggests that the data are informative about $ \rho\times P_{eff} $. The priors for $ \rho $ and $ P_{eff} $ are not completely flat, and the scale of the observed counts is itself informative about $ P_{eff} $. Therefore, we still expect, and observe, some contraction in the posteriors for $ \rho $ and $ P_{eff} $, individually.
	
	\begin{table}[htbp]
		\caption{Parameters and priors used in simulating 500 SIR outbreaks where the effective population size was a parameter in the model. The true population size was 100,000. Each outbreak was simulated from a MJP with SIR dynamics. The observed incidence was a negative binomial sample of the true incidence.}
		\label{tab:effpop_coverage_settings}
		\small\centering
		\begin{tabular}{cllr}
			\hline
			\textbf{Parameter} & \textbf{Interpretation} & \textbf{Prior} & \textbf{Median (95\% Interval)} \\ \hline
			$ R0-1 $ & Basic reproduction \# - 1 & LogNormal(0, 0.5) & $ \implies R0 = $ 2.00 (1.38, 3.66) \\ 
			$ 1/\mu $ & Mean infectious period & LogNormal(0.7, 0.35)& 2.01 (1.01, 4.00) \\
			$ \rho / (1-\rho) $ & Odds of case detection & LogNormal(0, 1.4) & $ \implies \rho =$ 0.5 (0.06, 0.94) \\
			$ \phi $ & Neg.Binom. overdispersion & Exponential(0.1) & 6.93 (0.25, 36.89)\\
			$ P_{eff} $ & Effective population size & Unif(5000, 50000) & 27500 (6125, 48875)\\
			\hline
		\end{tabular}\vspace{0.25in}
	\end{table}
	
	\begin{table}[htbp]
		\caption{Results for the models fit to the outbreaks simulated under SIR dynamics with random effective population sizes. Reported are the nominal coverage rates of 95\% credible intervals, the median (95\% CI) of the posterior median deviations (PMD), credible interval widths (CIW), and credible interval widths relative to the 95\% prior interval widths (Rel.CIW). The relative widths of the credible intervals for $ \rho\times P_{eff} $ are computed with respect to the induced prior resulting from the marginal priors for $ \rho $ and $ P_{eff} $.}
		\label{tab:effpop_coverage_results}
		\centering
		\small
		\begin{tabular}{ccccc}
			\hline
			\textbf{Parameter} & \textbf{Coverage} & \textbf{PMD} & \textbf{CIW} & \textbf{Rel.CIW} \\ 
			\hline
			$ R_0 $ & 0.94 & -0.02 (-0.9, 0.58) & 1.1 (0.47, 2.43) & 0.48 (0.21, 1.06) \\ 
			$ \mu $ & 0.95 & 0 (-0.36, 0.21) & 0.49 (0.29, 0.85) & 0.66 (0.4, 1.16) \\ 
			$ \rho $ & 0.95 & -0.01 (-0.36, 0.24) & 0.55 (0.14, 0.74) & 0.62 (0.16, 0.84) \\ 
			$ P_{eff} $ & 0.96 & 800 (-18900, 16100) & 32200 (14600, 41100) & 0.75 (0.34, 0.96) \\ 
			$ \rho\times P_{eff} $& 0.93 & 35 (-9000, 4200) & 6750 (640, 26750) & 0.18 (0.02, 0.72) \\ 
			$ \phi $ & 0.95 & 0.04 (-13.49, 8.95) & 9.1 (0.31, 46.85) & 0.25 (0.01, 1.28) \\ 
			\hline
		\end{tabular}\vspace{0.25in}
	\end{table}
	
	\newpage
	\section{Web Appendix G: Modeling the Spread of Ebola in West Africa}
	\label{sec:ebola_mods}
	
	We modeled the spread of Ebola in Guinea, Liberia, and Sierra Leone under SEIR transmission dynamics within each country, incorporating cross--border transmission via virtual migration of infectious individuals. The model was first fit to a dataset simulated from the true model under known parameters, and then to incidence data from the 2013--2015 West Africa outbreak, each time using both the LNA and ODE approximation computation. The priors and model fitting procedures were largely the same for each dataset and for each approximation. Hence, the following sections apply to all Ebola models. 
	
	\subsection{Model Parameters and Rates of State Transition}
	\label{subsec:ebola_mod_pars_rates}
	
	\begin{table}[htbp]
		\caption{Parameters and their interpretations. Subscripts, $ A,B $, indicate countries. All parameters governing rates of state transition for Liberia and Sierra Leone are zero until three weeks prior to the first detected case, when transmission was assumed to commence in each country.} 
		\label{tab:ebola_mod_pars}
		\centering
		\begin{tabular}{clc}
			\hline \textbf{Parameter} & \textbf{Interpretation} & \textbf{State Transition} \\
			\hline
			$ \beta_A(t) $ & Per--contact rate of transmission within country $ A $. & $ S^{E}_A \rightarrow E_A $\\
			$ \alpha_{AB}(t) $ & Per--contact rate of transmission from country $ A $ to $ B $. & $ S^{E}_A \rightarrow E_A $ \\
			$ \omega_A(t) $ & Rate at which latent individuals become infectious. & $ E_A \rightarrow I_A $ \\
			$ \mu_A(t) $ & Rate at which infectious individuals recover. & $ I_A \rightarrow R_A $ \\
			$ P_{eff,A} $& Effective population size. & ---\\
			$ \rho_A $ & Mean case detection rate. & --- \\
			$ \phi_A $ & Negative binomial overdispersion. & ---\\
			\hline
		\end{tabular}
	\end{table}
	
	\begin{table}[htbp]
		\caption{Rates of state transition. Subscripts for rates indicate model compartments and superscripts indicate countries, while subscripts for compartments and parameters indicate countries. All rates of state transition for Liberia and Sierra Leone are zero until three weeks prior to the first detected case, when transmission was assumed to commence in each country. } 
		\label{tab:ebola_mod_rates}
		\centering
		\begin{tabular}{cc}
			\hline \textbf{Rate} & \textbf{State Transition} \\
			\hline
			$ \lambda^{A}_{S^EE}(t) = \beta_A(t)\left (I_A + \alpha_{BA}(t)I_B + \alpha_{CA}(t)I_C\right )S^E_A $ & $ S^E_A\rightarrow E_A $\\
			$ \lambda^{A}_{EI}(t) = \omega_A(t)E_A $ & $ E_A\rightarrow I_A $ \\
			$ \lambda^A_{IR}(t) = \mu_A(t)I_A $ & $ I_A \rightarrow R_A $\\
			\hline
		\end{tabular}
	\end{table}
	
	\subsection{Priors for Model Parameters}
	\label{subsec:ebola_mod_priors}
	
	\nocite{chowell2014transmission,chretien2015mathematical,coltart2017ebola,scarpino2014epidemiological,dalziel2018unreported,velasquez2015time,dudas2017virus}
	
	\begin{sidewaystable}[htbp]
		\caption{Parameters for the West Africa Ebola outbreak model, prior distributions, 95\% prior intervals, and references that informed the choice of priors. Subscripts, $ G,L,S, $ indicate specific countries, or generic countries $ A,B $ if a prior is common to several parameters. Adjusted reproduction numbers are defined with respect to the effective population size as $ R_{adj} = \beta P_{eff} /\mu $. The mean hyper--parameters of priors for effective population sizes given in the table were used in modeling the West Africa outbreak. The mean hyper--parameters used in the simulated data example were 9.8, 10.5, and 10.6 for Guinea, Liberia, and Sierra Leone, respectively. The negative binomial overdispersion rate hyper--parameter in the simulated data example was also set to 1, reflecting that we expected less overdispersion in the setting where the model was not possibly misspecified.}
		\label{tab:ebola_priors}
		\scriptsize
		\centering
		\begin{tabular}{lllrr}
			\hline
			\textbf{Parameter} &  \textbf{Interpretation} & \textbf{Prior} & \textbf{Med. (90\% Prob. Interval)} & \textbf{References} \\ \hline
			$ R_{adj,A} -1 $ & Adjusted reproduction \#-1 &  LogNormal(log(0.5), $ 1.08^2 $) & $ \implies R_{adj,A} = 1.50 (1.08, 3.95)$ & Chowell (2014); Chretien (2015); Coltart (2017) \\
			$ \frac{\alpha_{AB}P_{eff,B}}{\mu_A}$& 			 Adjusted reproduction \#, $ A\rightarrow B $&Exponential(rate=40)& $ R_{adj,AB} = 0.017\ (0.001, 0.075) $& Low cross--border transmission; Dudas (2017)\\ 
			$ \omega_A/\mu_A$ & Relative latent vs. infectious period & LogNormal(0.0, 0.3$ ^2 $) & $ (1/\mu_A)\big/(1/\omega_A) $ = 1.00 (0.56, 1.80) & Chowell (2014); Velasquez (2015) \\
			$ 1/\mu_A $ & Infectious period duration &  LogNormal(0, 0.3$ ^2 $) & $ 7/\mu_A = $ 7 (3.9, 11.5) & Chowell (2014); Velasquez (2015) \\
			$ P_{eff,G} $ & Effective population size & LogNormal(9.6, 0.62$ ^2 $)& $ P_{eff,G}  = 14765\  (4380,\ 40938) $ & Scale of counts, see discussion below \\
			$ P_{eff,L} $ & Effective population size & LogNormal(9.9, 0.62$ ^2 $)& $ P_{eff,L}  = 19930\  (5912,\ 55260) $ & Scale of counts, see discussion below \\
			$ P_{eff,S} $ & Effective population size & LogNormal(10.7, 0.62$ ^2 $)& $ P_{eff,S}  = 44356\ (13158,\ 122984) $ & Scale of counts, see discussion below \\
			$ \rho_A $ &  Mean case detection prob. & LogitNormal(0.85, 0.75$ ^2 $) & $ \implies \rho_A = 0.7, (0.35, 0.89)$ & Very high and very low $ \rho $ unlikely; Scarpino (2014); Dalziel (2018)\\
			$ 1/\sqrt{\phi_A}$ &  Neg.Binomial overdispersion & Exponential(0.69) & $ \implies \ = 0.69 (0.05, 3)$ & --- \\
			\hline
		\end{tabular}
	\end{sidewaystable}
	
	\begin{sidewaystable}[htpb]
		\caption{Estimation scales and interpretations for Ebola model parameters. Subscripts, $ A,B $ indicate countries.}
		\label{tab:ebola_est_scales}
		\scriptsize
		\centering
		\begin{tabular}{clr}
			\hline
			\textbf{Parameter} &  \textbf{Interpretation} & \textbf{Support} \\ \hline
			$ \log(\beta_AP_{eff,A}/\mu_A-1) + \log(P_{eff,A}\rho_A) $ & Log adjusted effective endogenous reproduction \# - 1, offset by the magnitude of the observed outbreak & $ (-\infty, \infty) $ \\
			$ \log(\alpha_{AB}P_{eff,B}/\mu_A) $ & Log adjusted effective exogenous reproduction \#, $ A\rightarrow B $ & $ (-\infty, \infty) $ \\
			$ \log(\omega_A/\mu_A) $ & Log relative duration, latent vs. infectious period & $ (-\infty,\infty) $\\
			$ \log(1/\mu_A) $ & Log infectious period duration & $ (-\infty,\infty) $\\
			$ \log(P_{eff,A}\rho_A) $ & Log scale of detected outbreak & $ (-\infty,\infty) $ \\
			$ \logit(\rho_A) $ & Log odds of case detection & $ (-\infty,\infty) $ \\
			$ \log(1/\sqrt(\phi_A)) $ & Log overdispersion & $ (-\infty,\infty) $\\
			\hline
		\end{tabular}
	\end{sidewaystable}

	\subsection{Priors for Effective Population Sizes and Initial Compartment Counts}
	\label{subsec:effpop_initdist_priors}
	
	At the time when transmission was assumed to commence in each country, susceptible individuals in the population were considered to either be geographic or social proximate to the transmission process, in which case they could possibly become exposed, or were detached from the transmission process. We separate the $ S $ compartment into a compartment for individuals who are susceptible and connected to exposure, $ S^E $, and individuals who are susceptible but effectively removed, $ S^R $. The fraction of the initially susceptible population that could possibly become exposed corresponds the effective population size. We identify the scale of the prior for $ S^E $ by matching the total number of cases that were detected in each country to the number of cases that we would expect to detect given the effective population size under deterministic outbreak with SEIR dynamics. 
	
	The final size relation for the SEIR model \citep{miller2012note} relates the fraction of the population that becomes infected, $ \pi $, to the basic reproduction number via: 
	$$(1-\pi) = \exp^{-\pi R_0}.$$
	Tables \ref{tab:ebola_westafr_effpop_scenarios} and \ref{tab:ebola_synth_effpop_scenarios} give the effective fraction of the population that could possibly become exposed in each country in order for the expected number of detected cases under deterministic SEIR dynamics to match the observed number of cases under a range of values for the adjusted reproduction number, $ R_{adj} $, and mean case detection rate, $ \rho $. 
	
	In our Ebola models, we used a multivariate normal approximation to a multinomial distribution for the initial distribution of individuals. The hyperparameter for country $ A $, was set to $$ \balpha_A = (S_A = N_A - 30, E_A = 15, I_A = 10, R_A = 5), $$ and we compute the effective number of susceptibles as $ S^E_A = S - S_A^R$. The effective population size is $ P_{eff,A} = S_A^E + E_A + I_A + R_A $.
	
	\begin{table}[htbp]
		\caption{Rough estimates of effective population sizes under different reproduction numbers and detection rates that are needed to match the observed case counts from the West Africa Ebola outbreak to expected counts of detected cases under deterministic SEIR dynamics.}
		\label{tab:ebola_westafr_effpop_scenarios}
		\centering
		\begin{tabular}{lccc}
			\hline
			& \multicolumn{3}{c}{\textbf{Detection rate} ($ \rho $)}\\
			\cmidrule{2-4}$ \mathbf{R_0} $&  0.4 & 0.6 & 0.8\\
			\hline 
			\multicolumn{4}{c}{\textit{Guinea}}\\
			1.25 & 14400 & 9560 & 7200\\
			1.5 & 21700 & 14500 & 10900\\
			1.75 & 31500 & 21000 & 15800\\
			\hline
			\multicolumn{4}{c}{\textit{Liberia}}\\
			1.25 & 19800 & 13200 & 9940\\
			1.5 & 29900 & 19900 & 15000\\
			1.75 & 43500 & 29000 & 21700\\
			\hline
			\multicolumn{4}{c}{\textit{Sierra Leone}}\\
			1.25 & 45000 & 30000 & 22500\\
			1.5 & 67800 & 45200 & 33900\\
			1.75 & 98500 & 65700 & 49200\\
			\hline
		\end{tabular}
	\end{table}
	
	\begin{table}[htbp]
		\caption{Rough estimates of effective population sizes under different reproduction numbers and detection rates that are needed to match the observed case counts from a simulated Ebola outbreak to expected counts of detected cases under deterministic SEIR dynamics.}
		\label{tab:ebola_synth_effpop_scenarios}
		\centering
		\begin{tabular}{lccc}
			\hline
			& \multicolumn{3}{c}{\textbf{Detection rate} ($ \rho $)}\\
			\cmidrule{2-4}$ \mathbf{R_0} $&  0.4 & 0.6 & 0.8\\
			\hline 
			\multicolumn{4}{c}{\textit{Guinea}}\\
			1.25 & 17600 & 11700 & 8850\\
			1.5 & 26500 & 17700 & 13200\\
			1.75 & 38500 & 25700 & 19200\\
			\hline
			\multicolumn{4}{c}{\textit{Liberia}}\\
			1.25 & 37000 & 24700 & 18500\\
			1.5 & 55800 & 37200 & 27900\\
			1.75 & 81000 & 54000 & 40500\\
			\hline
			\multicolumn{4}{c}{\textit{Sierra Leone}}\\
			1.25 & 40500 & 27000 & 20200\\
			1.5 & 60900 & 40600 & 30500\\
			1.75 & 88500 & 59000 & 44200\\
			\hline
		\end{tabular}
	\end{table}
	
	\begin{table}[htbp]
		\caption{Priors for the initial compartment volumes at the times when transmission was assumed to commence in Guinea, Liberia, and Sierra Leone. The initial compartment volumes for each country are assigned independent truncated multivariate normal priors (\hyperref[subsec:lna_init_volumes]{Web Appendix B}). If the population size for country A is $ N_A $, and the initial state probability is denoted $ \bp_{0,A} = \bX_{0,A}/N_A  = (S_{0,A}, E_{0,A}, I_{0,A}, R_{0,A})/N_A$, the prior is a truncated multivariate normal approximation of a multinomial distribution with mean $ N_A\bp_{0,A}$ and covariance $ N_A(\bP_{0,A} - \bp_{0,A}\bp_{0,A}^T) $.} 
		\label{tab:ebola_joint_initdist_priors}
		\centering
		\begin{tabular}{lc}
			\hline \textbf{Country} & \textbf{Prior mean initial volumes} ($ \bX_0 $) \\
			\hline
			Guinea & ($ 11.8\times10^6 - 30$, 15, 10, 5) \\
			Liberia & ($ 4.4\times10^6 - 30$, 15, 10, 5) \\
			Sierra Leone & ($ 7.1\times10^6 - 30$, 15, 10, 5) \\
			\hline
		\end{tabular}
	\end{table}
	
	\subsection{Ebola Model MCMC Details for Models Fit via the LNA and ODE}
	\label{subsec:ebola_synth_lna}
	
	The model fitting procedure and priors were the same for both models, and were also the same for models fit via the LNA and ODE approximations. We ran five chains for each model, initialized at random parameter values, for 150,000 iterations per chain, the first 50,000 of which consisted of an adaptive tuning phase. The model parameters, not including the initial compartment volumes, were jointly updated via MVNSS (\hyperref[sec:mvnss]{Web Appendix C}). The empirical covariance for the MVNSS algorithm was adapted over the first 100,000 iterations using the gain factor sequence, $\gamma_n = 0.5(1 + 0.01n)^{-0.9}$. The contribution of isotropic Gaussian noise to the proposal was initialized at 0.001 and reduced throughout the adaptation phase according to the sequence $ \iota_n = 0.001(1 + 0.01n)^{-0.99} $. The covariance matrix was blocked by country, treating parameters belonging to different countries as independent in the MCMC proposal kernel. Migration parameters were blocked with parameters corresponding to the destination country. The initial compartment volumes were jointly updated in a separate EllipSS update than the rest of the latent LNA paths. The MCMC alternated between one ElliptSS and one MVNSS updates per MCMC iteration. The ElliptSS bracket width was reset after the first 5,000 MCMC iterations to $ \omega = 2\sqrt{2\log(10)}\sigma_{ElliptSS}$, where $ \sigma_{ElliptSS} $ was the standard deviation of the accepted angles over the initial iterations. The MCMC estimation scales for each country were parameterized as in Table \ref{tab:seir_params_est3} with the one addition being the log ratio of adjusted reproductive numbers, subtract one, for Guinea, $ \log\left ((R_{eff,G}^{(2)}-1)/(R_{eff,G}^{(1)}-1)\right ) $. Convergence was assessed visually by inspection of traceplots of posterior samples, and via potential scale reduction factors (PSRFs) \citep{brooks1998general} computed via the \texttt{coda} R package \citep{codapackage}. The LNA models each took roughly three days to run while run times for ODE models took between 1--2 hours. 
	
	\subsection{Ebola Model MCMC Details for Computation with Particle Marginal Metropolis--Hastings}
	\label{subsec:ebola_synth_pmmh_mcmcdetails}
	
	Computation using the MMTL approximation within PMMH was done using the \texttt{pomp} R package \cite{pompjss}. We used 1,000 particles in the PMMH algorithm. The time step for MMTL was set to 1/7, which, for example, corresponds to $ \tau $--leaping over one day increments given weekly incidence data. The MCMC was initialized and parameterized in the same way as LNA and ODE models. Parameter updates were made using a multivariate Metropolis algorithm with Gaussian proposals. The proposal covariance matrix was adapted over an initial run of 50,000 iterations, with adaptation starting after the first 100 iterations were completed. The gain factor sequence implemented in the package is of the form $ \gamma_n = n^\alpha $, where the cooling term, $ \alpha $, was set to 0.99975. MCMC was run for 300,000 iterations following the adaptive MCMC phase. The posterior samples from all five MCMC chains were combined after discarding the initial samples from the adaptation phase. We were unable to obtain a convergent and well mixing MCMC sampler, despite substantial expenditure of time and effort, even for the dataset simulated under the assumed model. Hence, PMMH was abandoned as a computational strategy in the analysis of data from the West Africa Ebola outbreak. The multivariate PSRF, computed using the \texttt{coda} package in \texttt{R} \citep{codapackage}, was 1.89 and four of the 24 parameters had univariate PSRF values above 1.05, indicating that the posterior had not been adequately explored. 
	
	\subsection{Single-Country Models Fit to Ebola Data From Guinea, Liberia, and Sierra Leone}
	\label{subsec:ebola_single_country_models}
	A typical first step in learning about the overall outbreak dynamics is to separately model the incidence data from each country. This is less challenging than fitting a multi-country model that incorporates cross--border transmission since each model will have fewer parameters and a smaller latent state space. Furthermore iterating through simplified models is helpful in identifying parameterizations that simplify the posterior geometry (see \hyperref[subsec:est_scale_discussion]{Web Appendix B}).
	
	\begin{figure}[htbp]
		\centering
		\includegraphics[width=\linewidth]{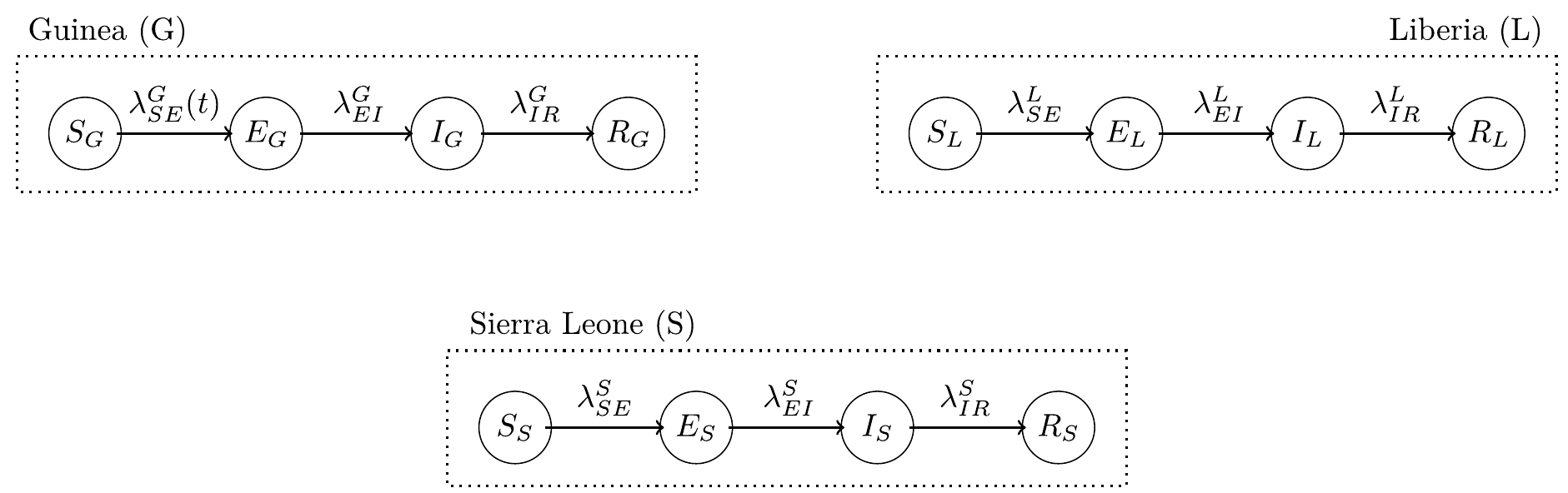}
		\caption[Diagram of single country SEIR models for the Ebola outbreak in West Africa.]{Diagram of state transitions for SEIR models fit to Ebola incidence data from Guinea, Liberia, and Sierra Leone. Dotted boxes denote countries, nodes in circles denote the model compartments: susceptible (S), exposed (E), infectious (I), recovered (R). Compartments  are subscripted with country indicators. The number of susceptible individuals is equal to the effective population size, estimated as a parameter in the model, minus the numbers of exposed, infected, and recovered individuals. Solid lines with arrows indicate stochastic transitions between model compartments, which occur continuously in time. Rates at which individuals transition between compartments are denoted by $ \lambda $ and are subscripted by compartments and superscripted by countries, e.g., $ \lambda_{SE}^L $ is the rate at which susceptible individuals become exposed in Liberia. Transmission in Liberia and Sierra Leone was assumed to commence at 10 and 19 weeks, respectively. Full expressions for the rates are given in Table \ref{tab:ebola_single_country_rates}.}
		\label{fig:ebola_single_diag}
	\end{figure}
	
	We fit separate SEIR models, diagrammed in Figure \ref{fig:ebola_single_diag}, to the incidence data from each country using both the LNA and ODE approximations. Transmission was modeled in Liberia beginning March 2, 2014, and in Sierra Leone from May 4, 2014, corresponding to three weeks prior to the first confirmed or probable cases in those countries. The force of infection in each country included a constant term for infectious contact from outside the population, but did not explicitly link exogenous transmission to the prevalence in other countries. The transition rates for the single-country models are given in Table \ref{tab:ebola_single_country_rates}. To account for the small scale of each outbreak relative to the population size in the country, we estimated the effective population size as a parameter in the model. The number of susceptible individuals was then equal to the effective population size, minus the numbers of exposed, infected, and recovered individuals. To complete the model specification, the observed incidence was modeled as a negative binomial sample of the true incidence in each inter--observation interval, as in (\ref{eqn:incidence_emitprob}).
	
	\begin{table}[htbp]
		\caption{Rates of state transition for single-country models for Ebola transmission. Subscripts for rates indicate model compartments. All rates of state transition for Liberia and Sierra Leone are zero until three weeks prior to the first detected case, when transmission was assumed to commence in each country.} 
		\label{tab:ebola_single_country_rates}
		\centering
		\begin{tabular}{cc}
			\hline \textbf{Rate} & \textbf{State Transition} \\
			\hline
			$ \lambda_{S^EE}(t) = \left (\alpha(t) + \beta(t)I\right )S^E $ & $ S^E\rightarrow E $\\
			$ \lambda_{EI}(t) = \omega(t)E $ & $ E\rightarrow I $ \\
			$ \lambda_{IR}(t) = \mu(t)I $ & $ I \rightarrow R $\\
			\hline
		\end{tabular}
	\end{table}
	
	The priors for single-country model parameters were the same as those used in fitting the multi-country model, given in Tables \ref{tab:ebola_priors} and \ref{tab:ebola_joint_initdist_priors}. The exception was the prior for the baseline rate of transmission from outside the population, where for each country we took $ 1000\alpha \sim $Exponential(rate=40), reflecting our assumption that there were probably fewer than a dozen (two dozen under the more diffuse prior regime) transmission events per 1,000 infected individuals outside the country. This was based in part on results published in \cite{dudas2017virus}, who estimated between one--half to two dozen reintroduction events, depending on the country, between April 2014 and May 2015. We also used the same parameterizations for the estimation scales on which our MCMC samplers explored the posteriors (Table \ref{tab:ebola_est_scales}), again with the exception of the baseline rate of exogenous transmission. For each single-country model, we parameterized the estimation scale for this parameter as $\log(1000\alpha)$. All other MCMC details --- number of iterations, adaptation tuning parameters, etc. --- were the same as those used in fitting the multi-country Ebola model. Posterior medians and credible intervals for all parameters are reported in Table \ref{tab:ebola_lna_vs_ode_ests}. 
	
	\subsection{Additional Results and Diagnostics for the Model Fit via the Linear Noise Approximation --- Simulated Data Example}
	\label{subsec:lna_ebola_synth_supp_res}
	
	\begin{sidewaysfigure}
		\centering
		\includegraphics[width=\linewidth]{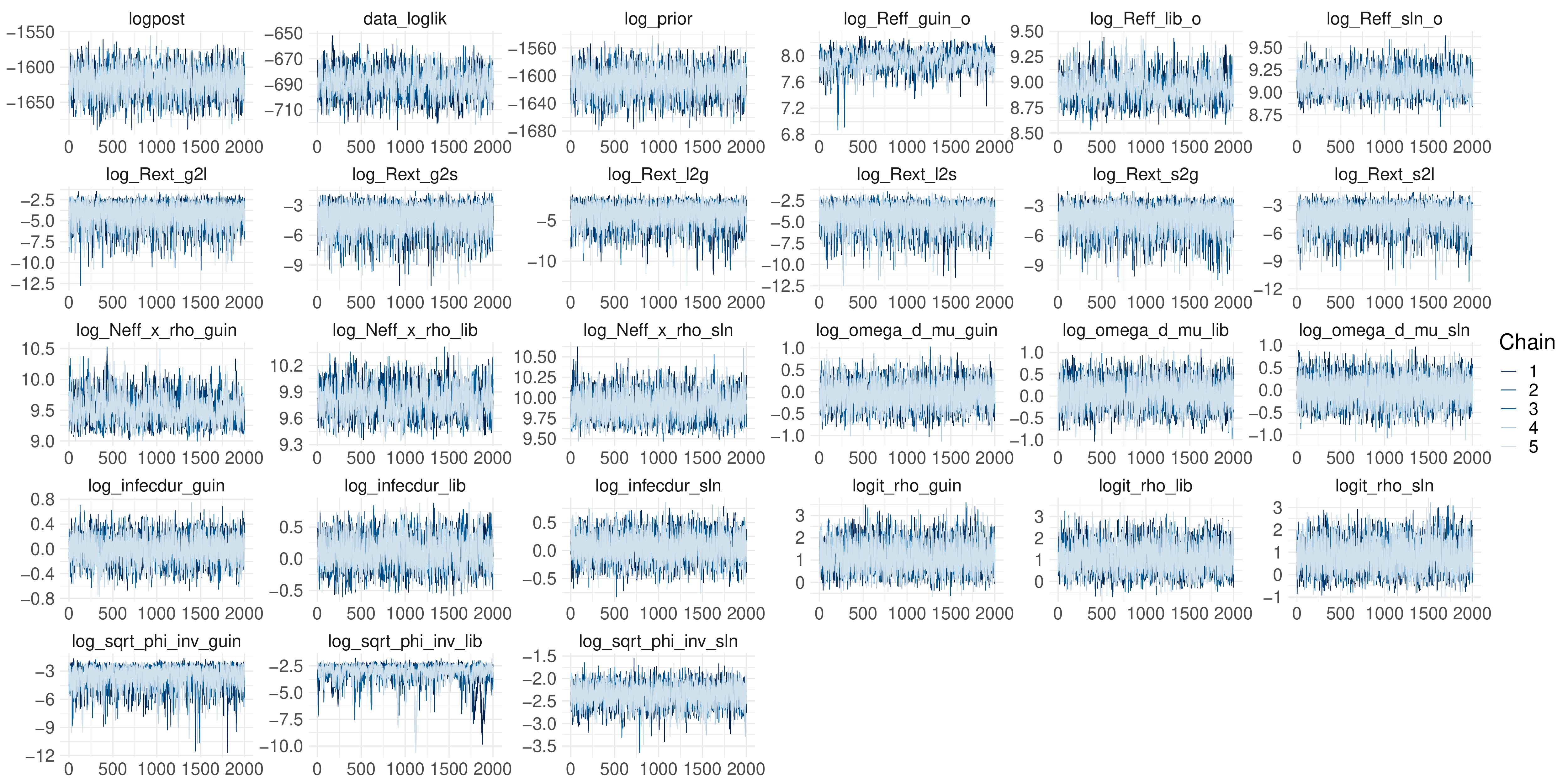}
		\caption{Traceplots for an Ebola LNA model to data from a simulated outbreak in West Africa. Traceplots were thinned to show every 25th MCMC iteration.}
		\label{fig:lna_synth_traces}
	\end{sidewaysfigure}

	\subsection{Additional Results and Diagnostics for the Model Fit via the Ordinary Differential Equation Approximation --- Simulated Data Example}
	\label{subsec:ode_ebola_synth_supp_res}
	
	\begin{sidewaysfigure}
		\centering
		\includegraphics[width=\linewidth]{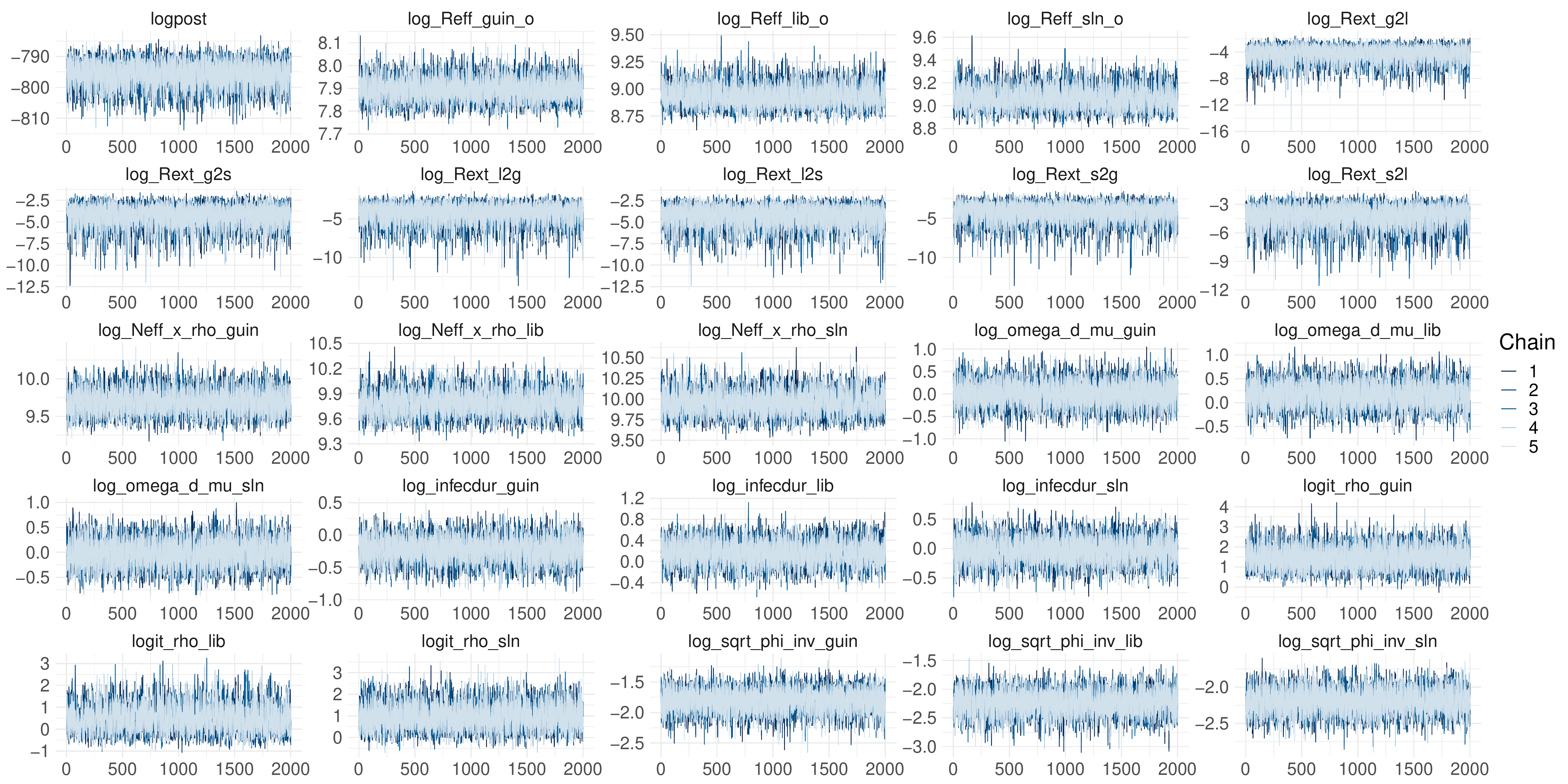}
		\caption{Traceplots for an Ebola ODE model to data from a simulated outbreak in West Africa. Traceplots were thinned to show every 25th MCMC iteration.}
		\label{fig:ode_synth_traces}
	\end{sidewaysfigure}
	
	\subsection{Additional Results for LNA and ODE Models Fit to Data from the West Africa Outbreak}
	\label{subsec:ebola_westafr_supp_res}
	
	\begin{sidewaystable}[ht]
		\caption{Posterior medians (95\% Bayesian credible intervals) of parameters for Ebola models fit to incidence data from the 2013--2015 outbreak in Guinea, Liberia, and Sierra Leone. Country--specific parameters are denoted by subscripts $ G,\ L,  $ and $ S $. Priors are given in Table \ref{tab:ebola_priors}. }
		\label{tab:ebola_lna_vs_ode_ests}
		\centering\scriptsize\def\arraystretch{2}
		\begin{tabular}{llrrrr}
			\hline
			& & \multicolumn{4}{c}{\textbf{Posterior median (95\% BCI)}}\\\cline{3-6}
			&& \multicolumn{2}{c}{\textit{Multi-country models}}& \multicolumn{2}{c}{\textit{Country-specific models}} \\ 
			\cmidrule(r){3-4}\cmidrule(l){5-6}
			\textbf{Parameter} & \textbf{Interpretation} & \multicolumn{1}{c}{LNA}& \multicolumn{1}{c}{ODE} & \multicolumn{1}{c}{LNA} & \multicolumn{1}{c}{ODE}\\\hline
			$ R_{adj}^G = \frac{\beta_G P_{eff,G}}{\mu_G} $& Adjusted basic reproduction number & 1.2 (1.1, 1.5) & 1.4 (1.2, 1.7) & 1.3 (1.1, 1.5) & 1.4 (1.2, 1.7) \\ 
			$ 7/ \omega_G $& Latent period, days & 6.4 (3, 14.5) & 12.2 (6.4, 22.2) & 7.7 (3.6, 16.1)& 12.8 (6.8, 23.4) \\
			$ 7/ \mu_G $& Infectious period, days & 6.4 (3.8, 11.6) & 9.9 (6.2, 16.1) &7.3 (4.3, 12.7)& 10.2 (6.4, 16.3) \\ 
			$ 1000\alpha_{G} $& Baseline rate of exogenous transmission $ \times $ 1,000 & \rule[0.5ex]{0.75in}{0.5pt} & \rule[0.5ex]{0.75in}{0.5pt} & 0.01 (0.0005, 0.07) & 0.01 (0.0006, 0.07) \\
			$ R_{ext}^{LG} = \frac{\alpha_{LG} P_{eff,G}}{\mu_L} $& Adjusted extrinsic reproduction number & 0.02 (0.0006, 0.09) & 0.02 (0.0006, 0.09) &\rule[0.5ex]{0.75in}{0.5pt}&\rule[0.5ex]{0.75in}{0.5pt} \\ 
			$ R_{ext}^{SG} = \frac{\alpha_{SG} P_{eff,G}}{\mu_S} $& Adjusted extrinsic reproduction number & 0.02 (0.0007, 0.09) & 0.02 (0.0007, 0.09) &\rule[0.5ex]{0.75in}{0.5pt}&\rule[0.5ex]{0.75in}{0.5pt} \\ 
			$ P_{eff,G}\times\rho_G $& Scale of detected outbreak & 10700 (6500, 21300) & 7700 (5400, 11400) & 9600 (6100, 17400) & 7500 (5400, 11100) \\
			$ 1/\sqrt{\phi_G} $& Negative binomial overdispersion & 0.4 (0.3, 0.6) & 0.6 (0.5, 0.7) & 0.5 (0.4, 0.6) & 0.6 (0.5, 0.7) \\ 
			\hline
			$ R_{adj}^L = \frac{\beta_L P_{eff,L}}{\mu_L} $& Adjusted basic reproduction number & 1.9 (1.4, 3.2) & 2.2 (1.6, 3.5) & 2 (1.5, 3.1) & 2.1 (1.6, 3.3) \\ 
			$ 7/ \omega_L $& Latent period, days & 7.8 (3.4, 18.6) & 11.2 (4.7, 20.1) & 9.2 (4.4, 18.4) & 10.6 (4.9, 19.2)\\ 
			$ 7/ \mu_L $& Infectious period, days & 7.5 (4.2, 13.7) & 9.2 (5.3, 15.3) & 8.5 (5.1, 14.4) & 9.1 (5.3, 15.2) \\
			$ 1000\alpha_{L} $& Baseline rate of exogenous transmission $ \times $ 1,000 & \rule[0.5ex]{0.75in}{0.5pt} & \rule[0.5ex]{0.75in}{0.5pt} & 0.01 (0.0004, 0.05) & 0.004 (0.0001, 0.02) \\
			$ R_{ext}^{GL} = \frac{\alpha_{GL} P_{eff,L}}{\mu_G} $& Adjusted extrinsic reproduction number & 0.02 (0.0006, 0.09) & 0.02 (0.0007, 0.09) &\rule[0.5ex]{0.75in}{0.5pt}&\rule[0.5ex]{0.75in}{0.5pt} \\ 
			$ R_{ext}^{SL} = \frac{\alpha_{SL} P_{eff,L}}{\mu_S} $& Adjusted extrinsic reproduction number  & 0.02 (0.0006, 0.09) & 0.02 (0.0006, 0.09) & \rule[0.5ex]{0.75in}{0.5pt}& \rule[0.5ex]{0.75in}{0.5pt} \\ 
			$ P_{eff,L}\times\rho_L $& Scale of detected outbreak & 6300 (4800, 9300) & 5600 (4600, 7900) & 5900 (4600, 8200)& 5700 (4400, 7800) \\ 
			$ 1/\sqrt{\phi_L} $& Negative binomial overdispersion & 0.3 (0.2, 0.5) & 0.4 (0.3, 0.6) & 0.4 (0.3, 0.6) & 0.4 (0.3, 0.6)\\
			\hline 
			$ R_{adj}^S = \frac{\beta_S P_{eff,S}}{\mu_S} $& Adjusted basic reproduction number & 1.3 (1.2, 1.4) & 1.4 (1.2, 1.7)& 1.3 (1.2, 1.5)  & 1.5 (1.3, 1.9) \\ 		
			$ 7/ \omega_S $& Latent period, days & 4.2 (2.4, 7.1) & 5.8 (3.2, 9.8) & 5 (2.7, 8.7) & 7.4 (4.1, 12.8) \\ 
			$ 7/ \mu_S $& Infectious period, days & 4.7 (3.1, 7) & 5.9 (3.8, 8.9) & 5.3 (3.5, 8.2) & 7.1 (4.5, 10.9)\\ 
			$ 1000\alpha_{S} $& Baseline rate of exogenous transmission $ \times $ 1,000 & \rule[0.5ex]{0.75in}{0.5pt} & \rule[0.5ex]{0.75in}{0.5pt} & 0.01 (0.0003, 0.05) & 0 (0, 0.1) \\
			$ R_{ext}^{GS} = \frac{\alpha_{GS} P_{eff,S}}{\mu_G} $& Adjusted extrinsic reproduction number & 0.02 (0.0007, 0.09) & 0.02 (0.0007, 0.1) & \rule[0.5ex]{0.75in}{0.5pt} & \rule[0.5ex]{0.75in}{0.5pt} \\ 
			$ R_{ext}^{LS} = \frac{\alpha_{LS} P_{eff,S}}{\mu_L} $& Adjusted extrinsic reproduction number & 0.01 (0.0003, 0.05) & 0.01 (0.0004, 0.08) &\rule[0.5ex]{0.75in}{0.5pt} &\rule[0.5ex]{0.75in}{0.5pt} \\ 		
			$ P_{eff,S}\times\rho_S $& Scale of detected outbreak & 27800 (20600, 40500) & 20300 (17500, 32900) & 25200 (18500, 37200) & 20300 (15300, 27900) \\ 
			$ 1/\sqrt{\phi_S} $& Negative binomial overdispersion & 0.1 (0.1, 0.2) & 0.4 (0.3, 0.5) & 0.3 (0.2, 0.4) & 0.5 (0.4, 0.6) \\ 
			\hline
		\end{tabular}
	\end{sidewaystable}
	
	\begin{figure}[htbp]
		\centering
		\includegraphics[width=\linewidth]{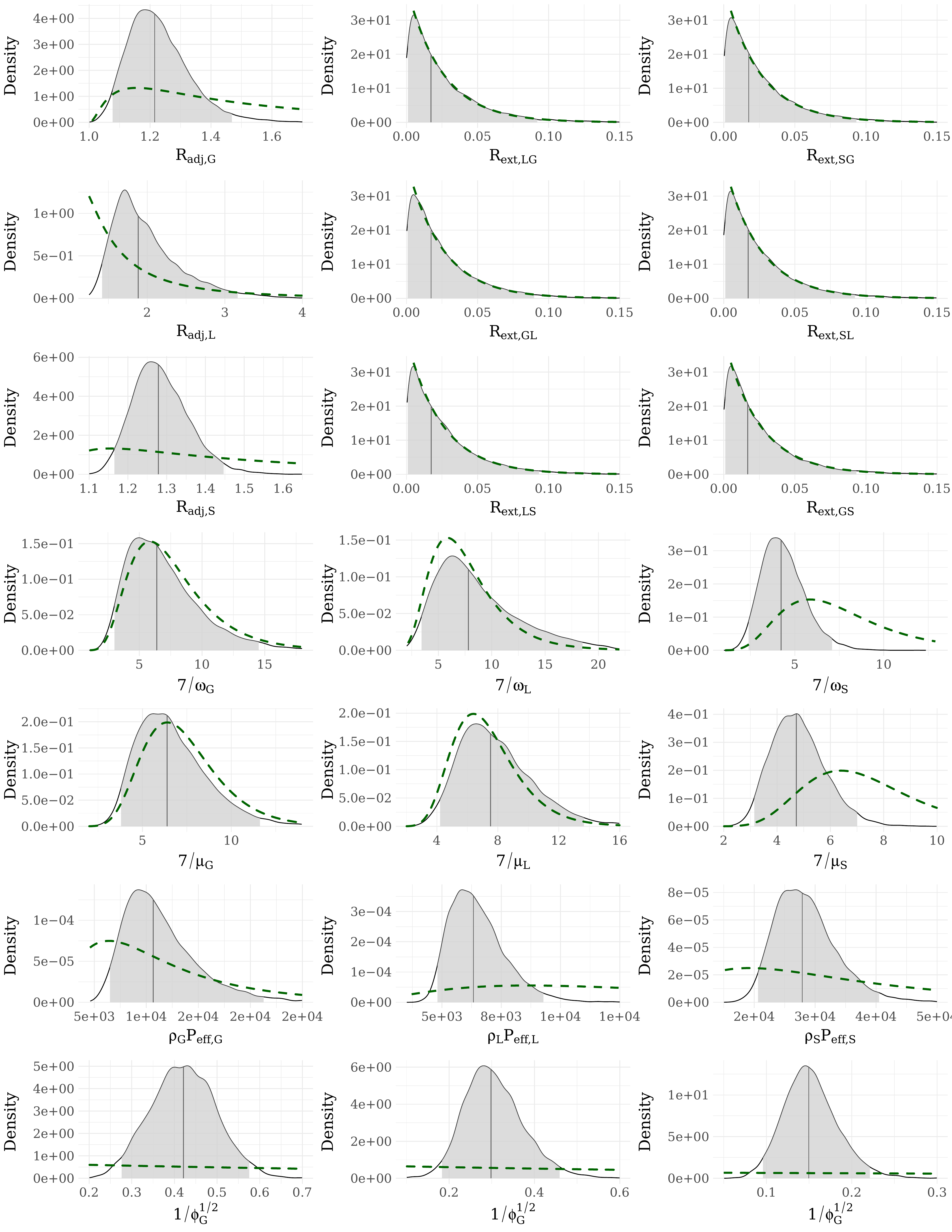}
		\caption{Posterior distributions of LNA model parameters for the model fit to data from the West Africa Ebola outbreak. We show posterior medians (solid gray lines), 95\% Bayesian credible intervals (light gray areas under the posterior densities), prior densities (induced priors for the reporting rate and latent period durations) over the posterior ranges (dashed green curves). Priors and interpretations of parameters are specified in Table \ref{tab:ebola_priors}.}
		\label{fig:ebolawestafrposts}
	\end{figure}
	
	\begin{figure}[htbp]
		\centering
		\includegraphics[width=\linewidth]{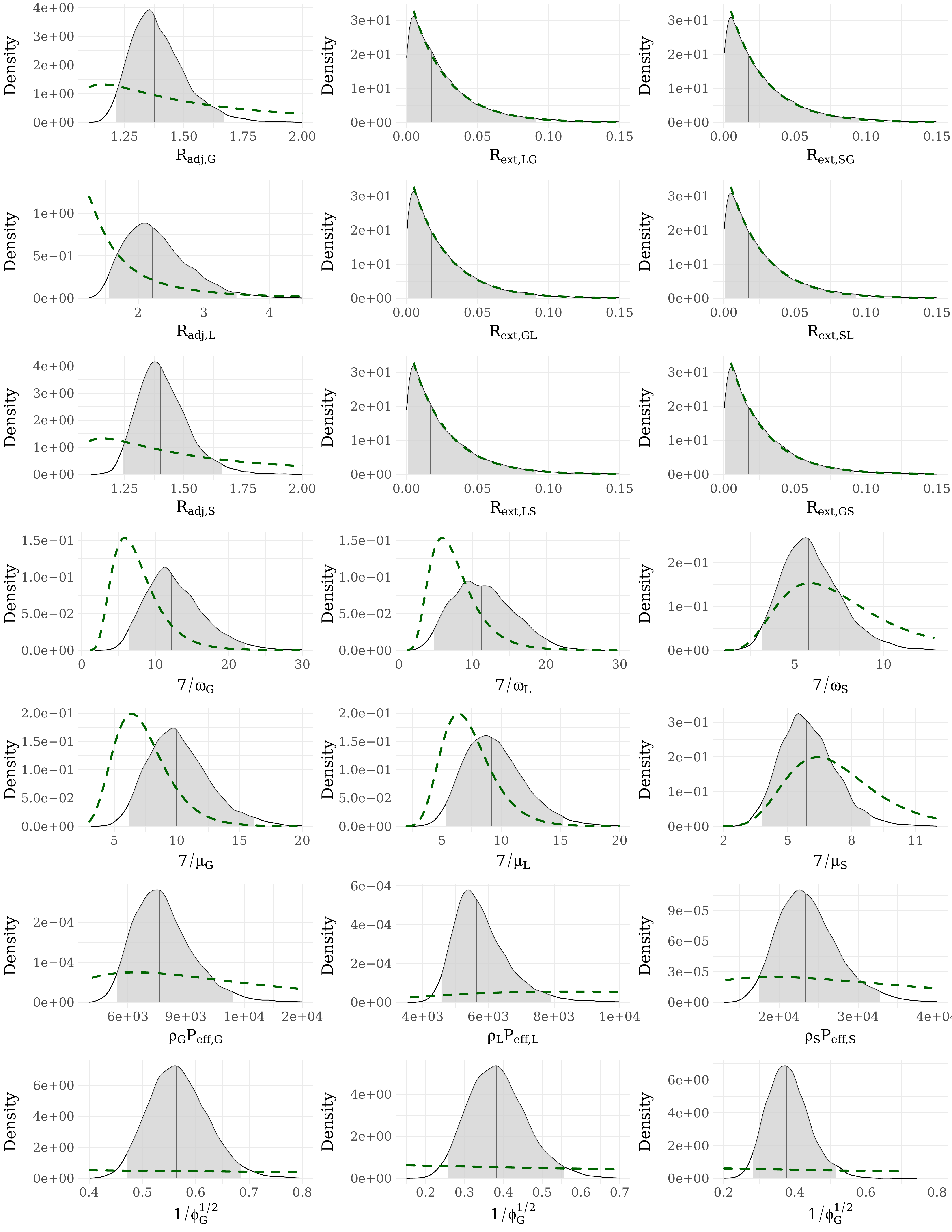}
		\caption{Posterior distributions of ODE model parameters for the model fit to data from the West Africa Ebola outbreak under tight priors. We show posterior medians (solid gray lines), 95\% Bayesian credible intervals (light gray areas under the posterior densities), prior densities (induced priors for the reporting rate and latent period durations) over the posterior ranges (dashed green curves). Priors and interpretations of parameters are specified in Table \ref{tab:ebola_priors}.}
		\label{fig:ebolawestafrpostsode}
	\end{figure}
	
	\begin{table}[htbp]
		\caption{Posterior estimates of initial numbers of exposed and infected individuals for the Ebola models fit to data from the West Africa outbreak. The effective number of susceptibles is equal to the effective population size minus the numbers of exposed, infected, and recovered individuals, but is not reported along with the number of recovered individuals since the effective population size is only weakly identified (\hyperref[sec:effpop_identifiability]{Web Appendix D}). Country--specific parameters are denoted by subscripts $ G,\ L,  $ and $ S $. Priors are given in Table \ref{tab:ebola_joint_initdist_priors}.}
		\label{tab:ebola_joint_initdist_res}
		\centering
		\begin{tabular}{lcc}
			\hline
			\textbf{Parameter} & \textbf{LNA} & \textbf{ODE} \\ 
			\hline
			$ E_{0,G} $& 8.9 (2.1, 17.6) & 12.4 (4.5, 20.2) \\ 
			$ E_{0,L} $& 9 (2, 16.9) & 0.6 (0, 3.3) \\ 
			$ E_{0,S} $& 6.4 (0.8, 14.1) & 14.6 (7.1, 21.5) \\ 
			$ I_{0,G} $& 6.9 (1.4, 13.4) & 7.9 (2, 14.4) \\ 
			$ I_{0,L} $& 6 (1, 12) & 0.3 (0, 1.6) \\ 
			$ I_{0,S} $& 7.3 (1.7, 13.2) & 11.1 (5.5, 16.7) \\
			\hline
		\end{tabular}
	\end{table}
	
	\begin{figure}[htbp]
		\centering
		\includegraphics[width=\linewidth]{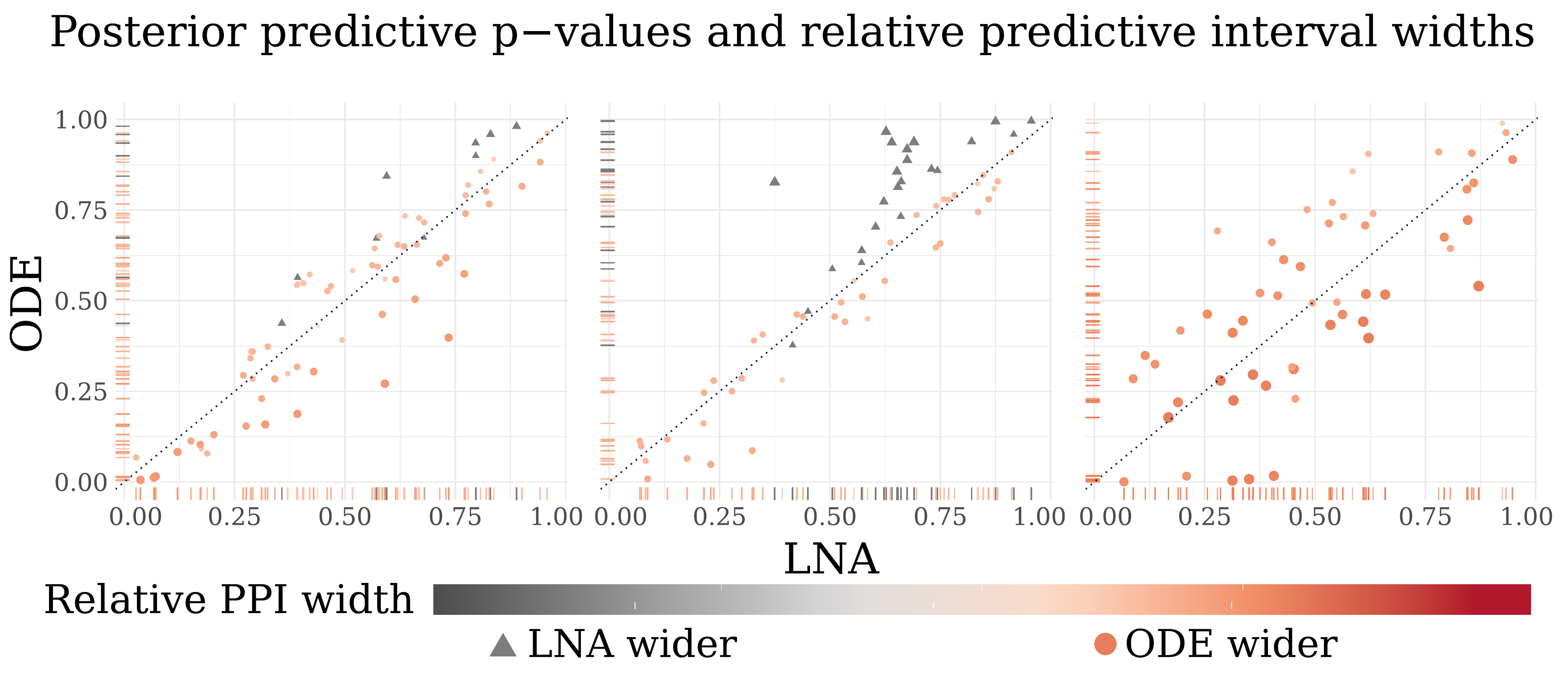}
		\caption{(Bottom panel) Comparison of posterior predictive p-values (PPPs) and relative posterior predictive interval (PPI) widths for LNA and ODE models fit to data from the West Africa Ebola outbreak. Each point corresponds to the observed incidence in a given week. The X--Y coordinates give the PPPs under each model. The size and color of each point corresponds to the relative PPI width, computed as $ (\widehat{\sigma}_{post,\ell}^{ODE} - \widehat{\sigma}_{post,\ell}^{LNA})/\widehat{\sigma}_{post,\ell}^{LNA} $, and the sign of the relative width is further emphasized by the shape of the point. Dots indicate that PPIs for the ODE model are wider, while triangles corresponds to observations for which the PPI produced by the LNA model was wider.}
		\label{fig:ebolappicomp}
	\end{figure}
	
	\begin{table}[htbp]
		\caption{Effective sample sizes and potential scale reduction factors for LNA and ODE Ebola multi-country model parameters. }
		\label{tab:ebola_jointmod_ess_psrfs}
		\centering
		\begin{tabular}{lcccc}
			\hline
			& \multicolumn{2}{c}{\textbf{LNA}} & \multicolumn{2}{c}{\textbf{ODE}}\\
			\cmidrule(r){2-3}\cmidrule(l){4-5}Parameter & ESS & PSRF & ESS & PSRF \\ 
			\hline
			$ \log(\beta_GP_{eff,G}/\mu_G-1) + \log(P_{eff,G}\rho_S) $& 1013 & 1.01 & 6449 & 1.00 \\ 
			$ \log(\beta_LP_{eff,L}/\mu_L-1) + \log(P_{eff,L}\rho_L) $& 559 & 1.02 & 6505 & 1.00 \\ 
			$ \log(\beta_SP_{eff,S}/\mu_A-1) + \log(P_{eff,S}\rho_S) $& 756 & 1.01 & 5604 & 1.00 \\ 
			$ \log(\alpha_{GL}P_{eff,L}/\mu_G) $& 2866 & 1.00 & 8626 & 1.00 \\ 
			$ \log(\alpha_{GS}P_{eff,S}/\mu_G) $& 1406 & 1.01 & 5012 & 1.00 \\ 
			$ \log(\alpha_{LG}P_{eff,G}/\mu_L) $& 3335 & 1.00 & 7964 & 1.00 \\ 
			$ \log(\alpha_{LS}P_{eff,S}/\mu_L) $& 863 & 1.01 & 4172 & 1.00 \\ 
			$ \log(\alpha_{SG}P_{eff,G}/\mu_S) $& 683 & 1.01 & 6304 & 1.00 \\ 
			$ \log(\alpha_{SL}P_{eff,L}/\mu_S) $& 1396 & 1.01 & 3475 & 1.00 \\ 
			$ \log(P_{eff,G}\rho_G) $& 6280 & 1.00 & 6581 & 1.00 \\ 
			$ \log(P_{eff,L}\rho_L) $& 6166 & 1.00 & 9172 & 1.00 \\ 
			$ \log(P_{eff,S}\rho_S) $& 7544 & 1.00 & 9312 & 1.00 \\ 
			$ \log(\omega_G/\mu_G) $& 6002 & 1.00 & 8624 & 1.00 \\ 
			$ \log(\omega_L/\mu_L) $& 7212 & 1.00 & 9266 & 1.00 \\ 
			$ \log(\omega_S/\mu_S) $& 7087 & 1.00 & 5961 & 1.00 \\ 
			$ \log(7 / \mu_G) $& 884 & 1.01 & 3818 & 1.00 \\ 
			$ \log(7 / \mu_L) $& 544 & 1.01 & 5112 & 1.00 \\ 
			$ \log(7 / \mu_S) $& 840 & 1.01 & 2736 & 1.00 \\ 
			$ \log(\rho_G/(1-\rho_G)) $& 1431 & 1.01 & 1696 & 1.00 \\ 
			$ \log(\rho_L/(1-\rho_L)) $& 783 & 1.01 & 2416 & 1.00 \\ 
			$ \log(\rho_S/(1-\rho_S)) $& 3503 & 1.00 & 4041 & 1.00 \\ 
			$ 1/\sqrt{\phi_G} $& 607 & 1.01 & 9627 & 1.00 \\ 
			$ 1/\sqrt{\phi_L} $& 794 & 1.03 & 7436 & 1.00 \\ 
			$ 1/\sqrt{\phi_S} $& 980 & 1.01 & 9384 & 1.00 \\ 
			\hline
		\end{tabular}
	\end{table}

\end{document}